\documentclass[sn-nature]{sn-jnl}
\usepackage{ulem}
\usepackage{graphicx}
\usepackage{multirow}
\usepackage{amsmath,amssymb,amsfonts}
\usepackage{amsthm}
\usepackage{mathrsfs}
\usepackage[title]{appendix}
\usepackage{xcolor}
\usepackage{textcomp}
\usepackage{manyfoot}
\usepackage{booktabs}
\usepackage{algorithm}
\usepackage{algorithmicx}
\usepackage{algpseudocode}
\usepackage{listings}
\usepackage{comment}
\usepackage{float}
\usepackage{xcolor}
\usepackage{lineno}
\usepackage{soul}

\theoremstyle{thmstyleone}

\theoremstyle{thmstyletwo}

\theoremstyle{thmstylethree}

\begin{document}

\title[Article Title]{Low Photon Number Non-Invasive Imaging Through Time-Varying Diffusers}

\author[1,2]{\fnm{Adrian} \sur{Makowski}}\email{adrian.makowski@fuw.edu.pl}

\author[1]{\fnm{Wojciech} \sur{Zwolinski}}\email{w.zwolinski2@uw.edu.pl}
\author[1]{\fnm{Pawel} \sur{Szczypkowski}}\email{pawel.szczypkowski@fuw.edu.pl}

\author[1]{\fnm{Bernard} \sur{Gorzkowski}}\email{bernard.gorzkowski@gmail.com}

\author[2]{\fnm{Sylvain} \sur{Gigan}}\email{sylvain.gigan@lkb.ens.fr}

\author*[1,]{\fnm{Radek} \sur{Lapkiewicz}}\email{radek.lapkiewicz@fuw.edu.pl}

\affil[1]{\orgdiv{Institute of Experimental Physics, Faculty of Physics, University of Warsaw}, \orgaddress{\street{Pasteura 5}, \postcode{02-093} \city{Warsaw}, \country{Poland}}}

\affil[2]{\orgdiv{Laboratoire Kastler Brossel, École normale supérieure (ENS)– Université Paris Sciences \& Lettres (PSL), CNRS, Sorbonne Université, Collège de France}, \orgaddress{\street{24 rue Lhomond}, \city{Paris}, \postcode{75005}, \country{France}}}

\abstract{

Optical imaging plays a crucial role in advancing science and technology, enabling applications in fields ranging from biomedicine to astronomy. However, imaging through scattering media such as biological tissues, fog, or turbulent atmosphere remains a major challenge. Light scattering and absorption in such media make imaging challenging; in the case of time-varying scatterers and low light regime imaging of incoherent objects has not been demonstrated so far. We present the first demonstration of such non-invasive imaging of dim objects hidden behind dynamic scattering layers, obtaining robust reconstruction even at extremely low photon counts per frame. We achieve this by developing a new data-processing approach. In our experiment, we utilize a photon number resolving camera to capture a sequence of frames, containing on average, fewer than one photon per pixel. We validate our approach in microscopy, where we reconstruct images of biological samples stained with standard fluorescent dyes. Beyond microscopy, our approach can be applied in different imaging techniques, such as endoscopy based on multicore fibers or ground-based astronomical observations.}

\maketitle
\newpage
\section{Introduction}\label{sec1}
Light scattering plays a crucial role in determining the appearance of most of the things we see. When scattering occurs multiple times within a medium, such as biological tissues, fog, or clouds, it is called a complex medium. The propagation of light through complex media results in random interference patterns known as speckle patterns\cite{Goodman2007}. For a long time, light scattering in complex media was considered uncontrollable. This fundamental challenge of light scattering affects not only imaging \cite{Yoon2020}, but also poses obstacles in communication, spectroscopy, and other optical and photonic technologies.

Imaging through complex media has long been seen as a formidable challenge. However, a groundbreaking experiment by Vellekoop and Mosk \cite{Vellekoop2007} demonstrated that by controlling wavefronts through spatial light modulators (SLMs), light can be focused through scattering media. 
Subsequent experiments by Popoff et al. \cite{Popoff2010, Popoff2010a} significantly advanced the field by showing that complex media can be described as a linear operator acting on electric field states. They experimentally measured this operator and introduced its matrix representation, known as the transmission matrix of the complex medium \cite{Popoff2010, Popoff2010a}. This development enabled light control behind the scatterers using SLMs.

These remarkable advancements led to a better understanding of complex media and paved the way for developing non-invasive imaging techniques through such media. Specifically, experiments by Bertolotti et al. or Katz et al. \cite{Bertolotti2012, Katz2014} utilized the phenomenon known as the memory effect. Within small regions, when an input wavefront reaching a complex medium is tilted within a certain angular range, the output wavefront is equally tilted, resulting in the translation of the speckle pattern behind the complex medium \cite{Freund1988, Freund1990, Judkewitz2015, Osnabrugge2017}. In imaging, within a memory effect region, light emitted from each point source behind the complex medium creates an almost identical speckle pattern that is shifted according to the position of the source origin.

Methods presented in \cite{Bertolotti2012, Katz2014, Hofer2018} utilize the memory effect to reconstruct objects hidden behind complex media by calculating the spatial autocorrelation of speckle images and employing iterative phase retrieval algorithms. An alternative group of methods relies on fluctuating, random speckle illumination and observing the sample's response \cite{Zhu2022, Weinberg2023}. Some of these methods use Non-negative matrix factorization (NMF) to disentangle the temporal and spatial information about the sample \cite{Moretti2020, Zhu2022, Rimoli2024}. More recently, well-established reflection-matrix techniques were applied to incoherent fluorescence imaging \cite{Weinberg2023}. Three-dimensional imaging of objects behind scattering layers is also possible with SLM-based approaches \cite{Baek2025}.

Another possible approach to overcome scattering is to use Time-of-Flight (ToF) analysis for imaging. By temporally gating detected photons, ToF methods can isolate photons that traversed the complex medium with no scattering \cite{Maccarone:15}, or exploit photon arrival times to improve image reconstruction in highly scattering environments \cite{Lyons2019, Lindell2020, Du2022}. ToF approaches can be applied to volumetric scattering; however, their spatial resolution is usually not sufficient for microscopy.

While several techniques for optical imaging in strongly scattering media have been developed and represent significant achievements, their widespread adoption remains limited. One limitation is that these techniques typically work with static diffusers. In live bioimaging, biological tissues change over time, necessitating methods that can handle dynamic scattering. Some approaches to overcome this issue involve using high-speed digital micromirror devices (DMDs) \cite{Conkey2012, Conkey2012a, Goorden2014, Mitchell2016, Hellman2019} as SLMs for fast transmission matrix measurement. However, all these methods require high signal levels, which hinder their practical applications. While recent studies have demonstrated light manipulation \cite{Mididoddi2025} and imaging \cite{Sunray2025} through dynamic scattering media, rapidly changing diffusers still require light to be acquired over short time intervals in order to approximate the scattering medium as static. This inevitably results in low light intensities being captured. This creates a growing need for robust imaging techniques that work in low light intensity scenarios through dynamic complex media.

There have been a few demonstrations of imaging objects that modulate a coherent optical field through dynamically scattering media at low photon flux levels, typically implemented using a DMD or an SLM to display objects from the training dataset \cite{Hao2025,Sun2021}. Moreover, almost all naturally occurring and everyday objects emit or reflect spatially incoherent light. Although these works demonstrate low-photon inference through dynamically perturbed scattering media, they do not provide a general solution for most photon-limited imaging techniques through dynamic scatterers.

We demonstrate that non-invasive imaging of incoherent objects behind dynamic scattering layers is possible even in the low photon number regime. There was no previous demonstration under such conditions. To achieve this, we develop a data-analysis approach that reconstructs the object from a sequence of ultra-dim frames recorded behind a scattering layer, without ballistic light. Our imaging method builds on concepts introduced in earlier works \cite{Bertolotti2012, Katz2014, Hofer2018}, but extends their capabilities to enable imaging through dynamically changing diffusers and under low light conditions.

The proposed approach has theoretically no fundamental lower bound on number of photons per frame, and remains effective in highly dynamic scenarios, as it does not require the scattering medium to remain static between frames. We present a proof-of-concept experimental demonstration of our method in a widefield microscope imaging fluorescent samples. We use a single-photon camera operating in an ultra low light regime, recording fewer than one photon per pixel per frame. Our method does not require wavefront shaping, transmission matrix characterization, or light focusing through the complex medium.

\section{Principles}\label{concept}

\begin{figure}[h!] 
\centering
\includegraphics[width=0.75\linewidth]{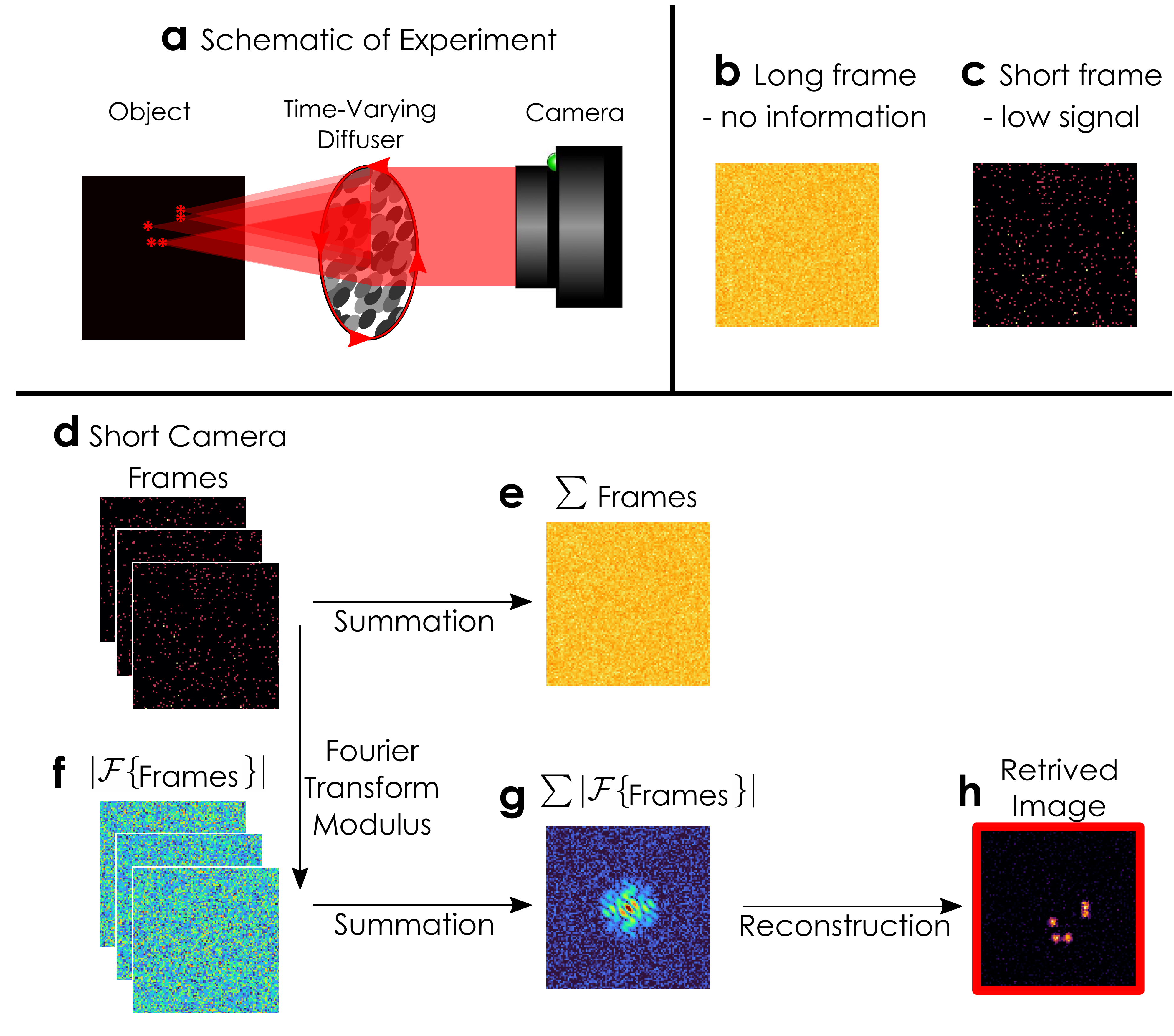}
\caption{Concept of the method. 
(a) A luminescent, spatially incoherent object is imaged through a time-varying diffuser and recorded with a high-speed, photon-counting camera. 
(b) A long exposure produces a homogeneous image that carries no spatial information about the object. 
(c) A single short-exposure frame preserves speckle structure but contains very few photons. 
(d) In the analysis, we use a series of short frames with a low signal.
(e) Summing many such low-signal frames similarly yields a homogeneous image without spatial content. 
(f) Instead, for each short-exposure frame, we compute the magnitude of its Fourier transform. 
(g) The root-mean-square (RMS) average of these Fourier magnitudes across the burst recovers the spatial-frequency content of the object. 
(h) After retrieving the missing Fourier phase using a phase-retrieval algorithm \cite{Fienup1982}, the object hidden behind the dynamic diffuser can be reconstructed.}

\label{fig:concept_general}
\end{figure}

 Our approach involves recording a movie of a luminescent object placed behind a changing diffuser (see Fig. \ref{fig:concept_general}a). When the camera integration time is long, the diffuser undergoes substantial changes during a single exposure (see Fig. \ref{fig:concept_general}b). As a result, the recorded speckle patterns average out, yielding a spatially homogeneous image that carries no information about the hidden object. If the exposure time is short, the frame can be approximated as corresponding to a single realization of the diffuser. However, such frames may contain very low photon counts,  sometimes down to a few detected photons per frame (see Fig. \ref{fig:concept_general}c).

In our method, we acquire a sequence of short exposure frames (see Fig. \ref{fig:concept_general}d). A naive integration of these frames would produce an image similar to that obtained with a single long-exposure measurement (see Fig. \ref{fig:concept_general}e), again lacking spatial information. Instead, for each frame we compute the modulus of its two-dimensional Fourier transform (see Fig. \ref{fig:concept_general}f). We then average these Fourier-modulus images (see Fig. \ref{fig:concept_general}g). Remarkably, the resulting integrated matrix resembles the Fourier-transform modulus of the object hidden behind the diffuser and can be obtained even when individual frames contain only single detected photons. With this modulus, the object image can be reconstructed (see Fig. \ref{fig:concept_general}h) using well established phase-retrieval algorithms.

This method enables effective imaging under dynamic scattering media conditions and achieves reliable results in low photon number regimes that allow tackling extreme low light and dynamic scattering conditions, which are completely beyond the reach of currently demonstrated techniques.

\begin{figure}[h!] 
\centering
\includegraphics[width=0.85\linewidth]{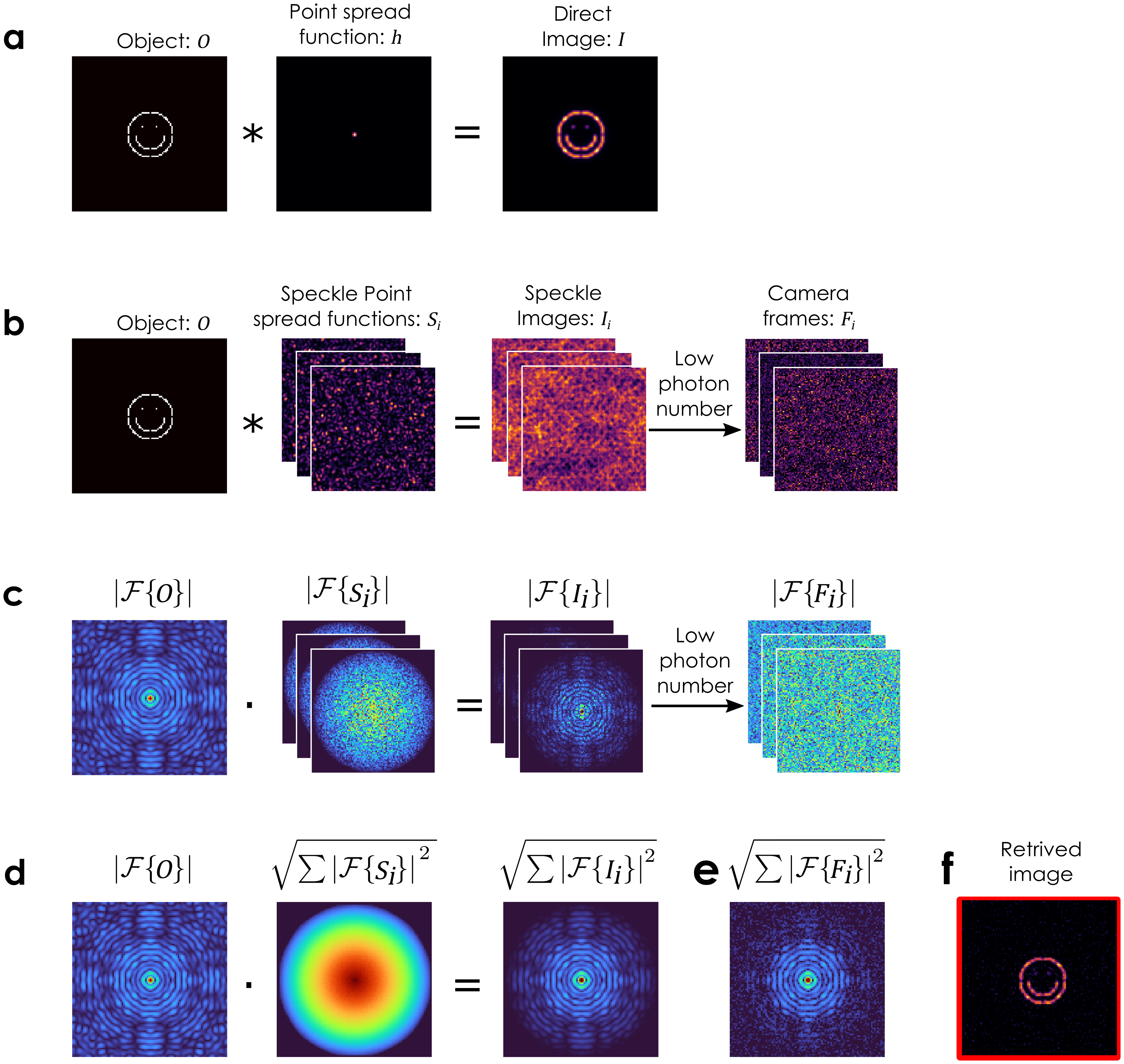}

\caption{Mathematical background of our method.
(a) Direct imaging in a scattering-free system. The image \(I\) formed on the camera is a convolution of the object \(O\) with the system Point Spread Function (PSF) \(h\).
(b) Under dynamic scattering, each speckle image \(I_i\) formed on the camera is a convolution of the object \(O\) with the instantaneous speckle Point Spread Function (PSF) \(S_i\). 
(c) Spatial Fourier transforms of the images shown in (b). 
(d) The root mean square (RMS) of the Fourier transform moduli, computed across many frames, converges to \(|\mathcal{F}\{O\}|\,|\mathcal{F}\{P\}|\), where \(P\) denotes the effective PSF corresponding to the scattering-free image formed by the imaging system. 
(e) The magnitude of the scattering-free Fourier transform is obtained, while its phase is lost. A phase retrieval algorithm \cite{Fienup1982} is then used to recover the missing phase information. 
(f) Reconstruction of the object after phase retrieval.}

\label{fig:concept}
\end{figure}

In a conventional, scattering-free imaging system, the image recorded by the camera \(I\) is a 2D mathematical convolution of the object \(O\) and the system’s Point Spread Function (PSF) \(h\) (see Fig. \ref{fig:concept}a). In the presence of scattering, however, the situation becomes more complex. The pattern formed on the camera, known as a speckle pattern, does not resemble an ordinary PSF. Nevertheless, for small objects located within the memory-effect range \cite{Freund1988, Judkewitz2015}, each point of the object produces the same speckle pattern, shifted relative to one another. As a result, the image captured by the camera is the convolution of the object with this speckle pattern, which acts as a new speckle Point Spread Function.

Under dynamic scattering conditions, each speckle image formed on camera is a convolution of the object \(O\) with the instantaneous speckle Point Spread Function (PSF) \(S_i\) of the varying diffuser (as illustrated in Figure \ref{fig:concept}b). We can represent each speckle image as:

\begin{equation}\label{convol}
I_i = O * S_i,
\end{equation}
where \(i = 1, 2, 3, \ldots, M\) is the image index, and \(M\) is the total number of images. Note that, each frame \(F_i\) captured by the camera is a Poisson-distributed measurement of the underlying noiseless speckle image.  

Under this model, the Fourier transforms satisfy:
\begin{equation}
\mathcal{F}\{I_i\} = \mathcal{F}\{O\} \mathcal{F}\{S_i\},
\end{equation}
where \(\mathcal{F}\{I_i\}\), \(\mathcal{F}\{O\}\), and \(\mathcal{F}\{S_i\}\) denote the spatial Fourier transforms of each image \(I_i\), the object \(O\), and the instantaneous speckle Point Spread Function \(S_i\), respectively (as shown in Fig. \ref{fig:concept}c). By computing the root mean square of the Fourier transform moduli across all images (Fig. \ref{fig:concept}d), we obtain:

\begin{equation}
\sqrt{\frac{1}{M}\sum_{i=1}^{M} \left| \mathcal{F}\{I_i\} \right|^2} = \left| \mathcal{F}\{O\} \right| \sqrt{\frac{1}{M}\sum_{i=1}^{M} \left| \mathcal{F}\{S_i\} \right|^2}.
\end{equation}

For many realizations of the diffuser, the root mean square of the Fourier transform moduli of speckle Point Spread Functions \(S_i\) converges \cite{Goodman2007} to Fourier transform of the effective Point Spread Function ($\mathcal{F}\{P\}$), where P denotes the Point Spread Function of the scattering-free imaging system (see Fig. \ref{fig:concept}d).

Among common averaging strategies, the RMS operator is statistically preferred under our model: the squared Fourier moduli follow an exponential distribution whose scale parameter is estimated by the sample mean of the Fourier powers. Therefore, the RMS-based estimator is the maximum-likelihood and asymptotically efficient estimator of the scattering-free Fourier magnitude, as derived in  \textit{Supplementary Information}, Section~2.1.4.

\begin{equation}\label{speckle_converge}
\sqrt{\frac{1}{M}\sum_{i=1}^{M} \left| \mathcal{F}\{S_i\} \right|^2}
\;\xrightarrow[M\to\infty]{}\;
\left| \mathcal{F}\{P\} \right|.
\end{equation}

For recorded frames affected by spatially uncorrelated detection noise, including Poisson shot noise, averaging the Fourier powers does not remove the noise contribution. Instead, the RMS average converges to the object-dependent Fourier magnitude combined with a frequency-independent noise floor:

\begin{equation}
\sqrt{\frac{1}{M}\sum_{i=1}^{M} \left| \mathcal{F}\{F_i\} \right|^2}
\;\xrightarrow[M\to\infty]{}\;\sqrt{
\left(\left| \mathcal{F}\{O\} \right| \, \left| \mathcal{F}\{P\} \right|\right)^2 +\nu_{\eta}},
\end{equation}

where $\nu_{\eta}$ denotes the frequency-independent contribution of the detection noise to the Fourier power spectrum. This flat offset can be estimated from the photon statistics, background measurements, or signal-free spectral regions and subtracted at the Fourier-power level before taking the square root.

Note that in this setting, imaging through the diffuser is different from direct imaging, because the effective numerical aperture NA is defined by the combination of the scatterer and the objective \cite{Vellekoop2010}.

In the \textit{Supplementary Information} Section 2.1.2, we provide a comprehensive statistical evaluation, estimating the required number of frames for image reconstruction, the achievable resolution, and the signal-to-noise ratio (SNR) of the recovered Fourier magnitude.

In practical experiments, the number of diffuser realizations may be limited. As a result, the RMS average of the Fourier moduli does not converge perfectly to the Fourier modulus of the direct image, and residual scattering artifacts may remain. To suppress these artifacts, we convolve the recovered Fourier modulus with a narrow two-dimensional Gaussian function. This procedure substantially improves reconstruction quality when a limited number of diffuser realizations is available and enables successful image reconstruction even in the case of a static diffuser. Further details of these artifacts and the influence of Gaussian filtering are provided in Sections 3.3 and 4.2 of the \textit{Supplementary Information}.

As a result of this analysis, we retrieve the magnitude of the Fourier transform of the scattering-free image of the object that is  (see Fig. \ref{fig:concept}e), and information about its complex phase is lost. As demonstrated previously \cite{Katz2014, Bertolotti2012, Hofer2018}, this phase can be retrieved using a phase retrieval algorithm. In our work, we utilized the combination of hybrid input-output (HIO) and the error-reduction (ER) algorithms \cite{Fienup1982}. Once the Fourier phases are estimated, the object obscured by the dynamic diffuser can be retrieved (like in Figure \ref{fig:concept}f). 

We detail the post-processing techniques and algorithms used in our analysis in \textit{Supplementary Information} Sections 2.1.4 and 3.

It is important to note that, even though phase retrieval algorithms are highly sensitive to noise, we do not apply them to noisy raw frames. Instead, we apply phase retrieval to the estimated high-SNR Fourier modulus of the object hidden behind a dynamically changing scatterer. For more details of data analysis, please refer to the \textit{Methods} Section \ref{sec:methods} and \textit{Supplementary Information} Section 3.

\section{Results}\label{results}

\begin{figure}[h!] 
\centering
\includegraphics[width=1.\linewidth]{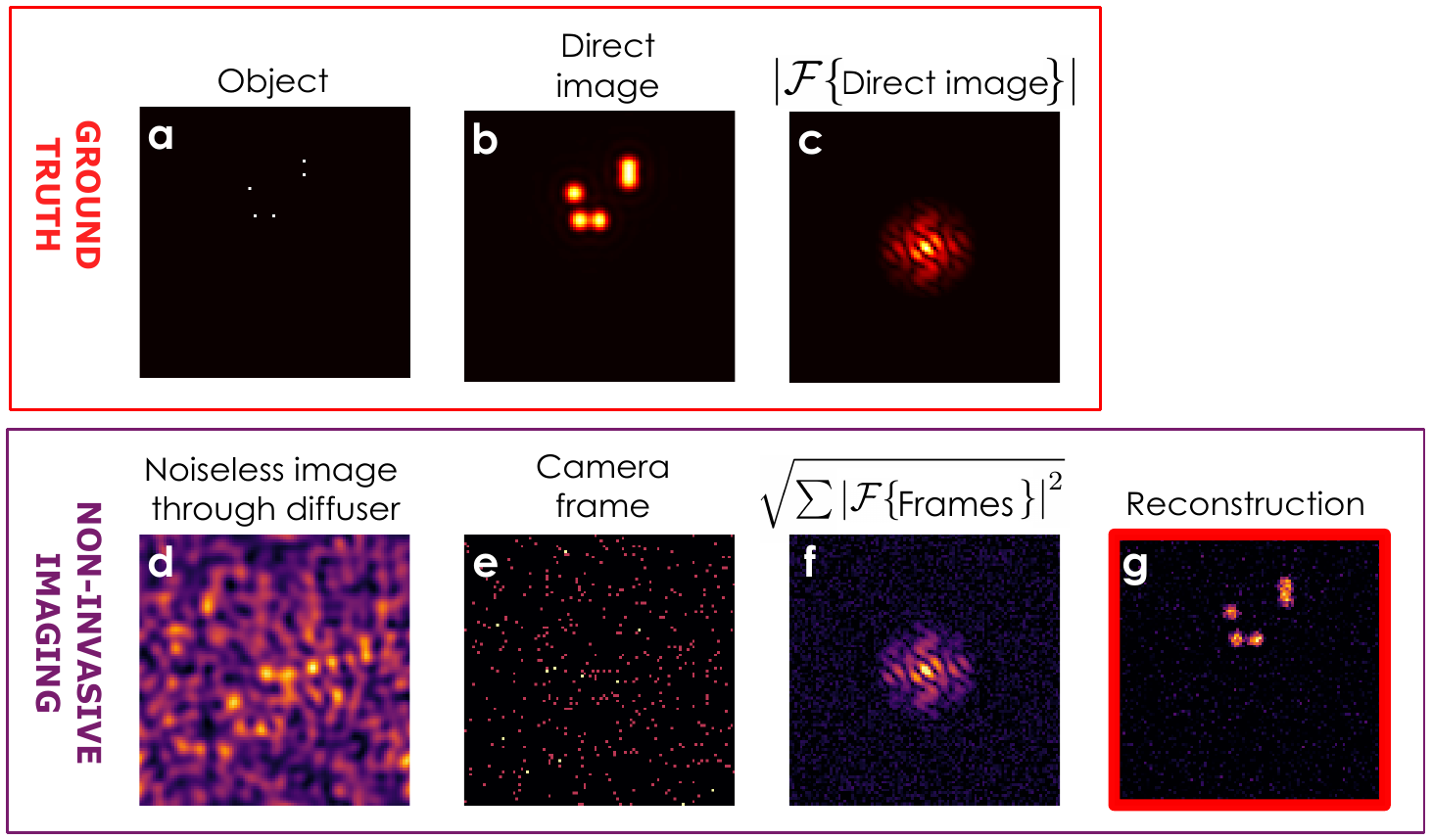}
\caption{Imaging in the photon counting regime -- simulation results. (a) Original object. (b) Image of the object captured with a finite resolution imaging system. (c) Modulus of the Fourier transform of the object (b). (d) Image of the object obscured by the diffuser - a single frame captured by the camera. (e) Single camera frame with Poisson noise. (f) The RMS averaging of Fourier transforms modulus of 10,000 frames. (g) Reconstructed object image through a dynamic diffuser. }
\label{fig:simulation}
\end{figure}

To demonstrate the effectiveness of our method in the single photon counting regime, we first conducted a computer simulation. Figure \ref{fig:simulation} presents the simulation results of imaging the object through the dynamic diffuser in a low photon number regime. Figure \ref{fig:simulation}a shows an object consisting of several point emitters. Imaging this object through an imaging system with finite resolution gives the image shown in Figure \ref{fig:simulation}b (\(I = O * PSF\)), with the Fourier transform modulus shown in Figure \ref{fig:simulation}c (\(\left| \mathcal{F}\{I\} \right| = \left| \mathcal{F}\{O\} \right| \left| \mathcal{F}\{PSF\} \right|\)).

Noiseless imaging of the object from Figure \ref{fig:simulation}a through a diffuser results in the speckle image shown in Figure \ref{fig:simulation}d (\(I_i = O * S_i\)). However, since each frame is short enough to capture the unchanged diffuser during its duration, the detected photon number is low. To account for this, we limit the number of photons detected during each frame from Figure \ref{fig:simulation}d (adding Poisson noise). Figure \ref{fig:simulation}e shows a simulation of a single frame from Figure \ref{fig:simulation}d with a limited number of photons.

In the simulation, we generated 10,000 frames, each with a different diffuser realization and an average of 400 photons per frame (0.04 photons per pixel). Figure \ref{fig:simulation}f shows the root mean square of the Fourier transform moduli of these 10,000 frames (\( \sqrt{\frac{1}{M}\sum_{i=1}^{M} \left| \mathcal{F}\{I_i\} \right|^2} = \left| \mathcal{F}\{O\} \right| \sqrt{\frac{1}{M}\sum_{i=1}^{M} \left| \mathcal{F}\{S_i\} \right|^2}\)). Notably, Figure \ref{fig:simulation}f shows excellent agreement with Figure \ref{fig:simulation}c. After applying the phase retrieval algorithm \cite{Fienup1982}, we obtain the reconstructed image (Figure \ref{fig:simulation}g), which corresponds to the direct image of the object in Figure \ref{fig:simulation}b.

\begin{figure}[h!] 
\centering
\includegraphics[width=0.8\linewidth]{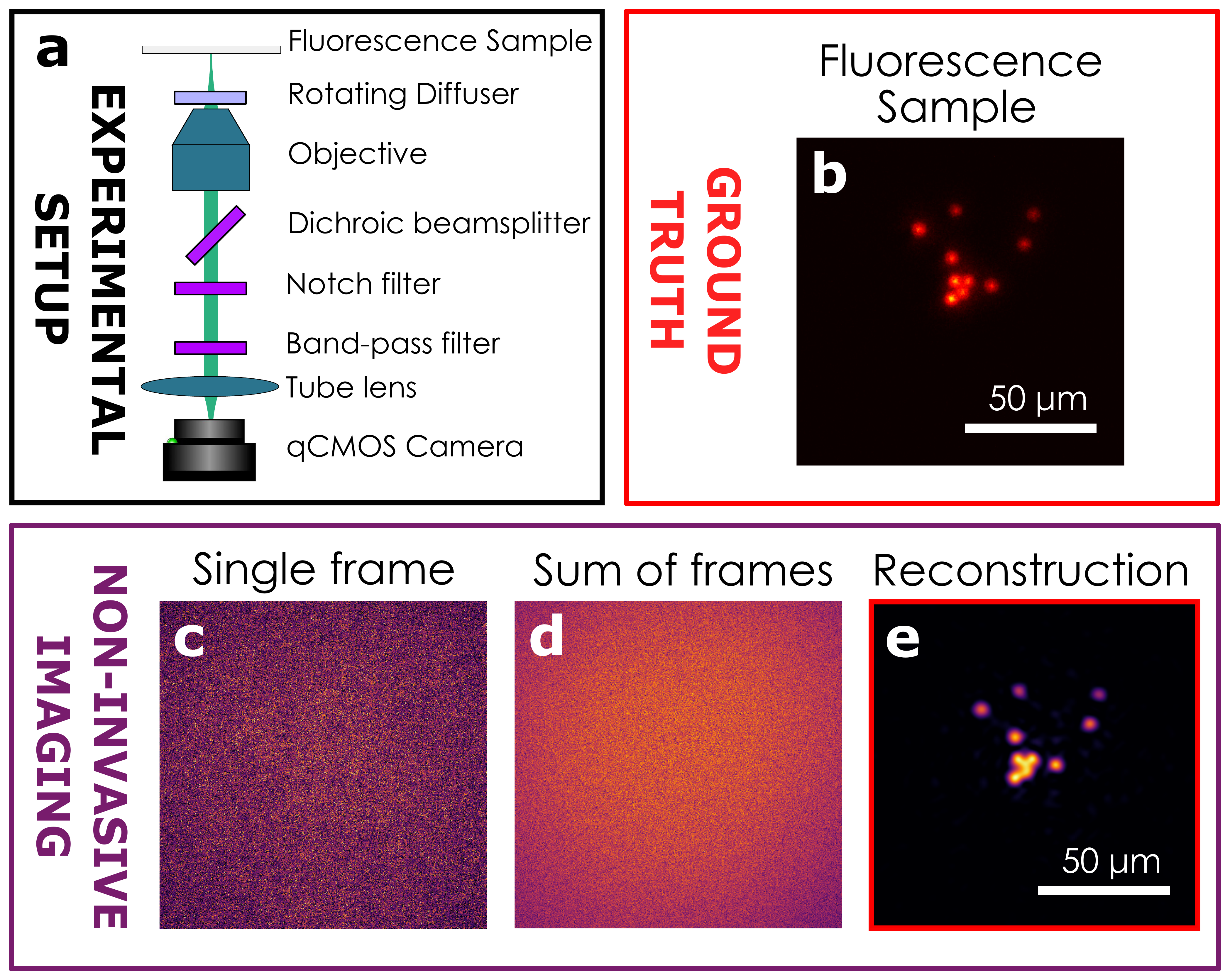}
\caption{Imaging of 5$\mu m$ fluorescence micro-spheres through the dynamic complex medium in wide-field microscope. (a) Schematic of the wide-field fluorescence microscope utilized in our experiments. (b) Direct image of the sample without the diffuser (ground truth). (c) Single frame captured by the qCMOS camera. (d) Average of all 1791 frames recorded. (e) Reconstruction result using the phase retrieval algorithm \cite{Fienup1982}. }
\label{fig:results_bio}
\end{figure}

In our experiment, we applied our method to fluorescence microscopy. As an experimental setup, we utilized a wide-field custom-made microscope (see Figure \ref{fig:results_bio}a). The details of the microscope construction can be found in \textit{Methods} Section \ref{sec:methods:setup} and \textit{Supplementary Information} Section 1. Our microscope allowed for the insertion and removal of a dynamic diffuser, placed just before the microscope objective (between the objective and the fluorescent sample). The diffuser was not continuously rotating during image acquisition. Instead, it was randomly rotated to a new angular position between consecutive frames and remained stationary during each camera exposure. Different diffuser realizations were therefore generated by changing the orientation of a 1-inch diffuser between frames. As a sensor in our microscope, we used the Hamamatsu ORCA-Quest 2 qCMOS camera, offering photon number resolution at each pixel.

We imaged a sample of 5$\mu m$ fluorescence micro-spheres. Details of samples preparation procedure are provided in \textit{Methods} Section \ref{sec:methods:sample}. Figure \ref{fig:results_bio}b shows a direct image of the sample without the diffuser. Figure \ref{fig:results_bio}c illustrates a single frame captured by the qCMOS camera. We recorded 1791 frames, each with an exposure time of 200.96 ms. Figure \ref{fig:results_bio}d shows the sum of all frames recorded in our experiment. As shown in Fig.~\ref{fig:results_bio}d, integrating the signal in real space results in the complete loss of spatial information. Therefore, we average the Fourier transform moduli of all the frames and apply a phase retrieval algorithm \cite{Fienup1982} to recover the phase information. The reconstruction result is presented in Figure \ref{fig:results_bio}e. Both the convergence of the algorithm and the overall accuracy of our method depend on the characteristics of the object, as described in the \textit{Supplementary Information} Section 2.1.6 and 4.

\begin{figure}[h!] 
\centering
\includegraphics[width=1\linewidth]{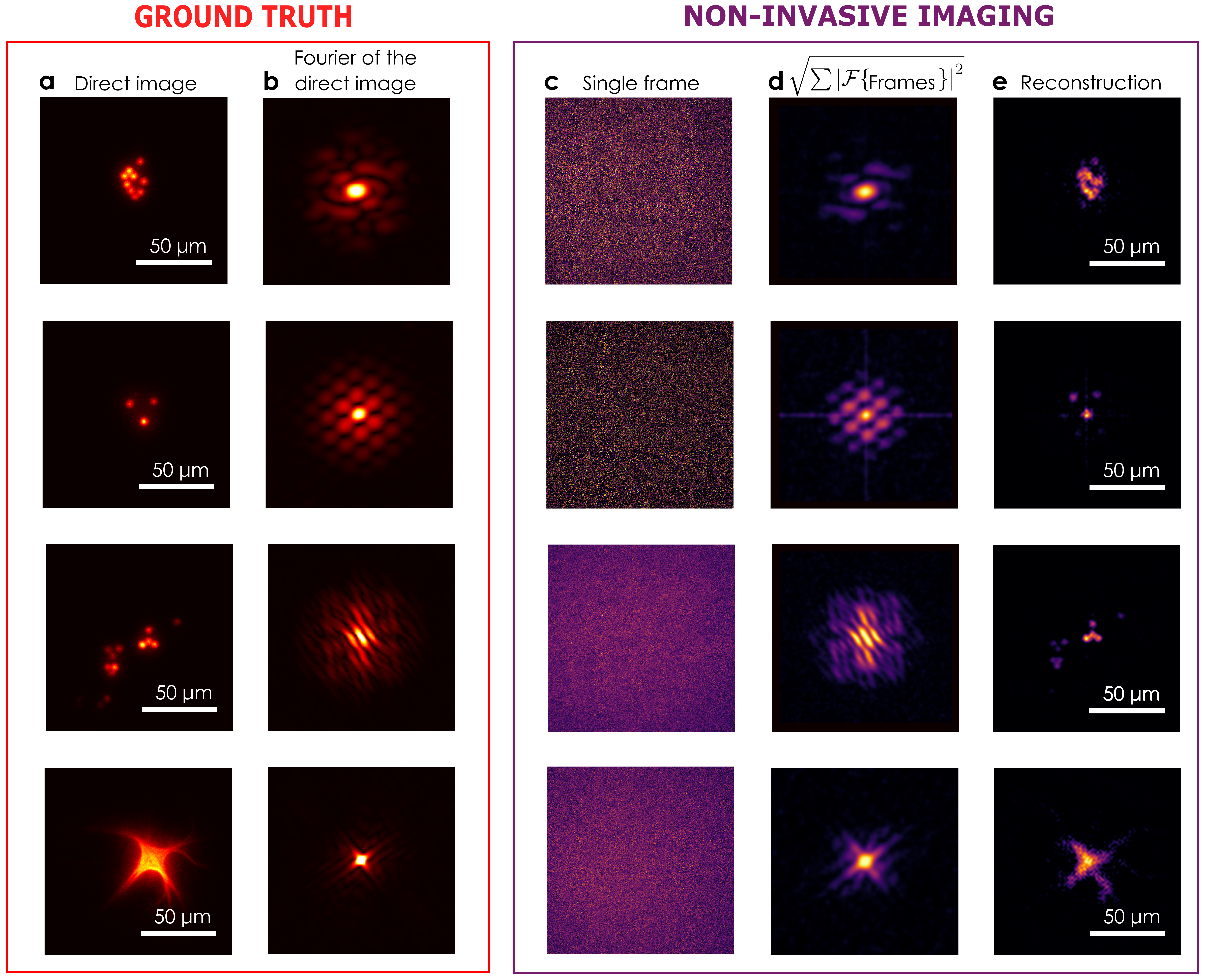}
\caption{ Results comparison of imaging through the dynamic complex medium. (\textbf{a}) Direct images of fluorescent samples without the diffuser. (\textbf{b}) The modulus of the Fourier transforms of the images in column (\textbf{a}). (\textbf{c})  Example camera frames of objects in column (\textbf{a}) obscured by the varying diffuser, with photon counts as low as 0.14 photons per pixel in 1152 frames, highlighted in the second row. (\textbf{d}) The RMS averaging of the Fourier transform moduli of the frames captured through the dynamic diffuser. (\textbf{e}) Recovered image computed using the phase retrieval algorithm \cite{Fienup1982}.  Rows 1 to 3 present fluorescent microspheres, while row 4 shows Alexa Fluor 568-labeled astrocytes.}
\label{fig:results_spheres}
\end{figure}

To demonstrate the versatility of our method, we applied it to fluorescent-microsphere structures and to astrocytes stained with the widely used fluorescent dye Alexa Fluor 568 (see Figure \ref{fig:results_spheres}). Fluorescent samples preparation procedures are detailed in Section \ref{sec:methods:sample}. The column \textbf{a} of Figure \ref{fig:results_spheres} shows direct images of the fluorescent samples in our microscope. Figure \ref{fig:results_spheres}\textbf{b} presents the modulus of the Fourier transforms of the images in the column \textbf{a}. Figure \ref{fig:results_spheres}\textbf{c} presents individual frames for different objects captured through the varying diffuser, with photon counts reaching as low as 0.14 photons per pixel in 1152 captured frames, as shown in the second row. Figure \ref{fig:results_spheres}\textbf{d} shows the RMS averaging of the Fourier transform modulus of the frames captured by the camera through the dynamic diffuser. Figure \ref{fig:results_spheres}\textbf{e} displays the recovered image imaged through the diffuser, computed using the column \textbf{d} from fig. \ref{fig:results_spheres} with the phase retrieval algorithm \cite{Fienup1982}. Rows 1 to 3 present results obtained with fluorescent microspheres, while row 4 shows data from Alexa Fluor 568-labeled astrocytes.

\section{Discussion}\label{discussion}

We successfully demonstrated imaging incoherent objects through a dynamic diffuser at signal levels low enough to detect single photons, conditions under which no imaging has been demonstrated so far. This scenario is highly relevant for deep-tissue imaging, where scattering and low photon counts are significant challenges. Our approach does not require physical alteration of the sample, nor does it necessitate characterizing the medium's transmission matrix or focusing light through it. Our approach does not require adaptive optics or wavefront shaping systems.

Our method is robust, as spatially uncorrelated noise can be progressively suppressed by accumulating a sufficiently large number of frames.

Moreover, our method is resilient to fluorescence bleaching and minor sample movements during multi-frame acquisition. Although bleaching reduces signal intensity, the spatial distribution of fluorophores in densely stained samples remains intact, preserving essential spatial information despite reduced Fourier amplitudes. By the RMS averaging the Fourier transform moduli across frames, we accurately recover the Fourier modulus of the object's scattering-free image. Similarly, since translational motion only introduces phase in the Fourier domain, our modulus-based approach naturally compensates for such shifts—only significant rotations or structural changes would impact the reconstruction. Our method works also in the case of static diffusers (see \textit{Supplementary Information} Section 2.1.6).

In Section 2 of the \textit{Supplementary Information}, we provide a detailed discussion of the theoretical foundations of our approach. We analyze the number of frames required for a given mean photon count per frame and show that, within the considered statistical model, image reconstruction remains possible at arbitrarily low photon counts per diffuser realization, provided that a sufficiently large number of statistically independent frames is accumulated. We also investigate the spatial resolution and signal-to-noise ratio of the proposed imaging method.

We validated our approach using a standard fluorescence microscope and applied it to various fluorescent samples. Notably, we successfully imaged astrocyte cells stained with the widely used dye Alexa Fluor 568, as well as various challenging objects constructed from fluorescent micro-spheres, further underscoring the versatility and effectiveness of our approach. 

Our approach has potential applications in various other fields of imaging. For instance, it can be utilized in astronomy for imaging distant, dim celestial objects through a fluctuating atmosphere \cite{Tyson2022}, or in endoscopy using multicore fibers \cite{Porat2016, Badt2022, Stasio2016}, where bends generate changing scattering scenarios. In such fibers, the memory effect is effectively unlimited if the light propagating in individual cores does not couple between neighboring cores, while differences in core lengths make the fiber behave as a phase object. This makes multicore fibers highly attractive for endoscopic imaging, as their overall diameter can remain comparable to that of standard optical fibers (typically ~125 µm cladding diameter). At the same time, this is orders of magnitude smaller than conventional flexible endoscopes, which typically have diameters ranging from a few millimeters up to ~10 mm. Consequently, multicore fiber-based approaches offer strong potential for minimally invasive imaging in both medical endoscopy and biological microscopy \cite{Porat2016, Badt2022, Stasio2016}.

The present study was demonstrated using a thin scattering layer. Extending the approach to thicker scattering media is expected to be more challenging, although not experimentally impossible. The main challenge is the limited range of the optical memory effect, as the technique presented in this work relies directly on it. As mentioned in Eq. \ref{convol}, we assume that the image recorded by the camera can be described as a convolution of the object with a speckle pattern that acts as the system point spread function for a given realization of the scattering medium during data acquisition. This assumption is valid only within the range of the memory effect. 
This directly limits the field of view of the method. In the diffusive regime, the angular field of view is determined by the memory effect range and can be estimated as \(\Delta \theta_{\mathrm{FOV}} \approx \lambda/(\pi L)\), where \(L\) is the thickness of the scattering medium \cite{Freund1988, Freund1990, Judkewitz2015, Osnabrugge2017}. Consequently, the field of view decreases with increasing thickness of the scattering medium. Beyond the memory effect range, the method in its current form is not applicable and would require more advanced data analysis approaches (such as \cite{Zhu2022}).

Another challenge of the proposed approach is the use of phase retrieval algorithms. These algorithms are highly nonlinear, and even small errors in the estimated Fourier modulus can affect the quality of the reconstructed image. In particular, reconstructing objects with complex grayscale structures is generally more challenging, whereas binary or high-contrast objects are comparatively straightforward. One potential way to mitigate this limitation is to employ bispectrum analysis \cite{Wu2016}, which may enable imaging without relying on a phase-retrieval algorithm.

A further limitation when imaging through thicker scattering media is the amount of detected signal. As the thickness of the scattering medium increases, light originating from a single object point is spread over a larger area, which may not be efficiently collected by the optical system. In addition, thicker scattering media exhibit stronger backscattering, further limiting the light collection efficiency. Nevertheless, the method remains inherently compatible with extremely low photon budgets.

Our approach has theoretically no fundamental lower bound on fluorescence intensity and weak photon fluxes can always be compensated by a larger number of frames. However, achieving this with a few photons per frame requires a substantial number of frames, as detailed in Section 2.1.6 of the Supplementary Information. Depending on the application, the required number of frames may become impractically large.

Single camera frames in our experiment contain low signals, reaching $0.14$ photons per pixel per frame (around $3\times10^5$ photons per field of view). These experiments are feasible thanks to recent advances in the development of single photon cameras. The camera that we utilized has a relatively low frame rate, capturing fewer than 5 frames per second for the full field of view required by the speckle size. Since our experiment requires thousands of camera frames, the total acquisition time for a single image reconstruction is typically a couple of minutes. Nonetheless, this acquisition duration is well-suited for many bioimaging applications, such as embryonic development, cell division, and synaptic modifications. 

A potential improvement in the data acquisition rate our setup is the use of SPAD arrays, which are increasingly being applied in imaging \cite{Lubin2019, Ghezzi2021, Bruschini2019, Slenders2021, Lukashchuk2024}. Recent advancements in this field have led to the development of megapixel SPAD arrays \cite{Bruschini2019}, making their practical application in widefield microscopy feasible. Cameras based on SPAD technology allow for single photon detection with high temporal resolution, providing a powerful tool for enhancing various bioimaging techniques and other applications. Some SPAD arrays offer sub-nanosecond resolution for tagging of each photon's arrival time \cite{Bruschini2019}. 

To push the temporal performance of the method towards its fundamental limits, one could combine a picosecond pulsed laser with an ultrafast SPAD array. In such a configuration, photons would be collected only within a narrow time window following each laser pulse, comparable to the fluorescence lifetime of the dye. The photons detected within this window could then be integrated into a single camera frame. In principle, the fluorescence lifetime sets the fundamental temporal limit of diffuser dynamics for this method, since the emission times of fluorescence photons are distributed over this interval.

\section{Methods}\label{sec:methods}
\subsection{Experimental setup}\label{sec:methods:setup}
We designed and constructed our experimental setup, which consisted of a fluorescence wide-field microscope. For the objective, we used Nikon S Plan Fluor ELWD 20x/0.45, and for the excitation beam, we utilized a Spectra-Physics Millennia PRO laser with a wavelength of 532 nm. We utilized several neutral density filters to attenuate the laser beam. The fluorescence sample was placed within the working distance of the primary objective, while a rotating diffuser (Edmund Holographic Diffuser of 5° Diffusing Angle) was positioned adjacent to the microscope objective (between the objective and the sample). To select the desired field of view in our experiment, we illuminated the sample from a transmissive side using another objective (Nikon S Plan Fluor ELWD 40x/0.60), which was usually out of focus to provide an adjustable excitation spot with a typical diameter of around 100~$\mu$m. The diffusing angle was chosen in accordance with the imaging system to produce scattered light that matches the camera's sensor size. In our experiment, it was important to have a speckle size that was at least two times larger than the pixel size to avoid aliasing. Therefore, the system magnification and the diffusing angle have to be adjusted accordingly.

To separate the fluorescence light from the excitation source, we employed a dichroic mirror (Edmund 552), a long-pass filter (Thorlabs FELH0550), and a notch filter Semrock NF01-532U-25. Additionally, to ensure good visibility of speckles, a band-pass filter (Thorlabs FBH590-10) was used. The tube lens in our setup was a Thorlabs TTL200-A with a focal length of 200 mm. The camera was positioned at the focal point of the tube lens. The experiments were performed using a Hamamatsu ORCA-Quest 2 qCMOS camera (C15550-22UP). The camera features a back-illuminated sensor with \(4096 \times 2304\) pixels and a pixel size of \(4.6 \mu\text{m} \times 4.6 \mu\text{m}\). According to the manufacturer specifications, the sensor provides a peak quantum efficiency of approximately 85\%, an ultra-low readout noise of 0.30 electrons rms in Ultra Quiet mode, and a dark current of 0.016 electrons pixel\(^{-1}\) s\(^{-1}\) at \(-20^\circ\)C.

Mounts for the fluorescence sample and the rotating diffuser with its motor were custom-designed and 3D-printed. 

For more details of the experimental setup, please refer to the \textit{Supplementary Information} Section 1.

\subsection{Data acquisition}

In our experiment, we utilized a galvo mirror (Thorlabs GVS211/M) positioned in the excitation laser beam path to modulate the illumination of the sample. The galvo mirror was connected to a microcontroller (Arduino Micro), which communicated with a computer via a serial port. A Python program controlled both the microcontroller and the camera.

Initially, the computer sent a command to the Arduino to illuminate the sample and then to the camera to begin the acquisition. The microcontroller ensured that the fluorescent sample was illuminated only during the camera's light detection period to limit the photo-bleaching of the sample. It was removing the excitation laser beam immediately after the camera sent a trigger indicating the end of the frame.

The microcontroller used in our experiment had an additional task: controlling the rotation of the diffuser. The microcontroller was connected to an electronic circuit (L2930) that enabled it to switch the 12V power supply required for motor (SKU GS06331-10) rotation on and off.

\subsection{Data analysis}

We performed several filtering techniques in our data processing to retrieve the object. The speckle image on the camera is not equally distributed across the sample, exhibiting a Gaussian envelope. By dividing by a 2D Gaussian function, we corrected for this uneven distribution.
Next, we applied a 2D Hann filter to smooth the edges of each frame, which helps mitigate edge effects in the Fourier domain.
After applying the Hann filter, we computed each frame's 2D Fourier transform and calculated its modulus. We then averaged these transformed frames with the RMS and convolved the resulting signal with a narrow 2D filter to remove speckle artifacts. Finally, we applied the phase retrieval algorithm (the combination of HIO and the ER algorithms) to the prepared Fourier amplitude, to obtain a reconstructed object.
\subsection{Sample preparation}\label{sec:methods:sample}
\subsubsection*{Fluorescence micro-spheres}
Fluorescent microsphere samples were prepared using a drop-casting method. A concentrated solution of fluorescent microspheres (PS-FluoRed-Particles; 4.99 µm/SD=0.16 µm; 530/607 nm) was significantly diluted with ethanol. The solution was subjected to a 15-minute sonication bath to prevent clustering of the fluorescent microspheres. A 50 µL drop of this diluted solution was then placed onto a microscope coverslip and allowed to air dry.
\subsubsection*{Alexa Fluor 568 stained astrocyte}
Three-week-old primary mixed neuronal cultures on glass slides were fixed in PBS with 4\% paraformaldehyde and 4\% sucrose for 8 minutes at $37^o$C. After washing three times with PBS containing 4\% sucrose, cells were permeabilized with 0.25\% Triton X-100 for 2 minutes. Blocking was performed in PBS with 3\% BSA and 0.1\% Triton X-100 for 1 hour at room temperature. Cultures were incubated overnight at $4^o$C with an anti-GFAP primary antibody (1:1000, \#ab7260), followed by three PBS washes. Secondary antibody incubation was conducted for 1 hour at room temperature using Alexa Fluor 568-conjugated anti-rabbit antibody (1:500, \#A-11011). After three PBS washes, samples were mounted on glass slides using Fluoromount-G solution (\#00-4958-02).

\backmatter
\bmhead{Data availability} 
The data used in this study is available from the corresponding author upon reasonable request.

\bmhead{Acknowledgements}
We would like to express our sincere gratitude to Diana Legutko for preparing the Alexa Fluor 568-stained astrocyte samples, and to Anat Daniel for insightful discussions and comments about the experimental setup. We want to acknowledge the support of the following funding agencies: Foundation for Polish Science (FIRST TEAM project FENG.02.02-IP.05-0253/23); Narodowe Centrum Nauki (grants 2023/49/N/ST7/04195 and 2022/47/B/ST7/03465); National Centre for Research and Development QuantERA II projects: QM3 (QuantERAII/02/QM3/03/2024) and EXTRASENS (QuantERAII/02/EXTRASENS/02/2024); HORIZON EUROPE Marie Skłodowska-Curie Actions (FLORIN ID 101086142); and the French Government Scholarship for Ph.D. Cotutelle/Codirection.`

\bmhead{Conflict of interest}
The authors declare no competing interests.

\bmhead{Contributions}
A.M. contributed to conceiving the project idea together with S.G. and R.L., designed and built the experimental setup, prepared the fluorescence microsphere samples, conducted the experiments, performed all simulations, developed and investigated the data analysis methods, analyzed the data, prepared all figures, and wrote the manuscript. W.Z. contributed to the theoretical framework, supported improvements in the data analysis, and authored the supplementary section on theoretical investigations. P.S. provided key input on constructing the experimental platform, contributed to data acquisition and theory, and wrote the supplementary section detailing the experimental setup. B.G. supported the theoretical investigations and co-authored relevant parts of the theoretical supplement. S.G. contributed to its initial conception and provided critical insights and guidance throughout the project. R.L. conceived the project, coordinated its development, and provided essential guidance and input across all stages. All authors reviewed and revised the manuscript and supplementary materials.

\bibliography{sn-bibliography}

\begin{thebibliography}{10}
\expandafter\ifx\csname url\endcsname\relax
  \def\url#1{\burl{#1}}\fi
\expandafter\ifx\csname urlprefix\endcsname\relax\def\urlprefix{URL }\fi
\providecommand{\bibinfo}[2]{#2}
\providecommand{\eprint}[2][]{\url{#2}}
\providecommand{\doi}[1]{\url{https://doi.org/#1}}
\bibcommenthead

\bibitem{Goodman2007}
\bibinfo{author}{Goodman, J.~W.}
\newblock \emph{\bibinfo{title}{Speckle phenomena in optics: theory and
  applications}}  (\bibinfo{publisher}{Roberts and Company Publishers},
  \bibinfo{year}{2007}).

\bibitem{Yoon2020}
\bibinfo{author}{Yoon, S.} \emph{et~al.}
\newblock \bibinfo{title}{Deep optical imaging within complex scattering
  media}.
\newblock \emph{\bibinfo{journal}{Nature Reviews Physics}}
  \textbf{\bibinfo{volume}{2}}, \bibinfo{pages}{141--158}
  (\bibinfo{year}{2020}).

\bibitem{Vellekoop2007}
\bibinfo{author}{Vellekoop, I.~M.} \& \bibinfo{author}{Mosk, A.~P.}
\newblock \bibinfo{title}{Focusing coherent light through opaque strongly
  scattering media}.
\newblock \emph{\bibinfo{journal}{Opt. Lett.}} \textbf{\bibinfo{volume}{32}},
  \bibinfo{pages}{2309--2311} (\bibinfo{year}{2007}).

\bibitem{Popoff2010}
\bibinfo{author}{Popoff, S.}, \bibinfo{author}{Lerosey, G.},
  \bibinfo{author}{Fink, M.}, \bibinfo{author}{Boccara, A.~C.} \&
  \bibinfo{author}{Gigan, S.}
\newblock \bibinfo{title}{Image transmission through an opaque material}.
\newblock \emph{\bibinfo{journal}{Nature communications}}
  \textbf{\bibinfo{volume}{1}}, \bibinfo{pages}{81} (\bibinfo{year}{2010}).

\bibitem{Popoff2010a}
\bibinfo{author}{Popoff, S.~M.} \emph{et~al.}
\newblock \bibinfo{title}{Measuring the transmission matrix in optics: An
  approach to the study and control of light propagation in disordered media}.
\newblock \emph{\bibinfo{journal}{Physical review letters}}
  \textbf{\bibinfo{volume}{104}}, \bibinfo{pages}{100601}
  (\bibinfo{year}{2010}).

\bibitem{Bertolotti2012}
\bibinfo{author}{Bertolotti, J.} \emph{et~al.}
\newblock \bibinfo{title}{Non-invasive imaging through opaque scattering
  layers}.
\newblock \emph{\bibinfo{journal}{Nature}} \textbf{\bibinfo{volume}{491}},
  \bibinfo{pages}{232--234} (\bibinfo{year}{2012}).

\bibitem{Katz2014}
\bibinfo{author}{Katz, O.}, \bibinfo{author}{Heidmann, P.},
  \bibinfo{author}{Fink, M.} \& \bibinfo{author}{Gigan, S.}
\newblock \bibinfo{title}{Non-invasive single-shot imaging through scattering
  layers and around corners via speckle correlations}.
\newblock \emph{\bibinfo{journal}{Nature photonics}}
  \textbf{\bibinfo{volume}{8}}, \bibinfo{pages}{784--790}
  (\bibinfo{year}{2014}).

\bibitem{Freund1988}
\bibinfo{author}{Freund, I.}, \bibinfo{author}{Rosenbluh, M.} \&
  \bibinfo{author}{Feng, S.}
\newblock \bibinfo{title}{Memory effects in propagation of optical waves
  through disordered media}.
\newblock \emph{\bibinfo{journal}{Physical review letters}}
  \textbf{\bibinfo{volume}{61}}, \bibinfo{pages}{2328} (\bibinfo{year}{1988}).

\bibitem{Freund1990}
\bibinfo{author}{Freund, I.}
\newblock \bibinfo{title}{Looking through walls and around corners}.
\newblock \emph{\bibinfo{journal}{Physica A: Statistical Mechanics and its
  Applications}} \textbf{\bibinfo{volume}{168}}, \bibinfo{pages}{49--65}
  (\bibinfo{year}{1990}).

\bibitem{Judkewitz2015}
\bibinfo{author}{Judkewitz, B.}, \bibinfo{author}{Horstmeyer, R.},
  \bibinfo{author}{Vellekoop, I.~M.}, \bibinfo{author}{Papadopoulos, I.~N.} \&
  \bibinfo{author}{Yang, C.}
\newblock \bibinfo{title}{Translation correlations in anisotropically
  scattering media}.
\newblock \emph{\bibinfo{journal}{Nature physics}}
  \textbf{\bibinfo{volume}{11}}, \bibinfo{pages}{684--689}
  (\bibinfo{year}{2015}).

\bibitem{Osnabrugge2017}
\bibinfo{author}{Osnabrugge, G.}, \bibinfo{author}{Horstmeyer, R.},
  \bibinfo{author}{Papadopoulos, I.~N.}, \bibinfo{author}{Judkewitz, B.} \&
  \bibinfo{author}{Vellekoop, I.~M.}
\newblock \bibinfo{title}{Generalized optical memory effect}.
\newblock \emph{\bibinfo{journal}{Optica}} \textbf{\bibinfo{volume}{4}},
  \bibinfo{pages}{886--892} (\bibinfo{year}{2017}).

\bibitem{Hofer2018}
\bibinfo{author}{Hofer, M.}, \bibinfo{author}{Soeller, C.},
  \bibinfo{author}{Brasselet, S.} \& \bibinfo{author}{Bertolotti, J.}
\newblock \bibinfo{title}{Wide field fluorescence epi-microscopy behind a
  scattering medium enabled by speckle correlations}.
\newblock \emph{\bibinfo{journal}{Optics express}}
  \textbf{\bibinfo{volume}{26}}, \bibinfo{pages}{9866--9881}
  (\bibinfo{year}{2018}).

\bibitem{Zhu2022}
\bibinfo{author}{Zhu, L.} \emph{et~al.}
\newblock \bibinfo{title}{Large field-of-view non-invasive imaging through
  scattering layers using fluctuating random illumination}.
\newblock \emph{\bibinfo{journal}{Nature communications}}
  \textbf{\bibinfo{volume}{13}}, \bibinfo{pages}{1447} (\bibinfo{year}{2022}).

\bibitem{Weinberg2023}
\bibinfo{author}{Weinberg, G.}, \bibinfo{author}{Sunray, E.} \&
  \bibinfo{author}{Katz, O.}
\newblock \bibinfo{title}{Noninvasive megapixel fluorescence microscopy through
  scattering layers by a virtual incoherent reflection matrix}.
\newblock \emph{\bibinfo{journal}{Science Advances}}
  \textbf{\bibinfo{volume}{10}}, \bibinfo{pages}{eadl5218}
  (\bibinfo{year}{2024}).

\bibitem{Moretti2020}
\bibinfo{author}{Moretti, C.} \& \bibinfo{author}{Gigan, S.}
\newblock \bibinfo{title}{Readout of fluorescence functional signals through
  highly scattering tissue}.
\newblock \emph{\bibinfo{journal}{Nature Photonics}}
  \textbf{\bibinfo{volume}{14}}, \bibinfo{pages}{361--364}
  (\bibinfo{year}{2020}).

\bibitem{Rimoli2024}
\bibinfo{author}{Rimoli, C.~V.} \emph{et~al.}
\newblock \bibinfo{title}{Demixing fluorescence time traces transmitted by
  multimode fibers}.
\newblock \emph{\bibinfo{journal}{Nature Communications}}
  \textbf{\bibinfo{volume}{15}}, \bibinfo{pages}{6286} (\bibinfo{year}{2024}).

\bibitem{Baek2025}
\bibinfo{author}{Baek, Y.}, \bibinfo{author}{de~Aguiar, H.~B.} \&
  \bibinfo{author}{Gigan, S.}
\newblock \bibinfo{title}{Three-dimensional holographic imaging of incoherent
  objects through scattering media}.
\newblock \emph{\bibinfo{journal}{Nature Communications}}
  \textbf{\bibinfo{volume}{16}}, \bibinfo{pages}{11653} (\bibinfo{year}{2025}).

\bibitem{Maccarone:15}
\bibinfo{author}{Maccarone, A.} \emph{et~al.}
\newblock \bibinfo{title}{Underwater depth imaging using time-correlated
  single-photon counting}.
\newblock \emph{\bibinfo{journal}{Opt. Express}} \textbf{\bibinfo{volume}{23}},
  \bibinfo{pages}{33911--33926} (\bibinfo{year}{2015}).

\bibitem{Lyons2019}
\bibinfo{author}{Lyons, A.} \emph{et~al.}
\newblock \bibinfo{title}{Computational time-of-flight diffuse optical
  tomography}.
\newblock \emph{\bibinfo{journal}{Nature Photonics}}
  \textbf{\bibinfo{volume}{13}}, \bibinfo{pages}{575--579}
  (\bibinfo{year}{2019}).

\bibitem{Lindell2020}
\bibinfo{author}{Lindell, D.~B.} \& \bibinfo{author}{Wetzstein, G.}
\newblock \bibinfo{title}{Three-dimensional imaging through scattering media
  based on confocal diffuse tomography}.
\newblock \emph{\bibinfo{journal}{Nature Communications}}
  \textbf{\bibinfo{volume}{11}}, \bibinfo{pages}{4517} (\bibinfo{year}{2020}).

\bibitem{Du2022}
\bibinfo{author}{Du, D.} \emph{et~al.}
\newblock \bibinfo{title}{A boundary migration model for imaging within
  volumetric scattering media}.
\newblock \emph{\bibinfo{journal}{Nature Communications}}
  \textbf{\bibinfo{volume}{13}}, \bibinfo{pages}{3234} (\bibinfo{year}{2022}).

\bibitem{Conkey2012}
\bibinfo{author}{Conkey, D.~B.}, \bibinfo{author}{Caravaca-Aguirre, A.~M.} \&
  \bibinfo{author}{Piestun, R.}
\newblock \bibinfo{title}{High-speed scattering medium characterization with
  application to focusing light through turbid media}.
\newblock \emph{\bibinfo{journal}{Optics express}}
  \textbf{\bibinfo{volume}{20}}, \bibinfo{pages}{1733--1740}
  (\bibinfo{year}{2012}).

\bibitem{Conkey2012a}
\bibinfo{author}{Conkey, D.~B.}, \bibinfo{author}{Brown, A.~N.},
  \bibinfo{author}{Caravaca-Aguirre, A.~M.} \& \bibinfo{author}{Piestun, R.}
\newblock \bibinfo{title}{Genetic algorithm optimization for focusing through
  turbid media in noisy environments}.
\newblock \emph{\bibinfo{journal}{Optics express}}
  \textbf{\bibinfo{volume}{20}}, \bibinfo{pages}{4840--4849}
  (\bibinfo{year}{2012}).

\bibitem{Goorden2014}
\bibinfo{author}{Goorden, S.~A.}, \bibinfo{author}{Bertolotti, J.} \&
  \bibinfo{author}{Mosk, A.~P.}
\newblock \bibinfo{title}{Superpixel-based spatial amplitude and phase
  modulation using a digital micromirror device}.
\newblock \emph{\bibinfo{journal}{Optics express}}
  \textbf{\bibinfo{volume}{22}}, \bibinfo{pages}{17999--18009}
  (\bibinfo{year}{2014}).

\bibitem{Mitchell2016}
\bibinfo{author}{Mitchell, K.~J.}, \bibinfo{author}{Turtaev, S.},
  \bibinfo{author}{Padgett, M.~J.}, \bibinfo{author}{{\v{C}}i{\v{z}}m{\'a}r,
  T.} \& \bibinfo{author}{Phillips, D.~B.}
\newblock \bibinfo{title}{High-speed spatial control of the intensity, phase
  and polarisation of vector beams using a digital micro-mirror device}.
\newblock \emph{\bibinfo{journal}{Optics express}}
  \textbf{\bibinfo{volume}{24}}, \bibinfo{pages}{29269--29282}
  (\bibinfo{year}{2016}).

\bibitem{Hellman2019}
\bibinfo{author}{Hellman, B.} \& \bibinfo{author}{Takashima, Y.}
\newblock \bibinfo{title}{Angular and spatial light modulation by single
  digital micromirror device for multi-image output and nearly-doubled
  {\'e}tendue}.
\newblock \emph{\bibinfo{journal}{Optics Express}}
  \textbf{\bibinfo{volume}{27}}, \bibinfo{pages}{21477--21496}
  (\bibinfo{year}{2019}).

\bibitem{Mididoddi2025}
\bibinfo{author}{Mididoddi, C.~K.} \emph{et~al.}
\newblock \bibinfo{title}{Threading light through dynamic complex media}.
\newblock \emph{\bibinfo{journal}{Nature Photonics}}
  \textbf{\bibinfo{volume}{19}}, \bibinfo{pages}{434--440}
  (\bibinfo{year}{2025}).

\bibitem{Sunray2025}
\bibinfo{author}{Sunray, E.}, \bibinfo{author}{Weinberg, G.},
  \bibinfo{author}{Laufer, B.} \& \bibinfo{author}{Katz, O.}
\newblock \bibinfo{title}{Matrix-based imaging through dynamic scattering}.
\newblock \emph{\bibinfo{journal}{Nature Communications}}
  \textbf{\bibinfo{volume}{16}}, \bibinfo{pages}{9413} (\bibinfo{year}{2025}).

\bibitem{Hao2025}
\bibinfo{author}{Hao, Y.}, \bibinfo{author}{Peng, Y.}, \bibinfo{author}{Zhang,
  T.} \& \bibinfo{author}{Chen, W.}
\newblock \bibinfo{title}{High-quality photon-limited imaging through dynamic
  and complex scattering media with a single-photon detector}.
\newblock \emph{\bibinfo{journal}{APL Photonics}} \textbf{\bibinfo{volume}{10}}
  (\bibinfo{year}{2025}).

\bibitem{Sun2021}
\bibinfo{author}{Sun, Y.}, \bibinfo{author}{Wu, X.}, \bibinfo{author}{Zheng,
  Y.}, \bibinfo{author}{Fan, J.} \& \bibinfo{author}{Zeng, G.}
\newblock \bibinfo{title}{Scalable non-invasive imaging through dynamic
  scattering media at low photon flux}.
\newblock \emph{\bibinfo{journal}{Optics and Lasers in Engineering}}
  \textbf{\bibinfo{volume}{144}}, \bibinfo{pages}{106641}
  (\bibinfo{year}{2021}).

\bibitem{Fienup1982}
\bibinfo{author}{Fienup, J.~R.}
\newblock \bibinfo{title}{Phase retrieval algorithms: a comparison}.
\newblock \emph{\bibinfo{journal}{Appl. Opt.}} \textbf{\bibinfo{volume}{21}},
  \bibinfo{pages}{2758--2769} (\bibinfo{year}{1982}).

\bibitem{Vellekoop2010}
\bibinfo{author}{Mosk, A.}
\newblock \bibinfo{title}{Exploiting disorder for perfect focusing}.
\newblock \emph{\bibinfo{journal}{Nature photonics}}
  \textbf{\bibinfo{volume}{4}}, \bibinfo{pages}{320--322}
  (\bibinfo{year}{2010}).

\bibitem{Tyson2022}
\bibinfo{author}{Tyson, R.~K.} \& \bibinfo{author}{Frazier, B.~W.}
\newblock \emph{\bibinfo{title}{Principles of adaptive optics}}
  (\bibinfo{publisher}{CRC press}, \bibinfo{year}{2022}).

\bibitem{Porat2016}
\bibinfo{author}{Porat, A.} \emph{et~al.}
\newblock \bibinfo{title}{Widefield lensless imaging through a fiber bundle via
  speckle correlations}.
\newblock \emph{\bibinfo{journal}{Opt. Express}} \textbf{\bibinfo{volume}{24}},
  \bibinfo{pages}{16835--16855} (\bibinfo{year}{2016}).

\bibitem{Badt2022}
\bibinfo{author}{Badt, N.} \& \bibinfo{author}{Katz, O.}
\newblock \bibinfo{title}{Real-time holographic lensless micro-endoscopy
  through flexible fibers via fiber bundle distal holography}.
\newblock \emph{\bibinfo{journal}{Nature Communications}}
  \textbf{\bibinfo{volume}{13}}, \bibinfo{pages}{6055} (\bibinfo{year}{2022}).

\bibitem{Stasio2016}
\bibinfo{author}{Stasio, N.}, \bibinfo{author}{Moser, C.} \&
  \bibinfo{author}{Psaltis, D.}
\newblock \bibinfo{title}{Calibration-free imaging through a multicore fiber
  using speckle scanning microscopy}.
\newblock \emph{\bibinfo{journal}{Opt. Lett.}} \textbf{\bibinfo{volume}{41}},
  \bibinfo{pages}{3078--3081} (\bibinfo{year}{2016}).

\bibitem{Wu2016}
\bibinfo{author}{Wu, T.}, \bibinfo{author}{Katz, O.}, \bibinfo{author}{Shao,
  X.} \& \bibinfo{author}{Gigan, S.}
\newblock \bibinfo{title}{Single-shot diffraction-limited imaging through
  scattering layers via bispectrum analysis}.
\newblock \emph{\bibinfo{journal}{Opt. Lett.}} \textbf{\bibinfo{volume}{41}},
  \bibinfo{pages}{5003--5006} (\bibinfo{year}{2016}).

\bibitem{Lubin2019}
\bibinfo{author}{Lubin, G.} \emph{et~al.}
\newblock \bibinfo{title}{Quantum correlation measurement with single photon
  avalanche diode arrays}.
\newblock \emph{\bibinfo{journal}{Opt. Express}} \textbf{\bibinfo{volume}{27}},
  \bibinfo{pages}{32863--32882} (\bibinfo{year}{2019}).

\bibitem{Ghezzi2021}
\bibinfo{author}{Ghezzi, A.} \emph{et~al.}
\newblock \bibinfo{title}{Multispectral compressive fluorescence lifetime
  imaging microscopy with a spad array detector}.
\newblock \emph{\bibinfo{journal}{Optics letters}}
  \textbf{\bibinfo{volume}{46}}, \bibinfo{pages}{1353--1356}
  (\bibinfo{year}{2021}).

\bibitem{Bruschini2019}
\bibinfo{author}{Bruschini, C.}, \bibinfo{author}{Homulle, H.},
  \bibinfo{author}{Antolovic, I.~M.}, \bibinfo{author}{Burri, S.} \&
  \bibinfo{author}{Charbon, E.}
\newblock \bibinfo{title}{Single-photon avalanche diode imagers in
  biophotonics: review and outlook}.
\newblock \emph{\bibinfo{journal}{Light: Science \& Applications}}
  \textbf{\bibinfo{volume}{8}}, \bibinfo{pages}{87} (\bibinfo{year}{2019}).

\bibitem{Slenders2021}
\bibinfo{author}{Slenders, E.} \emph{et~al.}
\newblock \bibinfo{title}{Confocal-based fluorescence fluctuation spectroscopy
  with a spad array detector}.
\newblock \emph{\bibinfo{journal}{Light: Science \& Applications}}
  \textbf{\bibinfo{volume}{10}}, \bibinfo{pages}{31} (\bibinfo{year}{2021}).

\bibitem{Lukashchuk2024}
\bibinfo{author}{Lukashchuk, A.} \emph{et~al.}
\newblock \bibinfo{title}{Photonic-electronic integrated circuit-based coherent
  lidar engine}.
\newblock \emph{\bibinfo{journal}{Nature Communications}}
  \textbf{\bibinfo{volume}{15}}, \bibinfo{pages}{3134} (\bibinfo{year}{2024}).

\end{thebibliography}


\begin{thebibliography}{10}
\expandafter\ifx\csname url\endcsname\relax
  \def\url#1{\burl{#1}}\fi
\expandafter\ifx\csname urlprefix\endcsname\relax\def\urlprefix{URL }\fi
\providecommand{\bibinfo}[2]{#2}
\providecommand{\eprint}[2][]{\url{#2}}
\providecommand{\doi}[1]{\url{https://doi.org/#1}}
\bibcommenthead

\bibitem{supp:Goodman2007}
\bibinfo{author}{Goodman, J.~W.}
\newblock \emph{\bibinfo{title}{Speckle phenomena in optics: theory and applications}}  (\bibinfo{publisher}{Roberts and Company Publishers}, \bibinfo{year}{2007}).

\bibitem{supp:Brillinger2001}
\bibinfo{author}{Brillinger, D.~R.}
\newblock \emph{\bibinfo{title}{Time Series: Data Analysis and Theory}}  (\bibinfo{publisher}{Society for Industrial and Applied Mathematics}, \bibinfo{year}{2001}).

\bibitem{supp:Peligrad2010}
\bibinfo{author}{Peligrad, M.} \& \bibinfo{author}{Wu, W.~B.}
\newblock \bibinfo{title}{Central limit theorem for fourier transforms of stationary processes}.
\newblock \emph{\bibinfo{journal}{The Annals of Probability}} \textbf{\bibinfo{volume}{38}} (\bibinfo{year}{2010}).

\bibitem{supp:Peligrad2019}
\bibinfo{author}{Peligrad, M.} \& \bibinfo{author}{Zhang, N.}
\newblock \bibinfo{title}{Central limit theorem for fourier transform and periodogram of random fields}.
\newblock \emph{\bibinfo{journal}{Bernoulli}} \textbf{\bibinfo{volume}{25}} (\bibinfo{year}{2019}).

\bibitem{supp:Katz2014}
\bibinfo{author}{Katz, O.}, \bibinfo{author}{Heidmann, P.}, \bibinfo{author}{Fink, M.} \& \bibinfo{author}{Gigan, S.}
\newblock \bibinfo{title}{Non-invasive single-shot imaging through scattering layers and around corners via speckle correlations}.
\newblock \emph{\bibinfo{journal}{Nature photonics}} \textbf{\bibinfo{volume}{8}}, \bibinfo{pages}{784--790} (\bibinfo{year}{2014}).

\bibitem{supp:Bertolotti2012}
\bibinfo{author}{Bertolotti, J.} \emph{et~al.}
\newblock \bibinfo{title}{Non-invasive imaging through opaque scattering layers}.
\newblock \emph{\bibinfo{journal}{Nature}} \textbf{\bibinfo{volume}{491}}, \bibinfo{pages}{232--234} (\bibinfo{year}{2012}).

\bibitem{supp:Hofer2018}
\bibinfo{author}{Hofer, M.}, \bibinfo{author}{Soeller, C.}, \bibinfo{author}{Brasselet, S.} \& \bibinfo{author}{Bertolotti, J.}
\newblock \bibinfo{title}{Wide field fluorescence epi-microscopy behind a scattering medium enabled by speckle correlations}.
\newblock \emph{\bibinfo{journal}{Optics express}} \textbf{\bibinfo{volume}{26}}, \bibinfo{pages}{9866--9881} (\bibinfo{year}{2018}).

\bibitem{supp:Fienup1982}
\bibinfo{author}{Fienup, J.~R.}
\newblock \bibinfo{title}{Phase retrieval algorithms: a comparison}.
\newblock \emph{\bibinfo{journal}{Appl. Opt.}} \textbf{\bibinfo{volume}{21}}, \bibinfo{pages}{2758--2769} (\bibinfo{year}{1982}).

\bibitem{supp:Nomerotski2023}
\bibinfo{author}{Nomerotski, A.} \emph{et~al.}
\newblock \bibinfo{title}{Intensified tpx3cam, a fast data-driven optical camera with nanosecond timing resolution for single photon detection in quantum applications}.
\newblock \emph{\bibinfo{journal}{Journal of Instrumentation}} \textbf{\bibinfo{volume}{18}}, \bibinfo{pages}{C01023} (\bibinfo{year}{2023}).
\newblock \urlprefix\url{https://doi.org/10.1088/1748-0221/18/01/C01023}.

\bibitem{supp:Ianzano2020}
\bibinfo{author}{Ianzano, C.} \emph{et~al.}
\newblock \bibinfo{title}{Fast camera spatial characterization of photonic polarization entanglement}.
\newblock \emph{\bibinfo{journal}{Scientific reports}} \textbf{\bibinfo{volume}{10}}, \bibinfo{pages}{6181} (\bibinfo{year}{2020}).

\bibitem{supp:Bruschini2019}
\bibinfo{author}{Bruschini, C.}, \bibinfo{author}{Homulle, H.}, \bibinfo{author}{Antolovic, I.~M.}, \bibinfo{author}{Burri, S.} \& \bibinfo{author}{Charbon, E.}
\newblock \bibinfo{title}{Single-photon avalanche diode imagers in biophotonics: review and outlook}.
\newblock \emph{\bibinfo{journal}{Light: Science \& Applications}} \textbf{\bibinfo{volume}{8}}, \bibinfo{pages}{87} (\bibinfo{year}{2019}).

\end{thebibliography}




\resetlinenumber[1]

\clearpage
\begingroup

\makeatletter
\def\toclevel@section{2}
\def\toclevel@subsection{3}
\def\toclevel@subsubsection{4}
\def\toclevel@paragraph{5}
\def\toclevel@subparagraph{6}

\renewcommand*{\theHsection}{supplement.\arabic{section}}
\renewcommand*{\theHsubsection}{
  supplement.\arabic{section}.\arabic{subsection}
}
\renewcommand*{\theHsubsubsection}{
  supplement.\arabic{section}.\arabic{subsection}.\arabic{subsubsection}
}
\renewcommand*{\theHfigure}{supplement.\arabic{figure}}
\renewcommand*{\theHtable}{supplement.\arabic{table}}
\renewcommand*{\theHequation}{supplement.\arabic{equation}}
\makeatother

\phantomsection
\pdfbookmark[1]{Supplementary Information}{supplementary-information}

\setcounter{section}{0}
\setcounter{subsection}{0}
\setcounter{subsubsection}{0}
\setcounter{figure}{0}
\setcounter{table}{0}
\setcounter{equation}{0}

\begin{center}
  {\Large\bfseries Supplementary Information\par}
  \vspace{0.8em}
  {\large
    Low Photon Number Non-Invasive Imaging Through
    Time-Varying Diffusers
    \par
  }
\end{center}

\vspace{1em}

\section{Experimental setup}\label{Setup}

\begin{figure}[h!] 
\centering
\includegraphics[width=0.7\linewidth]{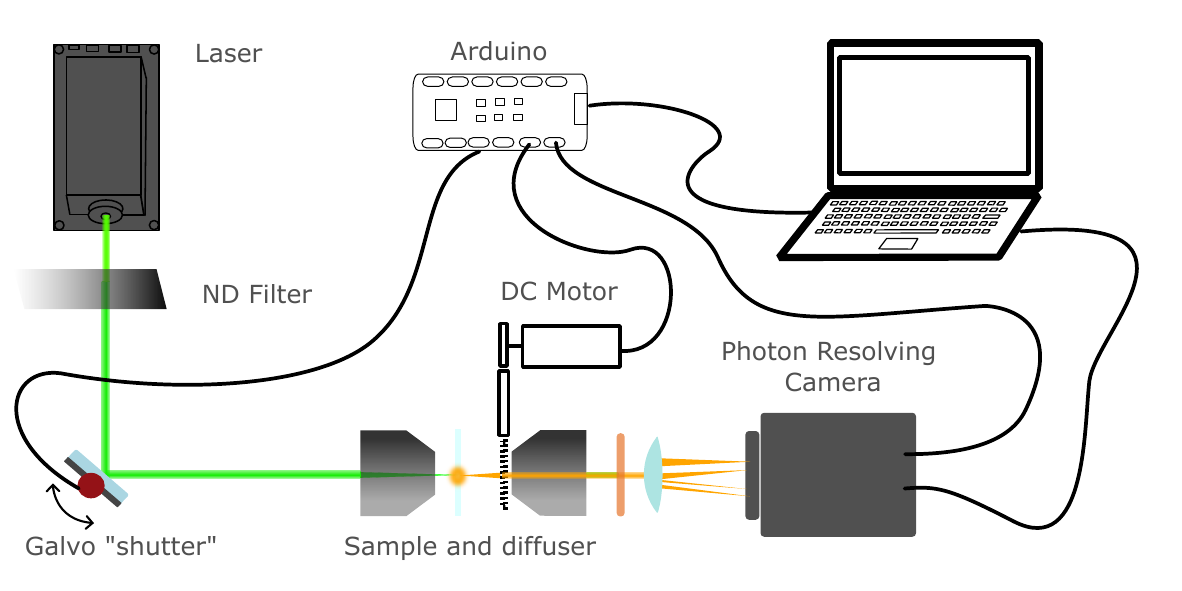}
\caption{Schematic of the experimental setup: The 532 nm excitation laser reflects off the galvo mirror, which acts as a shutter. During acquisition, the galvo mirror directs the light into the objective, illuminating the desired area on the fluorescent sample (beads or astrocytes). The fluorescence passes through a diffuser, which is rotated by a random angle between acquisitions. A second objective collects the scattered fluorescent light. The fluorescence is then separated from the remaining pump light using spectral filters. Finally, it is imaged onto a photon-resolving camera via a tube lens. Acquisition and illumination are synchronized using an Arduino board.}
\label{fig:scheme}
\end{figure}

For the experimental validation of our concept, we constructed a custom microscope setup. It was designed for stability during long experiments, flexibility, and simplicity. As the excitation source, we used a Spectra-Physics Millennia PRO laser with a wavelength of 532 nm. The laser was attenuated using a reflective neutral density filter (Thorlabs ND30A) and a gradient ND filter (Thorlabs NDL-10C-4). The beam was then directed by a galvanometric mirror (Thorlabs GVS011/M). This mirror acted as a shutter, directing the pump light to the rest of the setup only during camera exposure (“ON-state”). This reduced the bleaching of fluorophores during long experiments.

The 532 nm light passed through the objective on the transmissive side (Nikon S Plan Fluor ELWD 40x/0.60). The distance of the objective to the sample determined the illumination area. Green light excited the fluorescent beads or Alexa Fluor 568-stained astrocyte cultures. The resulting fluorescence propagated to a diffuser (Edmund Holographic Diffuser with a 5° diffusing angle). The diffuser was housed in a 3D-printed enclosure that allowed it to rotate freely. It was coupled to a DC motor with a set of gears. A slight loosening of the parts increased the randomization of the diffuser placement. The diffuser was positioned close to a second objective (Nikon S Plan Fluor ELWD 20x/0.45), which collected the scattered fluorescent light. The collected fluorescence passed through a dichroic mirror (Thorlabs DMLP550R). This mirror reflected most of the pump light. Residual 532 nm light was filtered using a notch filter (Semrock NF01-532U-25) and a bandpass filter (Thorlabs FBH590-10). The fluorescence was collected with a tube lens (Thorlabs TTL200-A). This lens projected the speckle pattern onto a QCMOS sensor (Hamamatsu Orca Quest). Data collection was managed by a computer connected to an Arduino board.

The Arduino was used to synchronize the camera, galvo mirror, and rotating diffuser. For each frame, a command was sent to the Arduino to switch on the DC motor for a random duration (0 to 0.5 seconds). After rotation, the galvo mirror was set to “ON-state,” and the camera exposure was triggered. The exposure time was read from the camera’s output electrical lines. Once the exposure ended, the galvo mirror was turned to “OFF-state” to avoid unnecessary bleaching of fluorescence samples. Frames were collected using Python code with Micro-Manager for easy camera access.

For our experiments, we collected 1000 to 10000 frames. The photon-resolving mode of the camera allowed for short exposures (up to 1 µs). However, it required long readout times (around 200 ms per frame). This extended the duration of a single measurement to several hours.

\section{Theoretical investigation}\label{Theory}
\subsection{Mathematical description}

\subsubsection{Noiseless case}
Let $I_i$ be the light intensity at the focal plane array, and let
$I_i((n_x,n_y))$ denote the light intensity at the pixel with position
$(n_x,n_y)$. Let the image have dimensions $N_x \times N_y$, with
$N=N_xN_y$ pixels. For simplicity, we use single-index notation
$(n_x,n_y)\rightarrow n$, remembering that the full two-dimensional notation
can be recovered at any point.

Speckle properties were described in \cite{supp:Goodman2007}. In particular, the
properties of interest to us are:
\begin{equation}
    \mathbb{E}[S_i(n)] = p ,
\end{equation}
\begin{equation}
    \mathbb{E}[S_i(n)S_i(m)] = p^2\left(1+g^2(m-n)\right).
\end{equation}
Here, $\mathbb{E}[\cdot]$ denotes the expectation over an ensemble of speckle
realizations, and $p$ is the average number of photons per pixel. The function
$g(m-n)$ is the complex correlation coefficient
\cite[p.~132, Eq.~(5.8)]{supp:Goodman2007}; it describes the spatial correlation of
the speckle field. In particular, $g(0)=1$ and $|g(\Delta n)|\leq 1$ by
construction.

The intensity on the $i$-th camera frame is described by
\begin{equation}
    I_i(n) = (O \ast S_i)(n)=\sum_m O(m)S_i(n-m).
\end{equation}
Therefore,
\begin{equation}
    \mathbb{E}[I_i(n)]
    =
    \sum_m O(m)\mathbb{E}[S_i(n-m)]
    =
    p\sum_m O(m).
\end{equation}
Since $p$ controls the average image intensity, we assume without loss of
generality that
\begin{equation}
    \sum_m O(m)=1.
\end{equation}
Thus,
\begin{equation}
    \mathbb{E}[I_i(n)]=p.
\end{equation}

Now we calculate the image spatial correlation:
\begin{equation}
    I_i(n_1)I_i(n_2)
    =
    \sum_{m_1}\sum_{m_2}
    O(m_1)O(m_2)
    S_i(n_1-m_1)S_i(n_2-m_2).
\end{equation}
Taking the ensemble average gives
\begin{equation}
    \mathbb{E}[I_i(n_1)I_i(n_2)]
    =
    \sum_{m_1}\sum_{m_2}
    O(m_1)O(m_2)
    \mathbb{E}\!\left[
        S_i(n_1-m_1)S_i(n_2-m_2)
    \right].
\end{equation}
Using the second-order speckle statistics,
\begin{equation}
    \mathbb{E}[I_i(n_1)I_i(n_2)]
    =
    p^2
    \sum_{m_1}\sum_{m_2}
    O(m_1)O(m_2)
    \left[
        1+
        g^2\!\left(n_2-n_1-(m_2-m_1)\right)
    \right].
\end{equation}
We introduce the autocorrelation of the object,
\begin{equation}
    R_O(l)=\sum_m O(m)O(m+l),
\end{equation}
and obtain
\begin{equation}
    \mathbb{E}[I_i(n_1)I_i(n_2)]
    =
    p^2
    \left[
        1+
        \sum_l R_O(l)g^2(n_2-n_1-l)
    \right].
\end{equation}
Equivalently,
\begin{equation}
    \mathbb{E}[I_i(n_1)I_i(n_2)]
    =
    p^2\left[1+(R_O\ast g^2)(n_2-n_1)\right].
\end{equation}
We define
\begin{equation}
    h_2(n)=(R_O\ast g^2)(n),
\end{equation}
so that
\begin{equation}
    \mathbb{E}[I_i(n_1)I_i(n_2)]
    =
    p^2\left[1+h_2(n_2-n_1)\right].
    \label{eq:E2I}
\end{equation}
Consequently,
\begin{equation}
    \operatorname{Cov}\!\left(I_i(n_1),I_i(n_2)\right)
    =
    p^2 h_2(n_2-n_1).
\end{equation}

The function $h_2$ is therefore the normalized autocovariance of the image
intensity:
\begin{equation}
    h_2(n_2-n_1)
    =
    \frac{
        \operatorname{Cov}\!\left(I_i(n_1),I_i(n_2)\right)
    }{p^2}.
\end{equation}
Since $h_2=R_O\ast g^2$, its Fourier transform factorizes into the product of
the object power spectrum and the speckle-correlation transfer function.

The normalization of $R_O$ follows from
\begin{equation}
    \sum_n R_O(n)
    =
    \sum_n\sum_m O(m)O(n+m)
    =
    \sum_m O(m)\sum_n O(n+m)
    =
    1.
    \label{eq:sumRO}
\end{equation}
Therefore,
\begin{equation}
    \min_{n'} g^2(n')
    \leq
    h_2(n)
    =
    \sum_m R_O(m)g^2(n-m)
    \leq
    \max_{n'} g^2(n').
\end{equation}
In particular, when $g^2$ is bounded by unity,
\begin{equation}
    -1 \leq h_2(n) \leq 1.
    \label{eq:defh_2}
\end{equation}

Now we apply the discrete Fourier transform:
\begin{equation}
    \mathcal{F}\{I_i\}((k_x,k_y))
    =
    \sum_{(n_x,n_y)}
    I_i((n_x,n_y))
    \exp\left(
        -i\frac{2\pi}{N_x}k_xn_x
        -i\frac{2\pi}{N_y}k_yn_y
    \right).
\end{equation}
In single-index notation, this becomes
\begin{equation}
    \mathcal{F}\{I_i\}(k)
    =
    \sum_n I_i(n)q^{kn},
\end{equation}
where
\begin{equation}
    q=\exp\left(-i\frac{2\pi}{N}\right).
\end{equation}
Here we also use the single-index notation $(k_x,k_y)\rightarrow k$, with the
understanding that the full two-dimensional notation can be restored when
needed.

The first moment in Fourier space is
\begin{equation}
    \mathbb{E}\!\left[\mathcal{F}\{I_i\}(k)\right]
    =
    \sum_n \mathbb{E}[I_i(n)]q^{kn}
    =
    pN\delta_{k,0}.
\end{equation}
Here, $pN$ is the average total number of photons in one frame. First-order
statistics only contain the DC component and do not allow us to retrieve the
object. We therefore consider second-order statistics.

The squared modulus of the Fourier transform is
\begin{equation}
    |\mathcal{F}\{I_i\}|^2(k)
    =
    \left(
        \sum_{n_1} I_i(n_1)q^{-kn_1}
    \right)
    \left(
        \sum_{n_2} I_i(n_2)q^{kn_2}
    \right),
\end{equation}
hence
\begin{equation}
    |\mathcal{F}\{I_i\}|^2(k)
    =
    \sum_{n_1}\sum_{n_2}
    I_i(n_1)I_i(n_2)q^{k(n_2-n_1)}.
\end{equation}
Taking the expectation and using Eq.~\eqref{eq:E2I}, we get
\begin{equation}
    \mathbb{E}\!\left[|\mathcal{F}\{I_i\}|^2(k)\right]
    =
    p^2
    \sum_{n_1}\sum_{n_2}
    \left[1+h_2(n_2-n_1)\right]
    q^{k(n_2-n_1)}.
\end{equation}
Separating the constant and covariance terms gives
\begin{equation}
    \mathbb{E}\!\left[|\mathcal{F}\{I_i\}|^2(k)\right]
    =
    p^2N^2\delta_{k,0}
    +
    p^2
    \sum_{n_1}\sum_{n_2}
    h_2(n_2-n_1)q^{k(n_2-n_1)}.
\end{equation}
Now we reindex the second sum with $l=n_2-n_1$. For each $l$ there are $N$
values of $n_1$ under periodic boundary conditions, so
\begin{equation}
    \mathbb{E}\!\left[|\mathcal{F}\{I_i\}|^2(k)\right]
    =
    p^2N^2\delta_{k,0}
    +
    p^2N
    \sum_l h_2(l)q^{kl}.
\end{equation}
We define
\begin{equation}
    H_2(k)
    =
    \mathcal{F}\{h_2\}(k)
    =
    \sum_l h_2(l)q^{kl}.
\end{equation}
Therefore,
\begin{equation}
    \boxed{
    \mathbb{E}\!\left[|\mathcal{F}\{I_i\}|^2(k)\right]
    =
    p^2N^2\delta_{k,0}
    +
    p^2N H_2(k)
    }.
    \label{eq:EFI2_H2}
\end{equation}

The function $H_2(k)$ is the discrete Fourier transform of the normalized
autocovariance of $I_i(n)$. Equivalently, up to the factor $p^2N$, it is the
spectral power density of the frame fluctuations after removing the DC
component. For $k\neq 0$,
\begin{equation}
    \mathbb{E}\!\left[|\mathcal{F}\{I_i\}|^2(k)\right]
    =
    p^2N H_2(k).
\end{equation}

Because
\begin{equation}
    h_2 = R_O\ast g^2,
\end{equation}
the convolution theorem gives
\begin{equation}
    H_2(k)
    =
    \mathcal{F}\{R_O\}(k)\mathcal{F}\{g^2\}(k).
\end{equation}
Since the Fourier transform of the autocorrelation is the power spectrum,
\begin{equation}
    \mathcal{F}\{R_O\}(k)
    =
    |\mathcal{F}\{O\}(k)|^2.
\end{equation}
Thus,
\begin{equation}
    H_2(k)
    =
    |\mathcal{F}\{O\}(k)|^2
    \mathcal{F}\{g^2\}(k).
\end{equation}
We denote the effective point-spread function of the imaging system by $P$ and
define it through
\begin{equation}
    |\mathcal{F}\{P\}(k)|^2
    =
    \mathcal{F}\{g^2\}(k).
\end{equation}
Then
\begin{equation}
    H_2(k)
    =
    |\mathcal{F}\{O\}(k)|^2
    |\mathcal{F}\{P\}(k)|^2.
\end{equation}
Therefore, the experimentally accessible Fourier-modulus quantity is
\begin{equation}
    H(k)
    =
    \sqrt{H_2(k)}
    =
    |\mathcal{F}\{O\}(k)|
    |\mathcal{F}\{P\}(k)|.
\end{equation}
This is the modulus of the object Fourier transform filtered by the effective
optical transfer function. In the absence of noise and after removal of the DC
component, $H(k)$ is the normalized Fourier-modulus quantity that we aim to
retrieve experimentally.

\subsubsection{Poissonian noise case}
\paragraph{Image space}

Let $d_i(n)$ be the number of photons registered at pixel $n$ during the
$i$-th frame. Conditional on the noiseless intensity image $I_i$, photon
counting at each pixel follows a Poisson distribution:
\begin{equation}
    d_i(n)\mid I_i(n) \sim \operatorname{Poisson}(I_i(n)).
\end{equation}

Poisson counting noise is independent between different pixels. Therefore, for
$n_1\neq n_2$, the fluctuations caused by photon counting at pixels $n_1$ and
$n_2$ are independent after conditioning on the noiseless intensity image
$I_i$.

For each pixel,
\begin{equation}
    \mathbb{E}[d_i(n)\mid I_i(n)] = I_i(n),
    \qquad
    \operatorname{Var}(d_i(n)\mid I_i(n)) = I_i(n).
\end{equation}
Using the law of total expectation,
\begin{equation}
    \mathbb{E}[d_i(n)]
    =
    \mathbb{E}\!\left[
        \mathbb{E}[d_i(n)\mid I_i(n)]
    \right]
    =
    \mathbb{E}[I_i(n)]
    =
    p.
\end{equation}

It is useful to write the measured image as the noiseless image plus an
additive shot-noise term:
\begin{equation}
    d_i(n)=I_i(n)+\eta_i(n),
\end{equation}
where
\begin{equation}
    \eta_i(n)=d_i(n)-I_i(n).
\end{equation}
For Poisson noise,
\begin{equation}
    \mathbb{E}[\eta_i(n)\mid I_i]=0.
\end{equation}
Moreover, because photon-counting fluctuations at different pixels are
conditionally independent,
\begin{equation}
    \mathbb{E}[\eta_i(n_1)\eta_i(n_2)\mid I_i]
    =
    I_i(n_1)\delta_{n_1,n_2}.
\end{equation}
After averaging over the ensemble of noiseless images,
\begin{equation}
    \mathbb{E}[\eta_i(n_1)\eta_i(n_2)]
    =
    \mathbb{E}[I_i(n_1)]\delta_{n_1,n_2}
    =
    p\delta_{n_1,n_2}.
    \label{eq:poisson_noise_covariance}
\end{equation}
The cross-term between the noiseless image and the shot noise vanishes:
\begin{equation}
    \mathbb{E}[I_i(n_1)\eta_i(n_2)]
    =
    \mathbb{E}\!\left[
        I_i(n_1)
        \mathbb{E}[\eta_i(n_2)\mid I_i]
    \right]
    =
    0.
\end{equation}

Using Eq.~\eqref{eq:E2I},
\begin{equation}
    \mathbb{E}[I_i(n_1)I_i(n_2)]
    =
    p^2\left[1+h_2(n_2-n_1)\right],
\end{equation}
we obtain
\begin{equation}
    \mathbb{E}[d_i(n_1)d_i(n_2)]
    =
    \mathbb{E}[I_i(n_1)I_i(n_2)]
    +
    \mathbb{E}[\eta_i(n_1)\eta_i(n_2)].
\end{equation}
Therefore,
\begin{equation}
    \boxed{
    \mathbb{E}[d_i(n_1)d_i(n_2)]
    =
    p^2\left[1+h_2(n_2-n_1)\right]
    +
    p\delta_{n_1,n_2}
    }.
    \label{eq:Edd_poisson}
\end{equation}
Equivalently,
\begin{equation}
    \boxed{
    \operatorname{Cov}\!\left(d_i(n_1),d_i(n_2)\right)
    =
    p^2h_2(n_2-n_1)
    +
    p\delta_{n_1,n_2}
    }.
    \label{eq:Covdd_poisson}
\end{equation}

This form separates two effects. The term $p^2h_2(n_2-n_1)$ is the spatial
covariance of the noiseless image. The term $p\delta_{n_1,n_2}$ is the
Poisson shot-noise covariance. It is diagonal in image space; therefore, it is
spatially uncorrelated.

More generally, if
\begin{equation}
    d_i(n)=I_i(n)+\eta_i(n),
\end{equation}
with
\begin{equation}
    \mathbb{E}[\eta_i(n)\mid I_i]=0
\end{equation}
and spatially uncorrelated noise
\begin{equation}
    \mathbb{E}[\eta_i(n_1)\eta_i(n_2)]
    =
    \sigma_\eta^2(n_1)\delta_{n_1,n_2},
\end{equation}
then the noise contributes only a diagonal term in image space. As shown below,
such a diagonal covariance produces a constant, $k$-independent offset in
Fourier power space.

\paragraph{Fourier space}

We use the discrete Fourier transform
\begin{equation}
    \mathcal{F}\{d_i\}(k)
    =
    \sum_n d_i(n)q^{kn},
\end{equation}
where
\begin{equation}
    q=\exp\left(-i\frac{2\pi}{N}\right).
\end{equation}
The first moment is
\begin{equation}
    \mathbb{E}\!\left[\mathcal{F}\{d_i\}(k)\right]
    =
    \sum_n \mathbb{E}[d_i(n)]q^{kn}
    =
    pN\delta_{k,0}.
    \label{eq:EFd}
\end{equation}

Now consider the squared modulus:
\begin{equation}
    |\mathcal{F}\{d_i\}|^2(k)
    =
    \left(
        \sum_{n_1}d_i(n_1)q^{-kn_1}
    \right)
    \left(
        \sum_{n_2}d_i(n_2)q^{kn_2}
    \right).
\end{equation}
Thus,
\begin{equation}
    |\mathcal{F}\{d_i\}|^2(k)
    =
    \sum_{n_1}\sum_{n_2}
    d_i(n_1)d_i(n_2)q^{k(n_2-n_1)}.
\end{equation}
Taking the expectation and using Eq.~\eqref{eq:Edd_poisson},
\begin{equation}
    \mathbb{E}\!\left[|\mathcal{F}\{d_i\}|^2(k)\right]
    =
    p^2
    \sum_{n_1}\sum_{n_2}
    \left[1+h_2(n_2-n_1)\right]q^{k(n_2-n_1)}
    +
    p
    \sum_{n_1}\sum_{n_2}
    \delta_{n_1,n_2}q^{k(n_2-n_1)}.
\end{equation}
The constant-intensity term gives
\begin{equation}
    p^2
    \sum_{n_1}\sum_{n_2}
    q^{k(n_2-n_1)}
    =
    p^2N^2\delta_{k,0}.
\end{equation}
The spatial-covariance term gives, after substituting $l=n_2-n_1$,
\begin{equation}
    p^2
    \sum_{n_1}\sum_{n_2}
    h_2(n_2-n_1)q^{k(n_2-n_1)}
    =
    p^2N
    \sum_l h_2(l)q^{kl}.
\end{equation}
Using
\begin{equation}
    H_2(k)
    =
    \mathcal{F}\{h_2\}(k)
    =
    \sum_l h_2(l)q^{kl},
\end{equation}
we obtain the covariance contribution
\begin{equation}
    p^2N H_2(k).
\end{equation}
The Poisson-noise term gives
\begin{equation}
    p
    \sum_{n_1}\sum_{n_2}
    \delta_{n_1,n_2}q^{k(n_2-n_1)}
    =
    p\sum_n 1
    =
    pN.
\end{equation}
Therefore,
\begin{equation}
    \boxed{
    \mathbb{E}\!\left[|\mathcal{F}\{d_i\}|^2(k)\right]
    =
    p^2N^2\delta_{k,0}
    +
    p^2N H_2(k)
    +
    pN
    }.
    \label{eq:EFd_abs2_poisson}
\end{equation}

For non-zero spatial frequencies, $k\neq 0$, the DC term disappears:
\begin{equation}
    \boxed{
    \mathbb{E}\!\left[|\mathcal{F}\{d_i\}|^2(k)\right]
    =
    p^2N H_2(k)
    +
    pN,
    \qquad k\neq 0.
    }
    \label{eq:EFd_abs2_poisson_nonzero}
\end{equation}
Thus, Poissonian shot noise adds a flat noise floor equal to $pN$, i.e. equal
to the average total number of detected photons per frame.

The same conclusion follows directly from the additive-noise representation.
Since
\begin{equation}
    \mathcal{F}\{d_i\}(k)
    =
    \mathcal{F}\{I_i\}(k)
    +
    \mathcal{F}\{\eta_i\}(k),
\end{equation}
we have
\begin{equation}
    \mathbb{E}\!\left[|\mathcal{F}\{d_i\}|^2(k)\right]
    =
    \mathbb{E}\!\left[|\mathcal{F}\{I_i\}|^2(k)\right]
    +
    \mathbb{E}\!\left[|\mathcal{F}\{\eta_i\}|^2(k)\right],
\end{equation}
because
\begin{equation}
    \mathbb{E}\!\left[
        \mathcal{F}\{I_i\}(k)
        \overline{\mathcal{F}\{\eta_i\}(k)}
    \right]
    =
    0.
\end{equation}
For any spatially uncorrelated noise with
\begin{equation}
    \mathbb{E}[\eta_i(n_1)\eta_i(n_2)]
    =
    \sigma_\eta^2(n_1)\delta_{n_1,n_2},
\end{equation}
the Fourier-domain noise power is
\begin{equation}
    \mathbb{E}\!\left[|\mathcal{F}\{\eta_i\}|^2(k)\right]
    =
    \sum_{n_1}\sum_{n_2}
    \sigma_\eta^2(n_1)
    \delta_{n_1,n_2}
    q^{k(n_2-n_1)}
    =
    \sum_n \sigma_\eta^2(n).
\end{equation}
This quantity is independent of $k$. If the noise variance is spatially
homogeneous, $\sigma_\eta^2(n)=\sigma_\eta^2$, then
\begin{equation}
    \mathbb{E}\!\left[|\mathcal{F}\{\eta_i\}|^2(k)\right]
    =
    N\sigma_\eta^2.
\end{equation}
For Poisson noise,
\begin{equation}
    \sum_n \sigma_\eta^2(n)
    =
    \sum_n \mathbb{E}[I_i(n)]
    =
    pN.
\end{equation}

Finally, because in the noiseless case
\begin{equation}
    H_2(k)
    =
    |\mathcal{F}\{O\}(k)|^2
    |\mathcal{F}\{P\}(k)|^2,
\end{equation}
the measured Fourier power satisfies
\begin{equation}
    \boxed{
    \mathbb{E}\!\left[|\mathcal{F}\{d_i\}|^2(k)\right]
    =
    p^2N^2\delta_{k,0}
    +
    p^2N
    |\mathcal{F}\{O\}(k)|^2
    |\mathcal{F}\{P\}(k)|^2
    +
    pN
    }.
\end{equation}
For $k\neq 0$,
\begin{equation}
    \boxed{
    \mathbb{E}\!\left[|\mathcal{F}\{d_i\}|^2(k)\right]-pN
    =
    p^2N
    |\mathcal{F}\{O\}(k)|^2
    |\mathcal{F}\{P\}(k)|^2
    }.
\end{equation}
Thus, after subtracting the flat Poisson noise floor, the same spectral
quantity as in the noiseless case is recovered:
\begin{equation}
    H_2(k)
    =
    \frac{
        \mathbb{E}\!\left[|\mathcal{F}\{d_i\}|^2(k)\right]
        -
        pN
        -
        p^2N^2\delta_{k,0}
    }{
        p^2N
    }.
\end{equation}
For $k\neq 0$, this reduces to
\begin{equation}
    H_2(k)
    =
    \frac{
        \mathbb{E}\!\left[|\mathcal{F}\{d_i\}|^2(k)\right]-pN
    }{
        p^2N
    }.
\end{equation}
Equivalently,
\begin{equation}
    H(k)
    =
    \sqrt{H_2(k)}
    =
    |\mathcal{F}\{O\}(k)|
    |\mathcal{F}\{P\}(k)|.
\end{equation}

\paragraph{Discussion}

For non-zero spatial frequencies, Eq.~\eqref{eq:EFd_abs2_poisson_nonzero}
gives
\begin{equation}
\mathbb{E}\!\left[
|\mathcal{F}\{d_i\}(k)|^2
\right]
=
p^2 N H_2(k)
+
pN,
\qquad k\neq 0.
\end{equation}
Using
\begin{equation}
H_2(k)
=
|\mathcal{F}\{O\}(k)|^2
|\mathcal{F}\{P\}(k)|^2,
\end{equation}
we obtain
\begin{equation}
\mathbb{E}\!\left[
|\mathcal{F}\{d_i\}(k)|^2
\right]
=
p^2 N
|\mathcal{F}\{O\}(k)|^2
|\mathcal{F}\{P\}(k)|^2
+
pN,
\qquad k\neq 0.
\end{equation}
Thus, the measured Fourier power contains two contributions:
\begin{itemize}
\item\(p^2 N H_2(k)\), the object-dependent spectral term, equal to
\(p^2N|\mathcal{F}\{O\}(k)|^2|\mathcal{F}\{P\}(k)|^2\);
\item\(pN\), a frequency-independent Fourier-power offset produced by
Poisson shot noise. Under the adopted normalization, this term is equal to
the expected total number of detected photons in one frame.
\end{itemize}

The Poisson contribution therefore does not modify the shape of the
object-dependent spectrum, but adds a flat power floor. If\(p\) is known or
estimated from the mean photon count per frame, an unbiased estimate of the
spectral quantity is obtained by subtracting this offset:
\begin{equation}
H_2(k)
=
\frac{
\mathbb{E}\!\left[
|\mathcal{F}\{d_i\}(k)|^2
\right]
-
pN
}{
p^2N
},
\qquad k\neq 0.
\end{equation}

The same interpretation applies to additive, zero-mean, spatially
uncorrelated noise. Let
\begin{equation}
d_i(n)=I_i(n)+\eta_i(n),
\qquad
\mathbb{E}[\eta_i(n)\mid I_i]=0,
\end{equation}
with spatial covariance
\begin{equation}
\mathbb{E}\!\left[
\eta_i(n_1)\eta_i^{*}(n_2)
\right]
=
\sigma_\eta^2(n_1)\delta_{n_1,n_2}.
\end{equation}
Its contribution to the Fourier power is
\begin{equation}
\mathbb{E}\!\left[
|\mathcal{F}\{\eta_i\}(k)|^2
\right]
=
\sum_n \sigma_\eta^2(n)
\equiv
\nu_\eta,
\end{equation}
where\(\nu_\eta\) denotes the total noise power per frame under the adopted
Fourier-transform normalization. Because \(\nu_\eta\) is independent of
$k$, spatially uncorrelated noise produces a flat offset in the Fourier
power spectrum.

Consequently,
\begin{equation}
\mathbb{E}\!\left[
|\mathcal{F}\{d_i\}(k)|^2
\right]
=
\mathbb{E}\!\left[
|\mathcal{F}\{I_i\}(k)|^2
\right]
+
\nu_\eta.
\end{equation}
The corresponding RMS estimator therefore converges to
\begin{equation}
\sqrt{
\mathbb{E}\!\left[
|\mathcal{F}\{d_i\}(k)|^2
\right]
}
=
\sqrt{
\mathbb{E}\!\left[
|\mathcal{F}\{I_i\}(k)|^2
\right]
+
\nu_\eta
},
\end{equation}
rather than to the noise-free Fourier magnitude itself.

Averaging over multiple frames reduces the statistical uncertainty of the
empirical Fourier-power estimate, but does not remove the offset
\(\nu_\eta\). In practice, this flat noise floor may be estimated from photon
statistics, dark or background measurements, or spectral regions in which
the object-dependent signal is negligible, and then subtracted before taking
the square root. By contrast, spatially correlated noise generally produces
a\(k\)-dependent Fourier-power contribution and therefore biases the shape
of the recovered spectrum.

\subsubsection{Asymptotic distribution of $\mathcal{F}\{d_i\}(k)$}

\paragraph{Gaussian approximation}
The exact finite-$N$ distribution of $\mathcal{F}\{d_i\}(k)$ would be difficult to write explicitly but luckily is not needed in present derivation. Here we only need a tractable approximation for the
distribution of individual Fourier coefficients. Such a distribution , because $\mathcal{F}\{d_i\}(k)$ is a Fourier sum of a
spatially correlated intensity field with additional Poisson counting noise.
Nevertheless, even without knowing the exact finite-$N$ distribution, the
Fourier coefficients have simple and useful asymptotic properties under
standard large-sample conditions.

For the complex optical field of fully developed speckle, Gaussianity is a
classical consequence of the random-phasor model: the field is modeled as a
circular complex Gaussian random process
\cite{supp:Goodman2007}. Since the discrete Fourier transform is a
linear operation, Fourier coefficients of the complex optical field are also
complex Gaussian.

In the present work, however, the camera records intensity, not the complex
field. The intensity speckle pattern is a nonlinear transformation of the
field, and its Fourier coefficients are not exactly Gaussian at finite $N$.
We therefore use a weaker and more appropriate large-$N$ statement. When the
camera field of view contains a large number of effectively independent
speckle grains, or more generally many weakly dependent spatial regions, the
non-DC and non-self-conjugate Fourier coefficients of the measured intensity
image are well approximated by circular complex normal variables. This is
consistent with the standard spectral-domain asymptotics for finite Fourier
transforms of sufficiently regular stationary processes and random fields
\cite{supp:Brillinger2001, supp:Peligrad2010, supp:Peligrad2019}.

The special frequencies $k=0$ and, for even $N$, $k=N/2$ are excluded from the
following approximation. At these frequencies the Fourier coefficient of a
real-valued image is real-valued, so it is not circular complex normal. For the
remaining frequencies, the real and imaginary parts behave as two quadrature
components of the same complex random variable.

\paragraph{Limiting distribution}

We define the normalized Fourier coefficient
\begin{equation}
    D_i(k)
    =
    \frac{1}{\sqrt{N}}\mathcal{F}\{d_i\}(k).
\end{equation}
For $k\neq 0$, this is equivalent to the centered definition
\begin{equation}
    D_i(k)
    =
    \frac{1}{\sqrt{N}}\mathcal{F}\{d_i-p\}(k),
\end{equation}
because the constant component $p$ contributes only to the DC frequency:
\begin{equation}
    \mathcal{F}\{d_i-p\}(k)
    =
    \mathcal{F}\{d_i\}(k),
    \qquad k\neq 0.
\end{equation}

Using Eq.~\eqref{eq:EFd}, for $k\neq 0$ we obtain
\begin{equation}
    \mathbb{E}[D_i(k)] = 0.
\end{equation}
For non-self-conjugate Fourier modes, the second non-conjugated moment
vanishes,
\begin{equation}
    \mathbb{E}\!\left[D_i(k)^2\right] = 0.
\end{equation}
This is the condition that makes the complex Gaussian approximation circular.
Using Eq.~\eqref{eq:EFd_abs2_poisson}, the conjugated second moment is
\begin{equation}
    \mathbb{E}\!\left[|D_i(k)|^2\right]
    =
    p+p^2H_2(k),
    \qquad k\neq 0.
\end{equation}
Thus, the only non-zero second-order parameter of the limiting circular
complex distribution is its variance.

Therefore, for $k\neq 0$ and $k\not\equiv -k \pmod N$, we use
\begin{equation}
    \boxed{
    D_i(k)
    \xrightarrow{d}
    \mathcal{CN}\!\left(
        0,\,
        p+p^2H_2(k)
    \right)
    }.
\end{equation}
Equivalently, for large but finite $N$,
\begin{equation}
    \boxed{
    \frac{1}{\sqrt{N}}\mathcal{F}\{d_i\}(k)
    \approx
    \mathcal{CN}\!\left(
        0,\,
        p+
        p^2|\mathcal{F}\{O\}(k)|^2|\mathcal{F}\{P\}(k)|^2
    \right)
    }.
\end{equation}
Here, $\mathcal{CN}(\mu,\sigma^2)$ denotes a circular complex normal
distribution with mean $\mu$ and variance
\begin{equation}
    \mathbb{E}\!\left[|D_i(k)-\mu|^2\right]=\sigma^2.
\end{equation}

This approximation has a direct interpretation. The term $p^2H_2(k)$ comes
from the spatial correlations of the noiseless speckle image, while the
additive term $p$ comes from Poisson shot noise. Therefore, shot noise
increases the variance of the complex Fourier coefficient, but does not change
the $k$-dependent spectral shape.

The approximation should be understood as a large-sample approximation. Its
accuracy improves as the number of effectively independent spatial
contributions within the camera frame increases, for example when the number of
camera pixels is large compared with the speckle correlation area. If the
number of photons is low, or if the speckle grains are strongly correlated over
a large fraction of the camera frame, the finite-$N$ distribution can deviate
from the circular complex normal model. This does not affect the second-moment
identity
\begin{equation}
    \mathbb{E}\!\left[|\mathcal{F}\{d_i\}|^2(k)\right]
    =
    p^2N^2\delta_{k,0}
    +
    p^2NH_2(k)
    +
    pN,
\end{equation}
which follows directly from the covariance structure and remains valid without
the Gaussian approximation. The complex normal model is used only as a
tractable distributional approximation for finite-sample Fourier coefficients
and for the variability of empirical spectral estimates.

\subsubsection{Estimation of \(H(k)\)}

This section focuses on \( k \neq 0 \) and \( k \neq \frac{N}{2} \) i.e.
on frequencies which the circular complex normal approximation from the
previous section applies.

Recall that
\begin{equation}
    D_i(k)
    =
    \frac{1}{\sqrt{N}}\mathcal{F}\{d_i\}(k),
\end{equation}
and
\begin{equation}
    D_i(k)
    \approx
    \mathcal{CN}\!\left(0,\sigma_k^2\right),
\end{equation}
with
\begin{equation}
    \sigma_k^2
    =
    p+p^2H_2(k)
    =
    p+p^2H^2(k).
    \label{eq:sigma_k_estimation}
\end{equation}
For a circular complex normal random variable, the squared modulus follows an
exponential distribution. Therefore,
\begin{equation}
    X_i(k)
    =
    |D_i(k)|^2
    \sim
    \operatorname{Exp}(\sigma_k^2),
\end{equation}
where we use the scale parametrization:
\begin{equation}
    f(x;\sigma_k^2)
    =
    \frac{1}{\sigma_k^2}
    \exp\!\left(-\frac{x}{\sigma_k^2}\right),
    \qquad x\geq 0.
\end{equation}
Thus,
\begin{equation}
    \mathbb{E}[X_i(k)] = \sigma_k^2,
    \qquad
    \operatorname{Var}(X_i(k)) = \sigma_k^4.
\end{equation}

In typical experiments, \(p\) can be estimated from the mean photon count per
frame, equivalently from the DC Fourier component. In the following derivation
we treat \(p\) as known.

\paragraph{Maximum likelihood estimator}

For a fixed frequency \(k\), let
\begin{equation}
    X_i(k)=|D_i(k)|^2,
    \qquad i=1,\ldots,M,
\end{equation}
be the powers measured from \(M\) independent frames. The likelihood is
\begin{equation}
    L(\sigma_k^2)
    =
    \prod_{i=1}^{M}
    \frac{1}{\sigma_k^2}
    \exp\!\left(-\frac{X_i(k)}{\sigma_k^2}\right).
\end{equation}
The log-likelihood is
\begin{equation}
    \ell(\sigma_k^2)
    =
    -M\ln(\sigma_k^2)
    -
    \frac{1}{\sigma_k^2}
    \sum_{i=1}^{M}X_i(k).
\end{equation}
Maximizing with respect to \(\sigma_k^2\) gives
\begin{equation}
    \widehat{\sigma}_k^2
    =
    \frac{1}{M}\sum_{i=1}^{M}|D_i(k)|^2.
\end{equation}
Using Eq.~\eqref{eq:sigma_k_estimation}, the corresponding estimator of
\(H_2(k)=H^2(k)\) is
\begin{equation}
    \widehat{H}_2^{\,\mathrm{unc}}(k)
    =
    \frac{
        \frac{1}{M}\sum_{i=1}^{M}|D_i(k)|^2
        -
        p
    }{
        p^2
    }.
    \label{eq:H2_unc_estimator}
\end{equation}
This is the unconstrained estimator. Since \(H_2(k)\geq 0\), the constrained
maximum likelihood estimator is
\begin{equation}
    \widehat{H}_2(k)
    =
    \max\!\left[
        0,\,
        \frac{
            \frac{1}{M}\sum_{i=1}^{M}|D_i(k)|^2
            -
            p
        }{
            p^2
        }
    \right].
    \label{eq:H2_constrained_estimator}
\end{equation}

Imposing the physical constraint \(H_2[\boldsymbol{k}]\ge 0\), the estimator of the Fourier modulus
\[
H[\boldsymbol{k}]
=
\sqrt{H_2[\boldsymbol{k}]}
\]
is
\begin{equation}
\widehat{H}[\boldsymbol{k}]
=
\begin{cases}
\dfrac{1}{p}
\sqrt{
\widehat{\sigma}_{\boldsymbol{k}}^2-p
},
&
\widehat{\sigma}_{\boldsymbol{k}}^2>p,
\\[10pt]
0,
&
\widehat{\sigma}_{\boldsymbol{k}}^2\le p.
\end{cases}
\quad \text{where}\quad   \widehat{\sigma}_{\boldsymbol{k}}^2=\frac{1}{M}\sum_{i=1}^{M}|D_i(k)|^2
\label{eq:H_hat_from_H2_hat}
\end{equation}

\paragraph{Uncertainty}

Since \(X_i(k)\) are exponentially distributed with scale \(\sigma_k^2\),
\begin{equation}
    \operatorname{Var}\!\left[
        \frac{1}{M}\sum_{i=1}^{M}X_i(k)
    \right]
    =
    \frac{\sigma_k^4}{M}.
\end{equation}
Therefore, for the unconstrained estimator of \(H_2(k)\),
\begin{equation}
    \operatorname{Var}\!\left[
        \widehat{H}_2^{\,\mathrm{unc}}(k)
    \right]
    =
    \frac{\sigma_k^4}{M p^4}
    =
    \frac{\left(p+p^2H^2(k)\right)^2}{M p^4}
    =
    \frac{\left(1+pH^2(k)\right)^2}{M p^2}.
\end{equation}

This estimator is the maximum likelihood estimator obtained from the
exponential model for the measured Fourier powers. Therefore, for each
frequency treated separately, it is asymptotically unbiased and efficient. In
particular, for large \(M\), its variance reaches the Cramér--Rao lower bound
for this elementwise estimation problem. In this sense, under the assumed
model and when no information is shared between different frequencies, no
regular estimator can asymptotically achieve a smaller variance for estimating
\(H_2(k)\).

Using the delta method for
\begin{equation}
    H(k)=\sqrt{H_2(k)},
\end{equation}
we obtain, for \(H(k)>0\),
\begin{equation}
    \operatorname{Var}\!\left[\widehat{H}(k)\right]
    \approx
    \frac{
        \left(1+pH^2(k)\right)^2
    }{
        4Mp^2H^2(k)
    }.
\end{equation}

We can compute the relative standard error \(RSE\) of the estimator, which we
interpret as the \textbf{error of estimation} for a given spatial frequency:
\begin{equation}
\operatorname{RSE}_H(k)
=
\frac{
\sqrt{\operatorname{Var}\!\left[\widehat{H}(k)\right]}
}{
H(k)
}
\approx
\frac{1}{2\sqrt{M}}
\left(
1+\frac{1}{pH^2(k)}
\right).
\label{eq:est_err}
\end{equation}

This expression makes the scaling of the estimation error explicit. The
relative standard error decreases as \(M^{-1/2}\), so averaging more
independent frames improves the estimate only with the usual square-root law.
The second factor depends on the local spectral strength \(H(k)\) and on the
photon budget \(p\). Frequencies for which \(H(k)\) is small are intrinsically
harder to estimate, because the object-dependent contribution \(p^2H^2(k)\)
becomes small compared with the Poisson floor \(p\). Accurate estimation of
such frequencies therefore requires either more frames \(M\), a larger number
of photons per pixel \(p\), or both.

The limiting cases show the same trade-off explicitly. When \(pH^2(k)\gg 1\),
the estimator is limited mainly by the finite number of frames and
\(\operatorname{RSE}_H(k)\approx 1/(2\sqrt{M})\). When \(pH^2(k)\ll 1\), the
Poisson floor dominates and
\begin{equation}
\operatorname{RSE}_H(k)
\approx
\frac{1}{2\sqrt{M}\,pH^2(k)}.
\end{equation}
Thus, weak Fourier components require a rapidly increasing number of frames to
be recovered with a fixed relative accuracy.

\subsubsection{Approximating variables for experiment parameters}
If we now recover our 2D indexing, we can see that:

\begin{equation}
	\begin{split}
		\mathcal{F}\{O(n_x, n_y)\}(k_x, k_y) &= \sum_{n_x=0}^{N_x} \sum_{n_y=0}^{N_y} O(n_x, n_y) \exp\left( -i \frac{2\pi}{N_x} n_x k_x  - i \frac{2\pi}{N_y} n_y k_y \right) \leq \\
		&\qquad \leq \sum_{n_x=0}^{N_x} \sum_{n_y=0}^{N_y} O(n_x, n_y) \leq 1
	\end{split}
\end{equation}

In our experiment, speckles had an average diameter width of 2.7, so we can approximate that:

\begin{equation}
	g(n_x, n_y) = \exp\left( -\frac{n_x^2 + n_y^2}{2\sigma_g^2} \right)
\end{equation}

\begin{equation}
	\sigma_g = 1.35
\end{equation}

Then,

\begin{equation}
	\mathcal{F}\{g(n_x, n_y)\}(k_x, k_y) \approx 2\pi \sigma_g^2 \exp\left( -\frac{k_x^2 + k_y^2}{2\sigma_{fg}^2} \right)
\end{equation}

where

\begin{equation}
	\sigma_{fg} = \frac{1}{2\pi \sigma_g}
\end{equation}

Then,

\begin{equation}
	H^2 = |\mathcal{F}\{g\}|^2 |\mathcal{F}\{O\}|^2
\end{equation}

\begin{equation}
	0 \leq H^2(k_x, k_y) \leq 4\pi^2 \sigma_g^4 \approx 131
\end{equation}

\subsubsection{Discussion}
As we can see in Eq.~\ref{eq:est_err}, the relative standard error for the approximation of the modulus Fourier transform \( H(k) = H(k_x, k_y) \) depends on its value. \( H(k_x, k_y) \) with large values, which are usually close to \( (0,0) \), will be estimated better. However, there are always a lot of small \( H(k_x, k_y) \) for high frequencies, they will be approximated less accurately as the term \( \frac{1}{p H^2} \) will dominate. Therefore, to recover the entire Fourier transform \( H \) with good accuracy, we need a lot of frames, even if the average number of photons per pixel \( p \) is large.

Now, this explains why simple images with few isolated points (e.g., Fig.~4 from the article) are reconstructed better than complex ones with smooth continuous images (e.g., glial cell from Fig.~3 in the article). The former have their Fourier intensity concentrated in isolated points, so the loss of high frequencies with low intensity does not result in a significant loss of object detail, unlike in complex images.

\paragraph{Limit on Resolution}

The quantity recovered from the measurements is
\begin{equation}
H^2(k)
=
\frac{
|\mathcal{F}\{O\}(k)|^2
|\mathcal{F}\{P\}(k)|^2
}{
p^2 N
}.
\end{equation}
Consequently, no information about the object Fourier modulus can be recovered
at spatial frequencies for which
\(\lvert\mathcal{F}\{P\}(k)\rvert=0\).
The ultimate resolution is therefore determined by the spatial-frequency
bandwidth of the speckle intensity patterns. Equivalently, its characteristic
linear scale is of the order of the speckle-grain size.

To resolve a spatial scale \(\Delta_x\), the Fourier modulus must be estimated
at frequencies of order \(k\sim 1/\Delta_x\). The mean signal-to-noise ratio
of this estimate is approximately
\begin{equation}
\operatorname{SNR}*{H}(k)
\approx
2\sqrt{M},
\frac{pH^2(k)}{1+pH^2(k)}.
\label{eq:snr_h}
\end{equation}
A frequency component may be regarded as statistically resolved when
\(\operatorname{SNR}*{H}(k)\gtrsim 1\). Thus, the number of detected photons
and the number of recorded frames determine how closely the reconstruction
approaches the physical bandwidth limit, but they do not extend that bandwidth.

An order-of-magnitude estimate of the number of frames required to recover a
target scale \(\Delta_x\) can be written in four equivalent forms. Let
\[
A_{\mathrm{px}}=d_{\mathrm{px}}^2,
\qquad
A_{\mathrm{cam}}=N A_{\mathrm{px}},
\qquad
A_S=\frac{A_{\mathrm{sp}}}{A_{\mathrm{px}}},
\qquad
P_{\mathrm{tot}}=Np,
\]
where \(A_{\mathrm{sp}}\) is the characteristic physical area of one speckle
grain and \(P_{\mathrm{tot}}\) is the mean total number of detected photons per
frame.

Using physical areas and lengths, the estimate is
\begin{equation}
\boxed{
M
\gtrsim
\left[
\frac{2}{p}
\frac{A_{\mathrm{cam}}A_{\mathrm{px}}}
{A_{\mathrm{sp}}\Delta_x^2}
\right]^2
}
\qquad
\boxed{
M
\gtrsim
\left[
\frac{2}{P_{\mathrm{tot}}}
\frac{A_{\mathrm{cam}}^2}
{A_{\mathrm{sp}}\Delta_x^2}
\right]^2
}.
\label{eq:frames_resolution_physical}
\end{equation}
The first form uses the mean photon number \(p\) per camera pixel and per
frame, whereas the second uses the total photon budget
\(P_{\mathrm{tot}}=Np\) per frame.

Equivalently, expressing all areas and distances in camera-pixel units gives
\begin{equation}
\boxed{
M
\gtrsim
\left[
\frac{2}{p}
\frac{N}
{A_S\left(\Delta_x/d_{\mathrm{px}}\right)^2}
\right]^2
}
\qquad
\boxed{
M
\gtrsim
\left[
\frac{2}{P_{\mathrm{tot}}}
\frac{N^2}
{A_S\left(\Delta_x/d_{\mathrm{px}}\right)^2}
\right]^2
}.
\label{eq:frames_resolution_pixel}
\end{equation}
Here, \(N\) is the camera area expressed as the number of pixels,
\(A_S\) is the speckle-grain area expressed in pixels, and
\(\Delta_x/d_{\mathrm{px}}\) is the target linear resolution expressed in
pixels.

At the physical resolution limit, the target scale is of the order of the
equivalent linear speckle size,
\begin{equation}
\Delta_x
\sim
\sqrt{A_{\mathrm{sp}}},
\qquad
\frac{\Delta_x}{d_{\mathrm{px}}}
\sim
\sqrt{A_S}.
\end{equation}
The above relations then estimate the number of frames required for photon
statistics to cease being the dominant resolution-limiting factor. They should
be interpreted as order-of-magnitude scaling relations rather than as exact
thresholds for successful phase retrieval. The actual reconstruction quality
also depends on the object structure, the spatial distribution of the
estimation error, the phase-retrieval constraints, and the initialization.

Increasing \(M\) or the photon budget beyond this regime improves the precision
of the estimated Fourier modulus, but it does not recover frequencies outside
the bandwidth supported by the speckle patterns.

\paragraph{2 photons per frame regime}
The theoretical analysis shows that the method presented in the article works for any number of photons per pixel. However, intuitive considerations suggest that all frames with fewer than 2 photons provide no information, as they contribute zero or a constant value to the calculated Fourier modulus. Naturally, this raises the question of how many frames are needed to recover an image in the limiting case of an average of 2 photons per frame $p=\frac{2}{N}$. Let $\epsilon$ denote the relative accuracy with which we want to determine $H(k)$. Then, recalling Eq.~\ref{eq:est_err}:

\begin{equation}
	v(k) \leq \frac{1}{2 \sqrt{M}} \left(1 + \frac{1}{p\ \text{max}(H(k)^2)}\right) < \epsilon
\end{equation}

\begin{equation}
	M_{2P} > \frac{1}{4\epsilon^2} \left(1 + \frac{1}{p\ \text{max}(H(k)^2)}  \right)^2 \gtrapprox\frac{1}{4\epsilon^2} \frac{1}{p^2 \text{max}(H(k)^2)^2}
\end{equation}

For the experimental data and assuming $\epsilon=10^{-2}$:

\begin{equation}
	\text{Max}(H(k)^2) \approx 131
\end{equation}

\begin{equation}
	p = \frac{2}{9 \times 10^6} \approx 2 \times 10^{-7}
\end{equation}

\begin{equation}
	M_{2P} \gtrapprox 3.6 \times 10^{12}
\end{equation}

This approximation of $M_{2P}$ provides the accuracy for the highest possible intensity points in Fourier space; the actual number of frames needed for reconstruction is a few orders of magnitude higher.

Although this is theoretically possible, gathering this number of frames is infeasible with the current measurement setup.

\paragraph{Partialy Static diffuser}
In reality, diffusers can possess both static and dynamic components, and the observer does not need to know in advance which scenario applies. Therefore, we performed a simulation which verified that our method remains robust regardless of whether the diffuser is static or dynamic. In Figure~\ref{fig:l1_dyn_stat} below, we present the results.

\begin{figure}[H]
    \centering
    \includegraphics[width=0.75\linewidth]{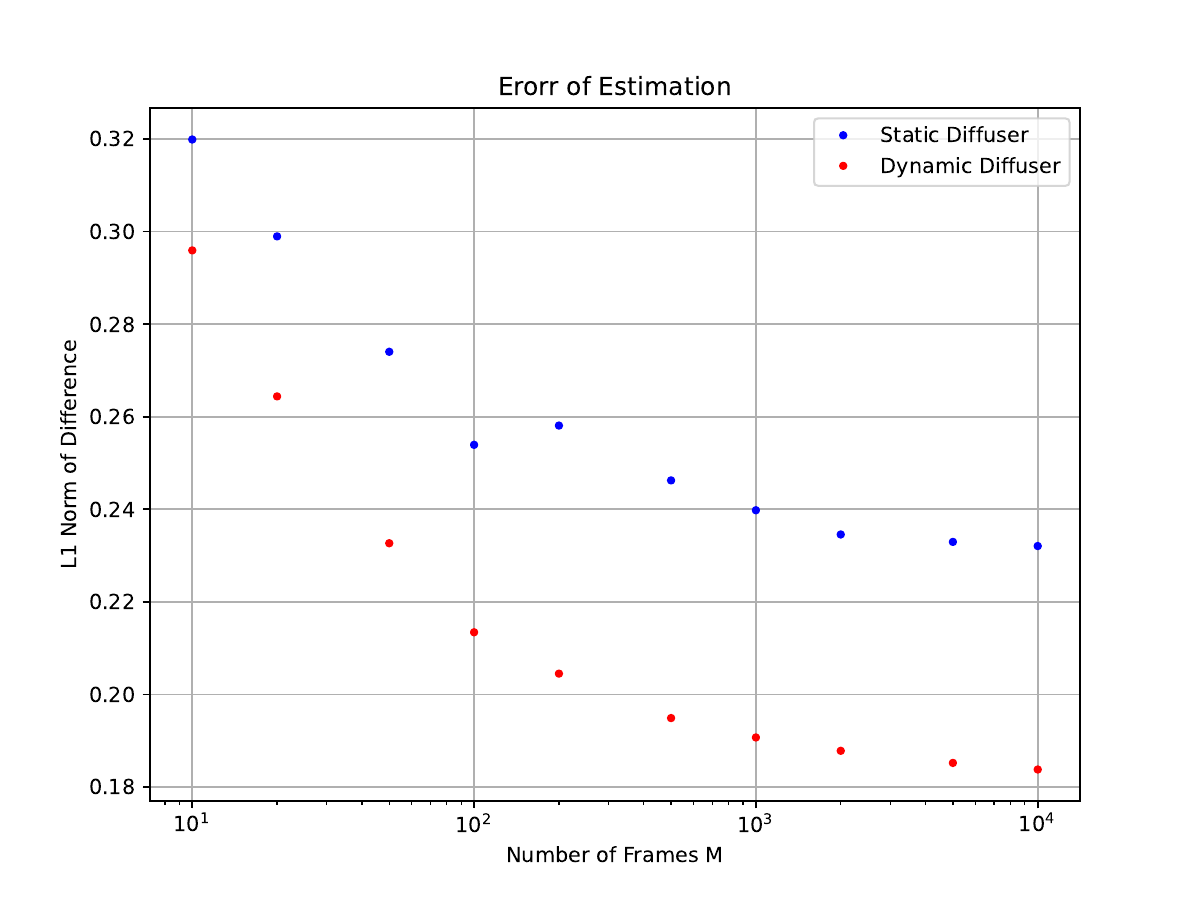}
    \caption{The normalized $L_1$ norm of the difference between the estimated Fourier modulus and the actual value is computed for a frame of size $1024 \times 1024$ pixels, with $p = 5$ photons per pixel. The estimation error is measured as a function of the number of frames $M$.}
    \label{fig:l1_dyn_stat}
\end{figure}
Note that for both cases (for dynamic and static diffusers), the $L_1$ norm converges with a number of frames. However, for dynamic scattering, the object estimation is more accurate and can be obtained with a lower number of frames.

\subsection{Low photon regime}

We can estimate a lower bound for the power that has to be produced by the fluorescent dyes if we want to achieve at least two photons per frame with the following considerations:

\begin{itemize}
    \item The duration of a frame must be much shorter than the characteristic time of change of the scattering medium: $$t_{frame} << t_{scatt}$$
    \item The energy arriving on the camera during a frame is at least the energy of two photons:
    $$\displaylines{
        2E_{photon} < E_{camera}\cr
        2\frac{hc}{\lambda} < P_{camera}t_{frame}
    } $$
    ($\lambda$ is the wavelength emitted by the dye, $P_{camera}$ the power received by the camera).
    \item Photons emitted by the fluorescent dyes arrive to the camera only if they are emitted within the solid angle given by the numerical aperture (assuming no scatterer):
    $$P_{camera} = \frac{\Omega}{4\pi}P_{fluo} = \frac{NA^2}{4}P_{fluo}$$
\end{itemize}

We arrive at the following inequality for the total power emitted by the fluorescent dyes:

$$P_{fluo} >> \frac{8hc}{NA^2\lambda t_{scatt}}$$

Given this inequality and the properties of the dyes used in a given setup, if the characteristic time of the dynamic scattering medium is known, one can estimate the theoretical minimum laser power required to apply our method.

\subsection{Effect of a Finite Ensemble of Speckles}

In experiments, there is not always a guarantee that the speckles produced by the scatterer are unique and uncorrelated with the speckles obtained in previous frames. For example, one might simulate a ``variable scatterer'' by using only a small fragment of a static scatterer. As a result, the total number of possible speckle realizations $L$ is roughly
\begin{equation}
L \approx\frac{\text{(total area of the scatterer)}}{\text{(area illuminated by the beam of light)}}.
\end{equation}
Hence, the set of distinct speckle realizations can be significantly limited.

\medskip

We now investigate how a finite number of effectively distinct speckle
realizations affects the outcome of the experiment.

Assume that, during the experiment, only \(L\) effectively distinct speckle
patterns can occur. We denote this finite dictionary of effective speckle
realizations by
\begin{equation}
\mathcal{S}_L
=
\left\{
S_1(\boldsymbol{x}),
S_2(\boldsymbol{x}),
\dots,
S_L(\boldsymbol{x})
\right\}.
\label{eq:finite_speckle_dictionary}
\end{equation}
For the \(l\)-th speckle pattern, the corresponding noiseless intensity frame is
\begin{equation}
I_l(\boldsymbol{x})
=
(O*S_l)(\boldsymbol{x}),
\qquad
l=1,\dots,L,
\label{eq:yl_def}
\end{equation}
and its discrete Fourier transform is denoted by
\begin{equation}
\widetilde{I}_l[\boldsymbol{k}]
=
\mathcal{F}\{I_l\}[\boldsymbol{k}].
\label{eq:Yl_def}
\end{equation}

We acquire a total of \(M\) frames. In each frame \((m)\), one element of the dictionary
\(\mathcal{S}_L\) is observed. We denote its index by
\begin{equation}
\iota_m \in \{1,\dots,L\},
\qquad
m=1,\dots,M.
\label{eq:iota_def}
\end{equation}
Thus,
\begin{equation}
S^{(m)}(\boldsymbol{x})
=
S_{\iota_m}(\boldsymbol{x}),
\qquad
I^{(m)}(\boldsymbol{x})
=
I_{\iota_m}(\boldsymbol{x}),
\qquad
\widetilde{I}^{(m)}[\boldsymbol{k}]
=
\widetilde{I}_{\iota_m}[\boldsymbol{k}].
\label{eq:frame_assignment}
\end{equation}
A particular realization of the experiment is therefore described by the index
sequence
\begin{equation}
\boldsymbol{\iota}_{1:M}
=
(\iota_1,\iota_2,\dots,\iota_M),
\label{eq:iota_sequence}
\end{equation}
which specifies which effective speckle realization was present in each
recorded frame.

The quantity directly measured in the experiment at spatial frequency
\(\boldsymbol{k}\) is the averaged Fourier power of the recorded
Poisson-corrupted frames:
\begin{equation}
\widehat{\sigma}_{\boldsymbol{k}}^2
=
\frac{1}{M}
\sum_{m=1}^{M}
\left|
\widetilde{D}^{(m)}[\boldsymbol{k}]
\right|^2 .
\label{eq:sigma_hat_finite_dictionary}
\end{equation}
Here \(\widetilde{D}^{(m)}[\boldsymbol{k}]\) is the DFT of the \(m\)-th
measured camera frame.

Conditioned on the noiseless frame \(y^{(m)}\), or equivalently on its full
Fourier transform \(\widetilde{I}^{(m)}\), the second-order Poisson result gives
\begin{equation}
\mathbb{E}_{\mathrm{P}}
\left[
\left|
\widetilde{D}^{(m)}[\boldsymbol{k}]
\right|^2
\,\middle|\,
\widetilde{I}^{(m)}
\right]
=
\widetilde{I}^{(m)}[\boldsymbol{0}]
+
\left|
\widetilde{I}^{(m)}[\boldsymbol{k}]
\right|^2 .
\label{eq:conditional_power_single_frame}
\end{equation}
This is the conditional form of Eq.~\eqref{eq:EFd_abs2_poisson_nonzero}. The conditioning has to
include the full noiseless frame, or at least both
\(\widetilde{I}^{(m)}[\boldsymbol{k}]\) and
\(\widetilde{I}^{(m)}[\boldsymbol{0}]\). If one conditions only on
\(\widetilde{I}^{(m)}[\boldsymbol{k}]\), the zero-frequency term remains as a
conditional expectation.

Applying the law of total expectation to
Eq.~\eqref{eq:sigma_hat_finite_dictionary} gives
\begin{align}
\mathbb{E}_{\mathrm{P}}
\left[
\widehat{\sigma}_{\boldsymbol{k}}^2
\,\middle|\,
\{\widetilde{I}_l\}_{l=1}^{L},
\boldsymbol{\iota}_{1:M}
\right]
&=
\frac{1}{M}
\sum_{m=1}^{M}
\mathbb{E}_{\mathrm{P}}
\left[
\left|
\widetilde{D}^{(m)}[\boldsymbol{k}]
\right|^2
\,\middle|\,
\widetilde{I}_{\iota_m}
\right]
\nonumber\\
&=
\frac{1}{M}
\sum_{m=1}^{M}
\left(
\widetilde{I}_{\iota_m}[\boldsymbol{0}]
+
\left|
\widetilde{I}_{\iota_m}[\boldsymbol{k}]
\right|^2
\right).
\label{eq:sigma_hat_conditional_mean_sequence}
\end{align}

Equivalently, let
\begin{equation}
n_l
=
\#\{m:\iota_m=l\},
\qquad
\sum_{l=1}^{L} n_l = M,
\label{eq:nl_def}
\end{equation}
be the number of frames acquired with the \(l\)-th speckle pattern. Then
Eq.~\eqref{eq:sigma_hat_conditional_mean_sequence} can be written as
\begin{equation}
\mathbb{E}_{\mathrm{P}}
\left[
\widehat{\sigma}_{\boldsymbol{k}}^2
\,\middle|\,
\{\widetilde{I}_l\}_{l=1}^{L},
\{n_l\}_{l=1}^{L}
\right]
=
\sum_{l=1}^{L}
\frac{n_l}{M}
\left(
\widetilde{I}_{l}[\boldsymbol{0}]
+
\left|
\widetilde{I}_{l}[\boldsymbol{k}]
\right|^2
\right).
\label{eq:sigma_hat_conditional_mean_counts}
\end{equation}

In a typical experiment, the number of acquired frames is large. If the frame
sequence samples the effective dictionary without a systematic preference for
any particular element, then the empirical occupation numbers satisfy
\(n_l/M\simeq 1/L\). Under this balanced-use approximation,
Eq.~\eqref{eq:sigma_hat_conditional_mean_counts} reduces to
\begin{equation}
\mathbb{E}_{\mathrm{P}}
\left[
\widehat{\sigma}_{\boldsymbol{k}}^2
\,\middle|\,
\{\widetilde{I}_l\}_{l=1}^{L}
\right]
\simeq
\frac{1}{L}
\sum_{l=1}^{L}
\left(
\widetilde{I}_{l}[\boldsymbol{0}]
+
\left|
\widetilde{I}_{l}[\boldsymbol{k}]
\right|^2
\right).
\label{eq:sigma_hat_conditional_mean_balanced}
\end{equation}

Thus, by increasing \(M\), we reduce the Poisson fluctuations around the average
defined by the finite dictionary \(\mathcal{S}_L\). If the dictionary itself is
regarded as fixed i.e we have a given set of scatterers, the statistical analysis can essentially end here: the data
cannot contain more independent speckle information than is present in given sum over unique speckle realisations. For example, if all \(s_l\) are the same, there is effectively
only a single speckle realization. In such cases, to recover $\lvert O[\boldsymbol{k}]\rvert$, one would use known techniques for a \emph{static} diffuser \cite{supp:Katz2014}.

\paragraph{Representative dictionary of speckles}

We now consider the case in which \(\mathcal{S}_L\) is interpreted as a
representative finite sample from the ideal speckle ensemble. In other words,
the \(L\) patterns \(s_l\) are independent speckle realizations, apart from the
usual weak residual correlations between random speckle patterns. This situation
arises, for example, when each realization corresponds to a different small
region selected from a large scatterer.

In a real experiment, we do not know in advance which particular set of
effectively distinct speckles will be realized. Therefore, to predict the
typical outcome of the experiment, we average not only over Poisson noise and
over the frame sequence, but also over possible dictionaries
\(\mathcal{S}_L\). Averaging Eq.~\eqref{eq:sigma_hat_conditional_mean_balanced}
over independently drawn dictionaries gives:

\begin{align}
\mathbb{E}_{\mathcal{S}_L,\boldsymbol{\iota},\mathrm{P}}
\left[
\widehat{\sigma}_{\boldsymbol{k}}^2
\right]
&\simeq
\mathbb{E}_{\mathcal{S}_L}
\left[
\frac{1}{L}
\sum_{l=1}^{L}
\left(
\widetilde{I}_{l}[\boldsymbol{0}]
+
\left|
\widetilde{I}_{l}[\boldsymbol{k}]
\right|^2
\right)
\right]
\nonumber\\
&=
\frac{1}{L}
\sum_{l=1}^{L}
\mathbb{E}_{\mathcal{S}_L}
\left[
\widetilde{I}_{l}[\boldsymbol{0}]
\right]
+
\frac{1}{L}
\sum_{l=1}^{L}
\mathbb{E}_{\mathcal{S}_L}
\left[
\left|
\widetilde{I}_{l}[\boldsymbol{k}]
\right|^2
\right].
\label{eq:mean_over_dictionary_step}
\end{align}
Because the dictionary elements are identically distributed samples from the
speckle ensemble,
\begin{equation}
\mathbb{E}_{\mathcal{S}_L}
\left[
\widetilde{I}_{l}[\boldsymbol{0}]
\right]
=
p,
\qquad
\mathbb{E}_{\mathcal{S}_L}
\left[
\left|
\widetilde{I}_{l}[\boldsymbol{k}]
\right|^2
\right]
=
p^2H_2[\boldsymbol{k}].
\label{eq:dictionary_element_moments}
\end{equation}
Therefore,
\begin{equation}
\mathbb{E}_{\mathcal{S}_L,\boldsymbol{\iota},\mathrm{P}}
\left[
\widehat{\sigma}_{\boldsymbol{k}}^2
\right]
=
p+p^2H_2[\boldsymbol{k}].
\label{eq:finite_dictionary_ensemble_mean}
\end{equation}
Thus, the finite size of the speckle dictionary does not bias the ensemble mean.
It changes the variance. There are two different sources of randomness. The first one is the Poisson
noise in each acquired frame. This contribution is resampled independently in
every frame, and therefore it is reduced by the number of frames \(M\). The
second one is the finite number of effectively distinct speckle realizations in
the dictionary \(\mathcal{S}_L\). This contribution is reduced only by the
number \(L\) of independent speckle patterns, not by the number of Poisson-noisy
frames recorded from them.

To see this, consider the measured quantity
\begin{equation}
\widehat{\sigma}_{\boldsymbol{k}}^2
=
\frac{1}{M}
\sum_{m=1}^{M}
\left|
\widetilde{D}^{(m)}[\boldsymbol{k}]
\right|^2 .
\label{eq:sigma_hat_random_dictionary_sampling}
\end{equation}

In the present model, each frame independently selects one element from the
finite dictionary \(\mathcal S_L\). Therefore, even when \(L=M\), the experiment
is not equivalent to using \(M\) fresh independent speckles: some dictionary
elements may be selected more than once, while others may not be selected at all.
The ideal independent-speckle limit is recovered only for \(L\to\infty\).

Let
\begin{equation}
A[\boldsymbol{k}]
=
\mathbb{E}_{\mathrm P}
\left[
\left|
\widetilde D[\boldsymbol{k}]
\right|^2
\,\middle|\,
\widetilde Y
\right]
=
\widetilde Y[\boldsymbol 0]
+
\left|
\widetilde Y[\boldsymbol{k}]
\right|^2 .
\label{eq:A_k_def}
\end{equation}
The law of total variance for a single frame gives
\begin{align}
\operatorname{Var}
\left(
\left|
\widetilde D[\boldsymbol{k}]
\right|^2
\right)
&=
\mathbb{E}_{\mathcal S_L}
\left[
\operatorname{Var}_{\mathrm P}
\left(
\left|
\widetilde D[\boldsymbol{k}]
\right|^2
\,\middle|\,
\widetilde Y
\right)
\right]
\nonumber\\
&\quad+
\operatorname{Var}_{\mathcal S_L}
\left(
A[\boldsymbol{k}]
\right).
\label{eq:single_frame_total_variance_random_dictionary}
\end{align}
Thus,
\begin{align}
&\mathbb{E}_{\mathcal S_L}
\left[
\operatorname{Var}_{\mathrm P}
\left(
\left|
\widetilde D[\boldsymbol{k}]
\right|^2
\,\middle|\,
\widetilde Y
\right)
\right]
\nonumber\\
&\qquad =
\operatorname{Var}
\left(
\left|
\widetilde D[\boldsymbol{k}]
\right|^2
\right)
-
\operatorname{Var}_{\mathcal S_L}
\left(
A[\boldsymbol{k}]
\right).
\label{eq:poisson_part_single_frame_random_dictionary}
\end{align}

For a fixed dictionary \(\mathcal S_L\), the frame-to-frame variability contains
both the Poisson noise and the random selection of a dictionary element. Since
each frame selects one of the \(L\) elements uniformly,
\begin{align}
&\operatorname{Var}_{\boldsymbol{\iota},\mathrm P}
\left(
\left|
\widetilde D[\boldsymbol{k}]
\right|^2
\,\middle|\,
\mathcal S_L
\right)
\nonumber\\
&\qquad =
\frac{1}{L}
\sum_{l=1}^{L}
\operatorname{Var}_{\mathrm P}
\left(
\left|
\widetilde D[\boldsymbol{k}]
\right|^2
\,\middle|\,
\widetilde Y_l
\right)
+
\operatorname{Var}_{l\in\mathcal S_L}
\left(
A_l[\boldsymbol{k}]
\right),
\label{eq:within_dictionary_frame_variance}
\end{align}
where
\[
A_l[\boldsymbol{k}]
=
\widetilde Y_l[\boldsymbol 0]
+
\left|
\widetilde Y_l[\boldsymbol{k}]
\right|^2 .
\]
Averaging over possible dictionaries gives
\begin{align}
&\mathbb{E}_{\mathcal S_L}
\left[
\operatorname{Var}_{\boldsymbol{\iota},\mathrm P}
\left(
\left|
\widetilde D[\boldsymbol{k}]
\right|^2
\,\middle|\,
\mathcal S_L
\right)
\right]
\nonumber\\
&\qquad =
\operatorname{Var}
\left(
\left|
\widetilde D[\boldsymbol{k}]
\right|^2
\right)
-
\frac{1}{L}
\operatorname{Var}_{\mathcal S_L}
\left(
A[\boldsymbol{k}]
\right).
\label{eq:within_dictionary_frame_variance_averaged}
\end{align}
The term subtracted with the factor \(1/L\) appears because the empirical
dictionary mean is itself random. Indeed,
\begin{equation}
\operatorname{Var}_{\mathcal S_L}
\left(
\frac{1}{L}
\sum_{l=1}^{L}
A_l[\boldsymbol{k}]
\right)
=
\frac{1}{L}
\operatorname{Var}_{\mathcal S_L}
\left(
A[\boldsymbol{k}]
\right).
\label{eq:dictionary_mean_variance}
\end{equation}

Using the law of total variance once more for the measured average, we obtain
\begin{align}
&\operatorname{Var}_{\mathcal S_L,\boldsymbol{\iota},\mathrm P}
\left(
\widehat{\sigma}_{\boldsymbol{k}}^2
\right)
\nonumber\\
&\quad =
\mathbb{E}_{\mathcal S_L}
\left[
\operatorname{Var}_{\boldsymbol{\iota},\mathrm P}
\left(
\widehat{\sigma}_{\boldsymbol{k}}^2
\,\middle|\,
\mathcal S_L
\right)
\right]
+
\operatorname{Var}_{\mathcal S_L}
\left[
\mathbb{E}_{\boldsymbol{\iota},\mathrm P}
\left(
\widehat{\sigma}_{\boldsymbol{k}}^2
\,\middle|\,
\mathcal S_L
\right)
\right]
\nonumber\\
&\quad =
\frac{1}{M}
\left[
\operatorname{Var}
\left(
\left|
\widetilde D[\boldsymbol{k}]
\right|^2
\right)
-
\frac{1}{L}
\operatorname{Var}_{\mathcal S_L}
\left(
A[\boldsymbol{k}]
\right)
\right]
+
\frac{1}{L}
\operatorname{Var}_{\mathcal S_L}
\left(
A[\boldsymbol{k}]
\right)
\nonumber\\
&\quad =
\frac{1}{M}
\operatorname{Var}
\left(
\left|
\widetilde D[\boldsymbol{k}]
\right|^2
\right)
+
\left(
1-\frac{1}{M}
\right)
\frac{1}{L}
\operatorname{Var}_{\mathcal S_L}
\left(
A[\boldsymbol{k}]
\right).
\label{eq:finite_dictionary_variance_random_sampling}
\end{align}
This expression has the correct limiting behavior for random sampling from a
finite dictionary. For \(L\to\infty\), the second term vanishes and the usual
independent-speckle result is recovered:
\begin{equation}
\operatorname{Var}
\left(
\widehat{\sigma}_{\boldsymbol{k}}^2
\right)
\to
\frac{1}{M}
\operatorname{Var}
\left(
\left|
\widetilde D[\boldsymbol{k}]
\right|^2
\right).
\label{eq:ideal_limit_L_infty}
\end{equation}
For finite \(L\), increasing \(M\) reduces the frame noise, but it does not
remove the finite-dictionary contribution.

The remaining term can be simplified as
\begin{align}
\operatorname{Var}_{\mathcal S_L}
\left(
A[\boldsymbol{k}]
\right)
&=
\operatorname{Var}_{\mathcal S_L}
\left(
\widetilde Y[\boldsymbol 0]
+
\left|
\widetilde Y[\boldsymbol{k}]
\right|^2
\right)
\nonumber\\
&=
\operatorname{Var}_{\mathcal S_L}
\left(
\widetilde Y[\boldsymbol 0]
\right)
+
\operatorname{Var}_{\mathcal S_L}
\left(
\left|
\widetilde Y[\boldsymbol{k}]
\right|^2
\right)
\nonumber\\
&\quad+
2\operatorname{Cov}_{\mathcal S_L}
\left(
\widetilde Y[\boldsymbol 0],
\left|
\widetilde Y[\boldsymbol{k}]
\right|^2
\right).
\label{eq:A_k_variance_expanded}
\end{align}
For \(\boldsymbol{k}\neq\boldsymbol 0\), the mixed covariance may be written as
\begin{align}
&\operatorname{Cov}_{\mathcal S_L}
\left(
\widetilde Y[\boldsymbol 0],
\left|
\widetilde Y[\boldsymbol{k}]
\right|^2
\right)
\nonumber\\
&\quad =
\kappa_{\mathcal S_L}
\left(
\widetilde Y[\boldsymbol 0],
\widetilde Y[\boldsymbol{k}],
\widetilde Y[-\boldsymbol{k}]
\right),
\label{eq:Y0_Yk2_covariance_cumulant}
\end{align}
because
\(\kappa_{\mathcal S_L}(\widetilde Y[\boldsymbol{k}])
=
\kappa_{\mathcal S_L}(\widetilde Y[-\boldsymbol{k}])=0\).
In the large-camera limit, the Fourier components become asymptotically joint
Gaussian, and this third-order cumulant vanishes. Therefore,
\begin{equation}
\operatorname{Var}_{\mathcal S_L}
\left(
A[\boldsymbol{k}]
\right)
\simeq
\operatorname{Var}_{\mathcal S_L}
\left(
\widetilde Y[\boldsymbol 0]
\right)
+
\operatorname{Var}_{\mathcal S_L}
\left(
\left|
\widetilde Y[\boldsymbol{k}]
\right|^2
\right).
\label{eq:A_k_variance_gaussian_limit}
\end{equation}

If the zero-frequency photon-budget fluctuations are negligible, or if the
frames are normalized so that \(\widetilde Y[\boldsymbol 0]\) is effectively
fixed, then
\begin{equation}
\operatorname{Var}_{\mathcal S_L}
\left(
A[\boldsymbol{k}]
\right)
\simeq
\operatorname{Var}_{\mathcal S_L}
\left(
\left|
\widetilde Y[\boldsymbol{k}]
\right|^2
\right).
\label{eq:A_k_variance_fixed_photon_budget}
\end{equation}
Using
\[
\mathbb{E}
\left[
\left|
\widetilde D[\boldsymbol{k}]
\right|^2
\right]
=
p+p^2H_2[\boldsymbol{k}]
\]
and
\[
\mathbb{E}_{\mathcal S_L}
\left[
\left|
\widetilde Y[\boldsymbol{k}]
\right|^2
\right]
=
p^2H_2[\boldsymbol{k}],
\]
we approximate
\[
\operatorname{Var}
\left(
\left|
\widetilde D[\boldsymbol{k}]
\right|^2
\right)
\simeq
\left(
p+p^2H_2[\boldsymbol{k}]
\right)^2
\]
and
\[
\operatorname{Var}_{\mathcal S_L}
\left(
\left|
\widetilde Y[\boldsymbol{k}]
\right|^2
\right)
\simeq
\left(
p^2H_2[\boldsymbol{k}]
\right)^2 .
\]
Substitution into Eq.~\eqref{eq:finite_dictionary_variance_random_sampling}
gives
\begin{align}
&\operatorname{Var}_{\mathcal S_L,\boldsymbol{\iota},\mathrm P}
\left(
\widehat{\sigma}_{\boldsymbol{k}}^2
\right)
\nonumber\\
&\quad \simeq
\frac{
\left(
p+p^2H_2[\boldsymbol{k}]
\right)^2
-
\frac{1}{L}
\left(
p^2H_2[\boldsymbol{k}]
\right)^2
}{M}
+
\frac{
\left(
p^2H_2[\boldsymbol{k}]
\right)^2
}{L}
\nonumber\\
&\quad =
\frac{
\left(
p+p^2H_2[\boldsymbol{k}]
\right)^2
}{M}
+
\left(
1-\frac{1}{M}
\right)
\frac{
\left(
p^2H_2[\boldsymbol{k}]
\right)^2
}{L}.
\label{eq:finite_dictionary_variance_final_random_sampling}
\end{align}
This form separates the two sources of uncertainty. The first term decreases as
\(1/M\): it is reduced by recording more frames, because Poisson noise is
resampled in every frame and the frame sequence is averaged. The second term
decreases only as \(1/L\): it is the variability of the finite-dictionary
average itself. Once the \(L\) effective speckle realizations have been fixed,
acquiring more frames only makes the experiment converge more accurately to that
particular finite-dictionary average.

Thus, for random sampling with replacement from a finite dictionary, \(L=M\)
does not yet reproduce the ideal independent-speckle result. The ideal result is
obtained only in the limit \(L\to\infty\).\\

The impact of a finite speckle dictionary on the accuracy of the reconstructed
image can be quantified by calculating the corrected relative standard error   of the estimator of $H[\boldsymbol{k}]$.

Recalling estimator form Eq.~\eqref{eq:H_hat_from_H2_hat} with the physical constraint \(H_2[\boldsymbol{k}]\ge 0\), we propagate the variance of \(\widehat{\sigma}_{\boldsymbol{k}}^2\) using the delta method.
\begin{equation}
H[\boldsymbol{k}]
=
\frac{1}{p}
\sqrt{
\sigma_{\boldsymbol{k}}^2-p
},
\end{equation}
we have
\begin{equation}
\frac{\partial H}{\partial \sigma_{\boldsymbol{k}}^2}
=
\frac{1}{2p^2H[\boldsymbol{k}]}.
\label{eq:dH_dsigma_RSE}
\end{equation}
Therefore,
\begin{equation}
\operatorname{Var}
\left(
\widehat{H}[\boldsymbol{k}]
\right)
\simeq
\frac{
\operatorname{Var}
\left(
\widehat{\sigma}_{\boldsymbol{k}}^2
\right)
}{
4p^4H^2[\boldsymbol{k}]
}.
\label{eq:var_H_delta_method}
\end{equation}

Using the finite-dictionary variance without the additional
zero-frequency contribution,
\begin{equation}
\operatorname{Var}
\left(
\widehat{\sigma}_{\boldsymbol{k}}^2
\right)
\simeq
\frac{
\left(
p+p^2H_2[\boldsymbol{k}]
\right)^2
}{M}
+
\left(
1-\frac{1}{M}
\right)
\frac{
p^4H_2^2[\boldsymbol{k}]
}{L},
\label{eq:var_sigma_finite_dictionary_no_p}
\end{equation}
and using \(H_2[\boldsymbol{k}]=H^2[\boldsymbol{k}]\), we obtain corrected RSE of estimtator (Eq.~\eqref{eq:est_err}):
\begin{align}
\operatorname{RSE}
\left(
\widehat{H}[\boldsymbol{k}]
\right)
&=
\frac{
\sqrt{
\operatorname{Var}
\left(
\widehat{H}[\boldsymbol{k}]
\right)
}
}{
H[\boldsymbol{k}]
}
\nonumber\\
&\simeq
\frac{1}{2p^2H_2[\boldsymbol{k}]}
\sqrt{
\frac{
\left(
p+p^2H_2[\boldsymbol{k}]
\right)^2
}{M}
+
\left(
1-\frac{1}{M}
\right)
\frac{
p^4H_2^2[\boldsymbol{k}]
}{L}
}.
\label{eq:RSE_H_finite_dictionary_expanded}
\end{align}
Equivalently,
\begin{equation}
\operatorname{RSE}
\left(
\widehat{H}[\boldsymbol{k}]
\right)
\simeq
\frac{1}{2}
\sqrt{
\frac{1}{M}
\left(
1+
\frac{1}{pH_2[\boldsymbol{k}]}
\right)^2
+
\left(
1-\frac{1}{M}
\right)
\frac{1}{L}
}.
\label{eq:RSE_H_finite_dictionary}
\end{equation}

This expression separates the photon-noise term, which decreases with the
number of recorded frames \(M\), from the finite-dictionary term, which is
controlled by the number \(L\) of effectively distinct speckle realizations.
In the limit \(M\to\infty\),
\begin{equation}
\operatorname{RSE}
\left(
\widehat{H}[\boldsymbol{k}]
\right)
\longrightarrow
\frac{1}{2\sqrt{L}}.
\label{eq:RSE_H_floor}
\end{equation}
Thus, no matter how many frames \(M\) are collected, the relative standard error
of the amplitude estimate cannot be reduced below the finite-dictionary floor
\(1/(2\sqrt{L})\).

For finite \(M\), weak spatial frequencies remain difficult to estimate because
of the factor
\[
1+
\frac{1}{pH_2[\boldsymbol{k}]}.
\]
More generally,
\[
\operatorname{RSE}
\left(
\widehat{H}[\boldsymbol{k}]
\right)
>1
\]
whenever
\begin{equation}
\frac{1}{M}
\left(
1+
\frac{1}{pH_2[\boldsymbol{k}]}
\right)^2
+
\left(
1-\frac{1}{M}
\right)
\frac{1}{L}
>4.
\label{eq:RSE_larger_than_one_condition}
\end{equation}
For \(M\) of the order of \(L\), this condition is reached for frequencies for
which
\begin{equation}
1+
\frac{1}{pH_2[\boldsymbol{k}]}
\gtrsim
2\sqrt{L}.
\label{eq:RSE_weak_frequency_condition}
\end{equation}
Such frequencies cannot be estimated precisely.

Numerical investigations of this effect are presented in
Fig.~\ref{sim1}. The error is measured in the Frobenius norm between
the Fourier transform of the reconstructed image and the amplitude of the
Fourier transform used as input to the phase retrieval algorithm. The error is
plotted as a function of the number \(L\) of possible unique speckle
realizations for different photon numbers: \(2\), \(4\), \(6\), \(8\), and
\(10\) million photons distributed over \(10^4\) camera pixels. As expected, the
error decreases with increasing \(L\), but it does not vanish for finite \(L\)
because the finite speckle dictionary imposes an irreducible statistical floor.

\subsection{Numerical comparison of Fourier-domain estimators}
\label{sec:estimator_comparison}

To compare the maximum-likelihood RMS estimator with alternative Fourier-domain
averaging strategies, we performed numerical simulations under the
photon-counting model derived above. The same object, speckle statistics,
photon budget, and number of frames were used for all estimators. The mean
photon number was fixed at \(p=1\) photon per camera pixel and per frame. The
number of independent speckle frames was increased up to \(M=8192\), and the
simulation was repeated for \(N_{\mathrm{seed}}=10\) independent random seeds.

\begin{figure}[h]
    \centering
    \includegraphics[width=0.65\linewidth]{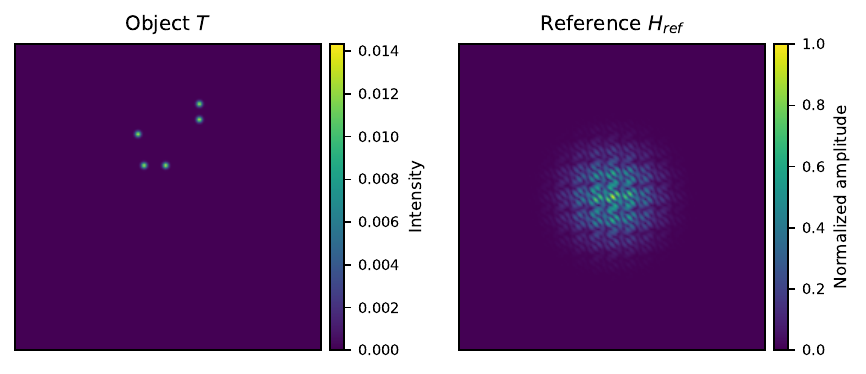}
    \caption{
    Simulation ground truth used for the estimator comparison. The panels show
    the simulated object \(O\) and the corresponding reference Fourier modulus
    \(H_{\mathrm{ref}}\).
    }
    \label{fig:estimator_ground_truth}
\end{figure}

For each recorded photon-count image \(d_i\), we used the normalized Fourier
coefficient
\begin{equation}
    D_i(k)
    =
    \frac{1}{\sqrt{N}}
    \mathcal{F}\{d_i\}(k),
\end{equation}
where \(N\) is the number of camera pixels. For non-zero spatial frequencies,
the statistical model is
\begin{equation}
    D_i(k) \sim \mathcal{CN}(0,\sigma_k^2),
    \qquad
    \sigma_k^2 = p + p^2 H^2(k),
    \label{eq:sigma2_count_model_estimator_comparison}
\end{equation}
with
\begin{equation}
    H(k)
    =
    \sqrt{H_2(k)}
    =
    |\mathcal{F}\{O\}(k)|
    |\mathcal{F}\{P\}(k)|.
\end{equation}
The reference spectrum \(H_{\mathrm{ref}}\) was calculated from the same
discrete Fourier convention and the same speckle statistics as used in the
simulation.

We compared three estimators of \(\sigma_k^2\). The RMS, or Fourier-power,
estimator was
\begin{equation}
    \widehat{\sigma}_{k,\mathrm{RMS}}^2
    =
    \frac{1}{M}
    \sum_{i=1}^{M}
    |D_i(k)|^2 .
    \label{eq:sigma2_rms_estimator_comparison}
\end{equation}
The estimator based on the arithmetic mean of the Fourier modulus was
\begin{equation}
    \widehat{\sigma}_{k,\mathrm{AM,raw}}^2
    =
    \frac{4}{\pi}
    \left[
        \frac{1}{M}
        \sum_{i=1}^{M}
        |D_i(k)|
    \right]^2 ,
    \label{eq:sigma2_am_estimator_raw_comparison}
\end{equation}
and the estimator based on the geometric mean of the Fourier power was
\begin{equation}
    \widehat{\sigma}_{k,\mathrm{GM,raw}}^2
    =
    \exp\left[
        \frac{1}{M}
        \sum_{i=1}^{M}
        \log\left(|D_i(k)|^2+\epsilon\right)
        +
        \gamma
    \right],
    \label{eq:sigma2_gm_estimator_raw_comparison}
\end{equation}
where \(\gamma\) is the Euler--Mascheroni constant and \(\epsilon\) is a small
positive numerical constant.

The RMS estimator is unbiased for \(\sigma_k^2\). For the arithmetic- and
geometric-mean estimators, we applied the finite-\(M\) multiplicative bias
correction expected under the circular complex Gaussian model. Thus,
\begin{equation}
    \widehat{\sigma}_{k,\mathrm{AM}}^2
    =
    \frac{
        \widehat{\sigma}_{k,\mathrm{AM,raw}}^2
    }{
        1 + \frac{4-\pi}{\pi M}
    },
    \label{eq:sigma2_am_bias_corrected_comparison}
\end{equation}
and
\begin{equation}
    \widehat{\sigma}_{k,\mathrm{GM}}^2
    =
    \frac{
        \widehat{\sigma}_{k,\mathrm{GM,raw}}^2
    }{
        e^\gamma
        \Gamma\!\left(1+\frac{1}{M}\right)^M
    }.
    \label{eq:sigma2_gm_bias_corrected_comparison}
\end{equation}
The RMS estimator was left unchanged.

For each estimator, the Fourier modulus was recovered as
\begin{equation}
    \widehat{H}(k)
    =
    \frac{1}{p}
    \sqrt{
        \max\left[
            \widehat{\sigma}_k^2-p,0
        \right]
    }.
    \label{eq:H_from_sigma2_estimator_comparison}
\end{equation}

\begin{figure}[h]
    \centering
    \includegraphics[width=\linewidth]{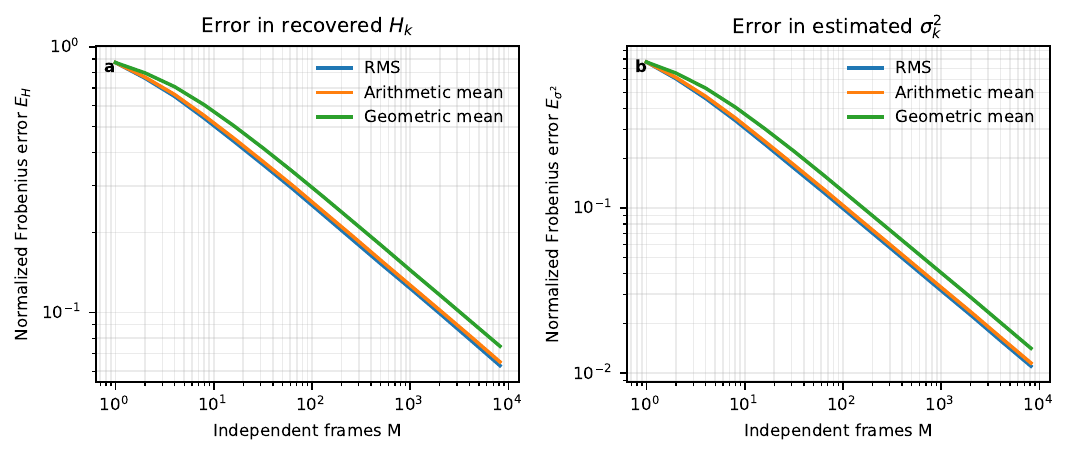}
    \caption{
    Convergence of Fourier-domain estimators.
    (a) Normalized Frobenius error \(E_H\) for the recovered Fourier modulus
    \(H\).
    (b) Normalized Frobenius error \(E_{\sigma^2}\) for the estimated variance
    parameter \(\sigma_k^2\).
    Solid curves show the mean over \(N_{\mathrm{seed}}=10\) independent
    simulation seeds; shaded regions denote one standard deviation.
    }
    \label{fig:estimator_convergence}
\end{figure}

The error was quantified using a normalized Frobenius shape distance. For the
recovered Fourier modulus, we used
\begin{equation}
    E_H(M)
    =
    \left\|
        \frac{\widehat{H}_{M}}
             {\|\widehat{H}_{M}\|_{F}}
        -
        \frac{H_{\mathrm{ref}}}
             {\|H_{\mathrm{ref}}\|_{F}}
    \right\|_{F},
    \label{eq:H_estimator_frobenius_error_comparison}
\end{equation}
and analogously for \(\sigma_k^2\),
\begin{equation}
    E_{\sigma^2}(M)
    =
    \left\|
        \frac{\widehat{\sigma}_{M}^{2}}
             {\|\widehat{\sigma}_{M}^{2}\|_{F}}
        -
        \frac{\sigma_{\mathrm{ref}}^{2}}
             {\|\sigma_{\mathrm{ref}}^{2}\|_{F}}
    \right\|_{F}.
    \label{eq:sigma2_estimator_frobenius_error_comparison}
\end{equation}
This metric compares the spectral shape and is insensitive to an overall
multiplicative scale factor.

Figure~\ref{fig:estimator_convergence} shows the convergence of the three
estimators. The RMS estimator gives the lowest error over the full simulated
range for both \(H\) and \(\sigma_k^2\). The arithmetic-mean estimator performs
only slightly worse, whereas the geometric-mean estimator gives the largest
error. This ordering agrees with the maximum-likelihood derivation: under the
exponential model for \(|D_i(k)|^2\), the sample mean of the Fourier power is
the maximum-likelihood estimator of \(\sigma_k^2\).

\begin{figure}[h]
    \centering
    \includegraphics[width=\linewidth]{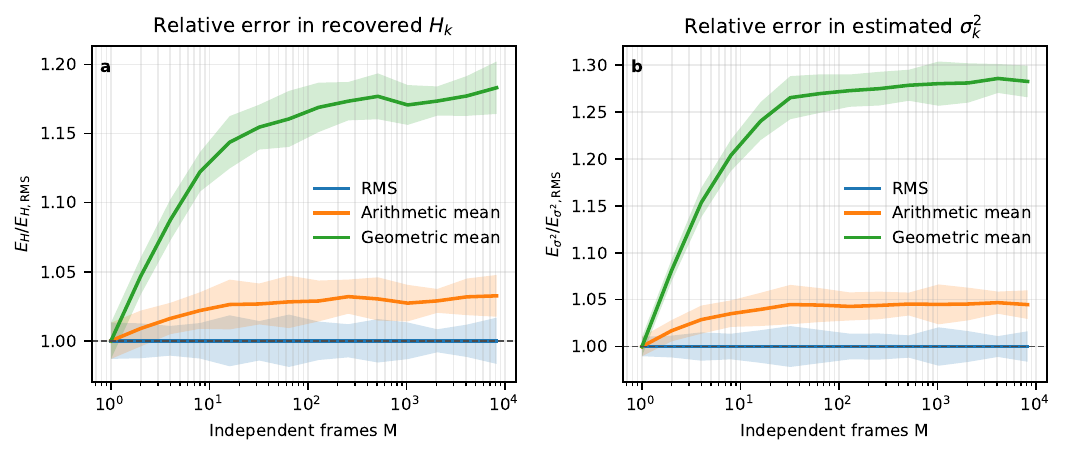}
    \caption{
    Error relative to RMS averaging.
    (a) Ratio \(E_H^{(q)} / E_H^{(\mathrm{RMS})}\).
    (b) Ratio
    \(E_{\sigma^2}^{(q)} / E_{\sigma^2}^{(\mathrm{RMS})}\).
    The dashed horizontal line indicates RMS performance. Values above one
    indicate a larger error than the RMS estimator at the same number of
    frames.
    }
    \label{fig:estimator_relative_to_rms}
\end{figure}

To make the relative difference between estimators more visible, we also
normalized the errors by the RMS error at the same number of frames:
\begin{equation}
    R_H^{(q)}(M)
    =
    \frac{
        E_H^{(q)}(M)
    }{
        E_H^{(\mathrm{RMS})}(M)
    },
    \qquad
    R_{\sigma^2}^{(q)}(M)
    =
    \frac{
        E_{\sigma^2}^{(q)}(M)
    }{
        E_{\sigma^2}^{(\mathrm{RMS})}(M)
    },
    \label{eq:relative_to_rms_error_comparison}
\end{equation}
where \(q\) denotes the arithmetic-mean or geometric-mean estimator. These
ratios are shown in Fig.~\ref{fig:estimator_relative_to_rms}. Values above one
correspond to a larger error than RMS averaging.

At the largest simulated number of frames, $M=8192$, the normalized
Fourier-modulus errors were
\begin{equation}
\begin{aligned}
E_H^{(\mathrm{RMS})} &= (6.31 \pm 0.07)\times 10^{-2},\\
E_H^{(\mathrm{AM})} &= (6.51 \pm 0.06)\times 10^{-2},\\
E_H^{(\mathrm{GM})} &= (7.46 \pm 0.08)\times 10^{-2}.
\end{aligned}
\end{equation}
The corresponding errors for $\sigma_k^2$ were
\begin{equation}
\begin{aligned}
E_{\sigma^2}^{(\mathrm{RMS})} &= (1.101 \pm 0.013)\times 10^{-2},\\
E_{\sigma^2}^{(\mathrm{AM})} &= (1.150 \pm 0.010)\times 10^{-2},\\
E_{\sigma^2}^{(\mathrm{GM})} &= (1.412 \pm 0.009)\times 10^{-2}.
\end{aligned}
\end{equation}
Thus, relative to RMS averaging, the final $(H)$-error increased by
$3.25\%$ for the arithmetic-mean estimator and by $18.31\%$ for the
geometric-mean estimator.

\newpage
\section{Data analysis}\label{Data}
Our data analysis follows the methodologies presented in \cite{supp:Bertolotti2012, supp:Katz2014, supp:Hofer2018}, but adapted for the single-photon regime. All data analysis and simulations were conducted using Python with standard libraries such as NumPy, Torch, and SciPy.
\subsection{Removal of Gaussian envelope}
In our data processing, we employed several filtering techniques to retrieve the Fourier amplitude of the object. One of the key steps was correcting for the uneven brightness distribution caused by vignetting in our imaging system, which manifests as reduced brightness at the edges of the image compared to the center. We addressed this by dividing each frame by a 2D Gaussian function that models the intensity envelope.

To compute the Gaussian correction, we employed a Fourier-based approach using Python. First, the Fourier transform of each frame was computed using a 2D Fast Fourier Transform (FFT). A mask, with a size determined by \texttt{Gaussian\_filter\_length}, was applied in the Fourier domain to isolate the central portion of the spectrum corresponding to the Gaussian envelope. The inverse FFT was then performed to obtain the spatial-domain Gaussian, which was subsequently used to normalize the original frame by dividing the image intensity by the Gaussian envelope.

Here is a simplified representation of the process in Python:

\begin{verbatim}
A_FL = np.fft.fftshift(np.fft.fft2(I))  # Perform FFT and shift
NN1, NN2 = np.shape(I)  # Get the shape of the image
T = np.zeros((NN1, NN2))  # Create an empty mask
# Apply Gaussian filter to the central part of the spectrum
T[(NN1//2)-Gaussian_filter_length+1:(NN1//2)+Gaussian_filter_length+1, 
  (NN2//2)-Gaussian_filter_length+1:(NN2//2)+Gaussian_filter_length+1
  ] = 2
A_FL = A_FL * T  # Multiply the FFT by the mask
A_FL = np.abs(np.fft.ifft2(A_FL))  # Inverse FFT 
#to get the Gaussian envelope
I_New = I[:, :] / A_FL[:, :]  # Normalize the image by the Gaussian
\end{verbatim}
\subsection{Hann filtering}
In our image processing, the edges of each frame can introduce artifacts, especially when applying Fourier transforms. These edge effects arise because the Fourier transform assumes the image is periodic. If there are sudden discontinuities at the image boundaries, this assumption is violated, leading to unwanted high-frequency components, in the Fourier domain. Such artifacts degrade the quality of the reconstructed image and must be mitigated for accurate analysis.

To address this issue, we applied a two-dimensional (2D) Hann filter to the frames, which smooths the edges by gradually tapering pixel intensities towards zero near the boundaries. The Hann window is a tapering function that prevents sharp intensity transitions at the edges, ensuring smooth transitions and reducing artifacts caused by edge discontinuities.

The Hann window is defined as follows:

\[
w(n) = 0.5 \left( 1 + \cos \left( \frac{2\pi n}{N-1} \right) \right)
\]

where \( n \) is the pixel index and \( N \) is the total number of pixels in a given image dimension. By multiplying the image by the Hann window along both the x and y axes, we ensure that the pixel intensity tapers smoothly towards zero at the edges.

The following Python code demonstrates the application of the 2D Hann filter:

\begin{verbatim}
# Smoothing of image edges with Hann windowing
hann_x = hann(I_New.shape[0])  # Create Hann window for x-dimension
hann_y = hann(I_New.shape[1])  # Create Hann window for y-dimension
hann_window = np.outer(hann_x, hann_y)  # Generate 2D Hann window
I_New = I_New * hann_window  # Apply Hann window to the image
\end{verbatim}

This filtering process reduces the sharp edge discontinuities in the frames, minimizing high-frequency artifacts and producing cleaner results when applying the Fourier transform.

\subsection{Fourier modulus estimation and low-pass filtering}
After applying the Hann filter to reduce high-frequency noise, we computed the 2D Fourier transform of each frame and calculated its modulus. The mean square of these transformed frames was computed. Then, the background offset was subtracted from the processed 2D array. Finally, we calculated the square root to recover the Fourier transform modulus of the object.

It is important to note that the diffuser used in our setup has finite dimensions. This results in a limited number of diffuser realizations, meaning that the relationship:

\[
\frac{1}{N}\sum_{i=1}^{N} \left| \mathcal{F}\{S_i\} \right|^2 \neq \left| \mathcal{F}\{P\} \right|^2
\]
where \( S_i \) represents the speckle point spread function (PSF) of the \(i\)-th frame, and \( P \) is the PSF without a diffuser. This inequality indicates that the summation of the Fourier transform moduli over multiple frames does not fully converge to the PSF of the undisturbed object, primarily due to the limited number of diffuser realizations. Consequently, residual artifacts from the diffuser may persist in the reconstructed image. These artifacts typically appear as rapidly varying fluctuations in the recovered Fourier modulus.

To mitigate these residual artifacts, we convolve the resulting signal with a narrow 2D Gaussian filter, which effectively removes speckle effects without significantly altering the squared Fourier modulus of the object, as the object typically lacks the high frequencies present in the speckles.

In Section 4.2 of the \textit{Supplementary Information}, we present a numerical analysis of the influence of such low-pass filtering on the estimation of the Fourier modulus of the direct image.

\subsection{Phase retrieval algorithm }

Finally, we applied a phase retrieval algorithm to the prepared Fourier amplitude, utilizing the combination of hybrid input-output (HIO) and the error-reduction (ER) algorithm \cite{supp:Fienup1982}. This allowed us to reconstruct the object.

The phase retrieval process was implemented using the following Python code. The function below runs multiple trials of the phase retrieval process with various settings, including the use of a GPU for acceleration:

\begin{verbatim}
def run_PR(Fixed_Amp, Ptrial=10, Method='HIO', iterations=30, beta_max=3,
beta_min=0.8, Beta_step=-0.01, useGPU=False):
    ReconstructionResultStore = []

    for k in range(Ptrial):
        if Method == 'HIO':
            print('HIO Method')
            Reconstruct_Field, recons_err, recons_err2 
            = HIO_BasicPhaseRetrieval(
                np.fft.ifftshift(Fixed_Amp), beta_max, Beta_step, 
                beta_min, iterations,
                np.random.rand(np.shape(Fixed_Amp)[0],
                np.shape(Fixed_Amp)[1]), useGPU)

        Reconstruct_Image_Amp = np.abs(Reconstruct_Field)
        plt.imshow(Reconstruct_Image_Amp)
        plt.colorbar()
        plt.show()

        ReconstructionResultStore.append(Reconstruct_Image_Amp)
        print(f'End of No. {k} trial\n\n')

    return ReconstructionResultStore
\end{verbatim}

\begin{verbatim}
def HIO_BasicPhaseRetrieval(sautocorr_temp, beta_start, beta_step,
beta_stop, N_iter, init_guess, useGPU=False):
    if useGPU:
        sautocorr = torch.from_numpy(sautocorr_temp).cuda()
        g1 = torch.from_numpy(init_guess).cuda()

        BETAS = torch.arange(beta_start, beta_stop-beta_step, beta_step)
        with torch.no_grad():
            for beta in BETAS:
                for _ in range(N_iter):
                    G_uv = torch.fft.fft2(g1)
                    g1_tag = torch.real(torch.fft.ifft2(sautocorr * G_uv
                    / torch.abs(G_uv)))
                    g1 = g1_tag * ((g1_tag >= 0).int())
                    + ((g1_tag < 0).int()) 
                    * (g1 - beta * g1_tag)

            for _ in range(N_iter):
                G_uv = torch.fft.fft2(g1)
                g1_tag = torch.real(torch.fft.ifft2(sautocorr * G_uv 
                / torch.abs(G_uv)))
                g1 = g1_tag * ((g1_tag >= 0).int())

        recons_err = torch.mean(torch.mean(
        (torch.abs(torch.fft.fft2(g1)) - sautocorr) ** 2))
        recons_err2 = torch.sqrt(torch.mean(torch.mean(
        (torch.abs(torch.fft.fft2(g1)) ** 2 - sautocorr ** 2) ** 2)))

        g1 = g1.cpu().numpy()
        recons_err = recons_err.cpu().numpy()
        recons_err2 = recons_err2.cpu().numpy()
    else:
        sautocorr = sautocorr_temp
        g1 = init_guess

        BETAS = np.arange(beta_start, beta_stop - beta_step, 
        beta_step)
        for beta in BETAS:
            for _ in range(N_iter):
                G_uv = np.fft.fft2(g1)
                g1_tag = np.real(np.fft.ifft2(sautocorr 
                * G_uv / np.abs(G_uv)))
                g1 = g1_tag * (g1_tag >= 0).astype(int) 
                + (g1_tag < 0).astype(int) 
                * (g1 - beta * g1_tag)

        recons_err = np.mean(np.mean((np.abs(np.fft.fft2(g1)) 
        - sautocorr) ** 2))
        recons_err2 = np.sqrt(np.mean(np.mean(
        (np.abs(np.fft.fft2(g1)) ** 2 - sautocorr ** 2) ** 2)))

    return [g1, recons_err, recons_err2]
\end{verbatim}

\section{Simulations}\label{Simulations}

\subsection{Photon Distribution Across Various Frame Number}

\begin{figure}[h!]
    \centering
    \includegraphics[width=0.7\textwidth]{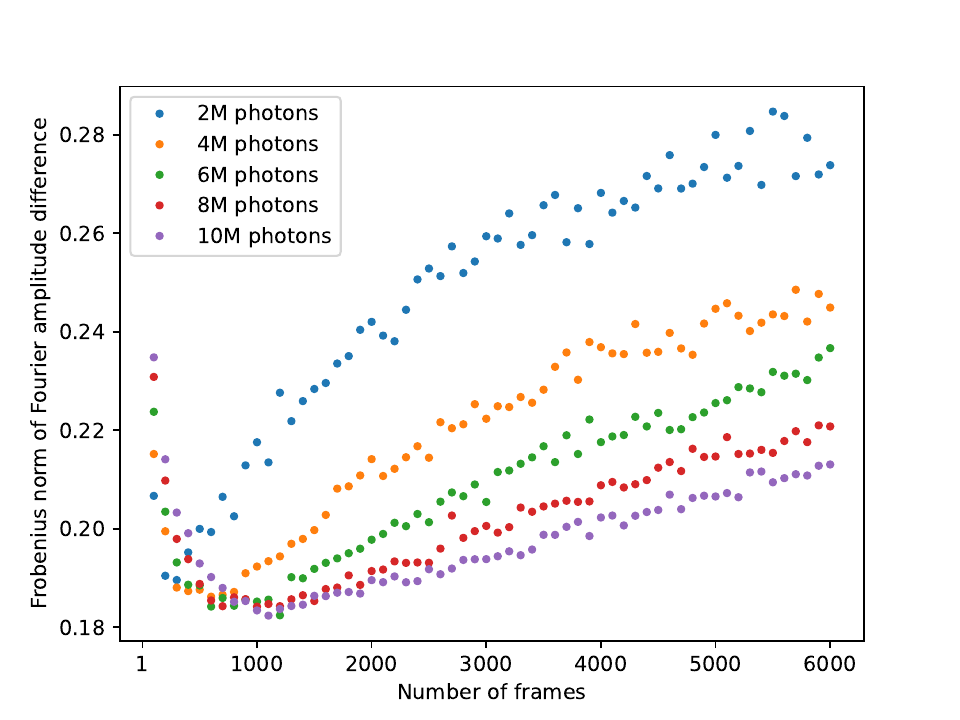} 
    \caption{The impact of distributing a constant number of photons across varying frame numbers on the accuracy of the reconstructed image. This analysis illustrates a comparison of simulation results for different photon numbers (2, 4, 6, 8, and 10 million) distributed across 10 thousand camera pixels. The X-axis represents the number of frames used in the simulations, while the Y-axis shows the corresponding error values. }
    \label{sim1}
\end{figure}

We simulated a constant number of photons, distributing them across varying numbers of camera frames. Each frame was simulated with a different realization of the diffuser. For each frame, we calculated the modulus of the Fourier transform, averaged with root mean square (RMS) these moduli, and then compared the result to the Fourier transform modulus of a direct image without a diffuser, taken with a large number of photons.

Note that no additional filtering is applied to the simulated data, in contrast to the processing used for experimental data (see Section \ref{Data}). In particular, we do not remove the Gaussian envelope from individual frames, since the simulated data do not contain such an envelope, unlike the experimental measurements, and therefore this step is not required (see Section \ref{Data}). Similarly, we do not apply Hann filtering, since the simulated data are inherently periodic and this step is not necessary, whereas for experimental data non-periodicity makes Hann windowing important to avoid edge effects (see Section \ref{Data}). Moreover, we do not apply smoothing of the recovered Fourier modulus using a convolution with a narrow Gaussian filter (i.e., low-pass filtering of the reconstructed object spectrum), as this operation significantly improves reconstruction quality, especially in the case of an insufficient number of diffuser realizations, in both experiments and simulations, and would otherwise bias the results that we aim to present in this section.

To quantitatively assess the difference between the two Fourier transform results, we normalized both matrices using the Frobenius norm, subtracted one from the other, and calculated the Frobenius norm of the resulting difference matrix:

\[
\text{Error} = \left\|\frac{F_{\text{RMS}}}{\|F_{\text{RMS}}\|_F} - \frac{F_{\text{direct}}}{\|F_{\text{direct}}\|_F}\right\|_F
\]

where \( F_{\text{RMS}} \) is the Fourier transform modulus of the averaged frames and \( F_{\text{direct}} \) is the Fourier transform modulus of the direct image. Identical matrices would result in a norm value of 0, indicating perfect similarity.

The simulation results, shown in Figure \ref{sim1}, compare the error values for 2, 4, 6, 8, and 10 million photons distributed across 10 thousand simulated camera pixels. The X-axis represents the number of frames, while the Y-axis indicates the corresponding error values.

As shown in Fig. \ref{sim1}, the results indicate the existence of an optimal number of frames for a given total number of photons. For a large number of frames, increasing the number of frames at fixed photon budget reduces the signal-to-noise ratio per frame, and the recovered Fourier modulus of the object becomes suboptimal. In contrast, for a small number of frames the number of independent diffuser realizations is limited, and the algorithm does not converge to the correct Fourier modulus estimate.

Improved reconstruction quality for both low frame numbers and a limited number of diffuser realizations could, in principle, be achieved by smoothing the recovered Fourier modulus via convolution with a narrow Gaussian filter. However, in these simulations we do not apply such filtering, in order to avoid artificially improving the reconstruction quality at low numbers of diffuser realizations and to preserve a direct comparison of the raw algorithm output.

The results of this simulation can be interpreted as indicating the existence of an optimal excitation pump intensity for fluorescent dyes in our method. One can assume a model in which each dye molecule can emit a finite number of photons before photobleaching occurs. By controlling the excitation pump power, one can therefore determine whether these photons are distributed over a smaller or larger number of frames prior to photobleaching.

It is worth noting that a high value of the $\text{Error}$ function does not necessarily imply that the object image cannot be recovered by the phase retrieval algorithm. Although the algorithm is highly nonlinear and sensitive to perturbations, a large $\text{Error}$ value does not directly translate into failure of object reconstruction.

The value of the $\text{Error}$ function depends directly on the object itself, and there is no specific threshold above which the phase retrieval algorithm ceases to work. We use the $\text{Error}$ function only as a quantitative measure of reconstruction quality. Figure \ref{sim1-1} compares the Fourier moduli obtained for $\text{Error}=0.183$ (Fig. \ref{sim1-1}c) and $\text{Error}=0.248$ (Fig. \ref{sim1-1}e), together with the corresponding reconstructed object images (Figs. \ref{sim1-1}d and \ref{sim1-1}f), against the direct image of the object (Fig. \ref{sim1-1}b) and its Fourier modulus (Fig. \ref{sim1-1}a).

\begin{figure}[h!]
    \centering
    \includegraphics[width=\textwidth]{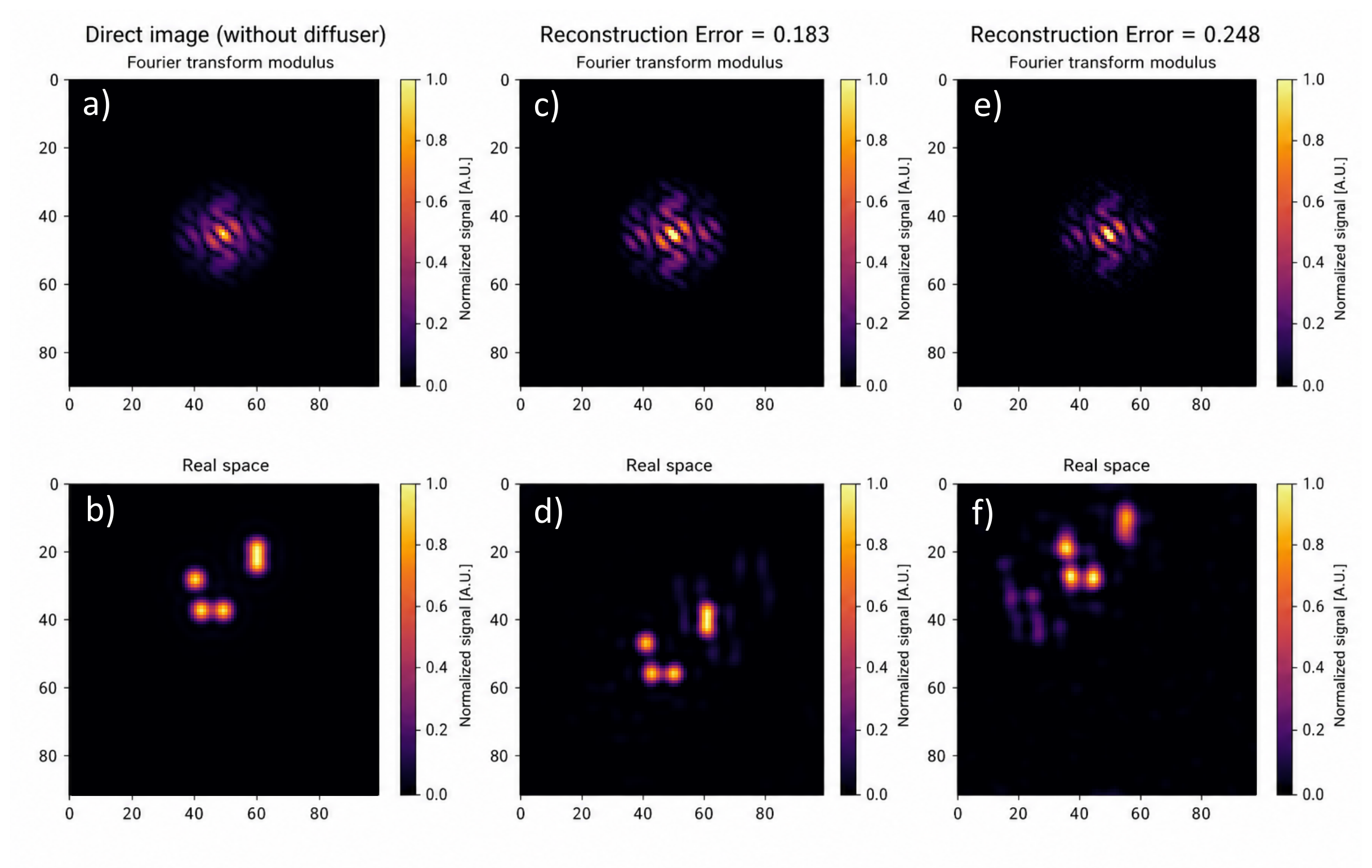} 
    \caption{Comparison of reconstruction results for different values of the $\text{Error}$ function. (a) Fourier modulus of the direct image.  (b) Direct image of the object. (c) Fourier modulus for $\text{Error}=0.183$. (d) Object image recovered using the phase retrieval algorithm from the Fourier modulus shown in (c). (e) Fourier modulus for $\text{Error}=0.248$. (f) Object image recovered from the Fourier modulus shown in (e) using the phase retrieval algorithm.}
    \label{sim1-1}
\end{figure}

\subsection{Simulation with Limited Diffuser Realizations}

\begin{figure}[h!]
    \centering
    \includegraphics[width=0.6\textwidth]{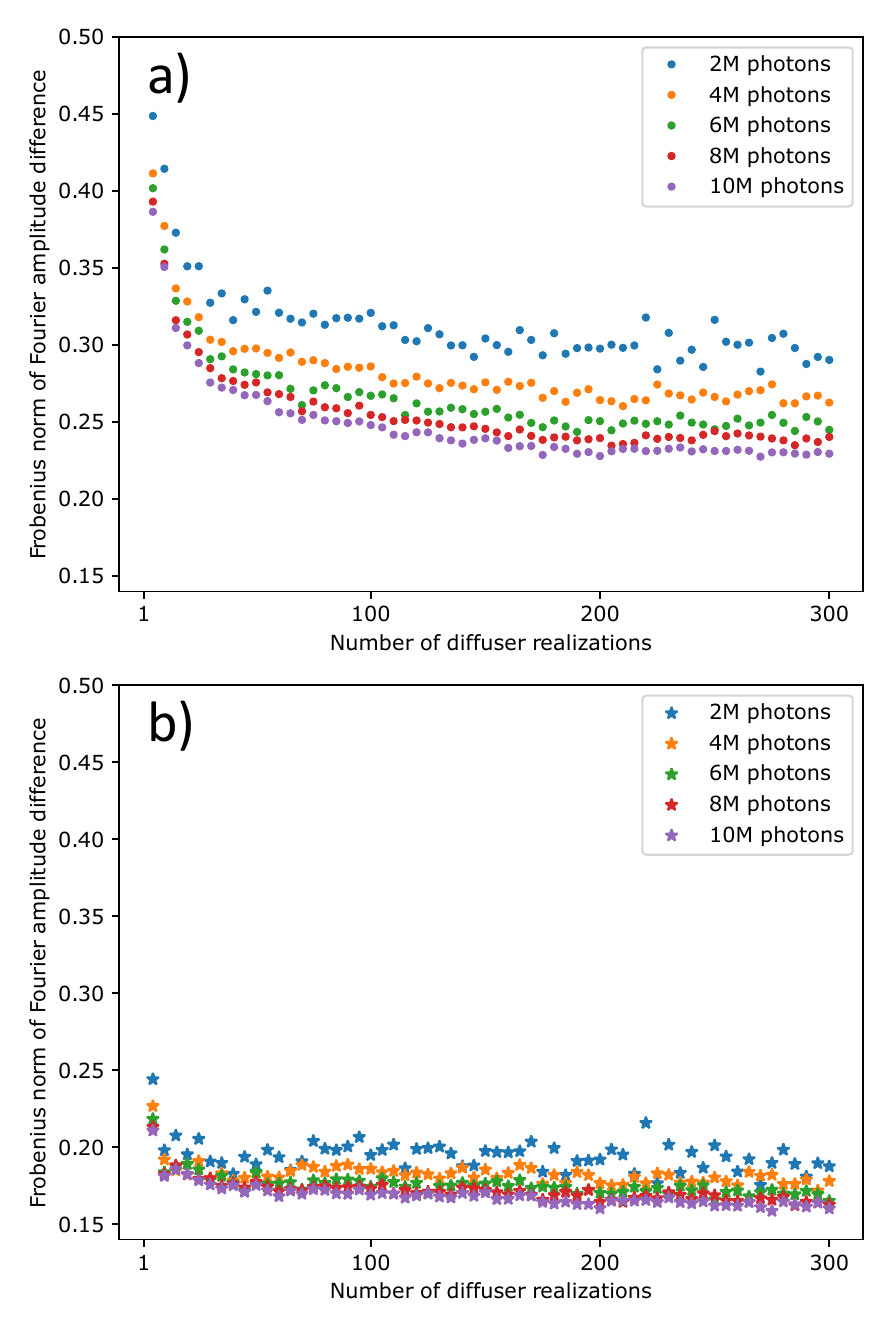} 
    \caption{Comparison of simulation results with a limited number of diffuser realizations for different photon counts (2, 4, 6, 8, and 10 million) distributed across 10 thousand camera pixels. (a) Results obtained without Gaussian smoothing. (b) Results obtained after smoothing the recovered Fourier modulus by convolution with a 2D Gaussian kernel $(\sigma = 0.8)$. The X-axis represents the number of diffuser realizations, while the Y-axis shows the corresponding error values.}
    \label{sim2}
\end{figure}

\begin{figure}[h!]
    \centering
    \includegraphics[width=\textwidth]{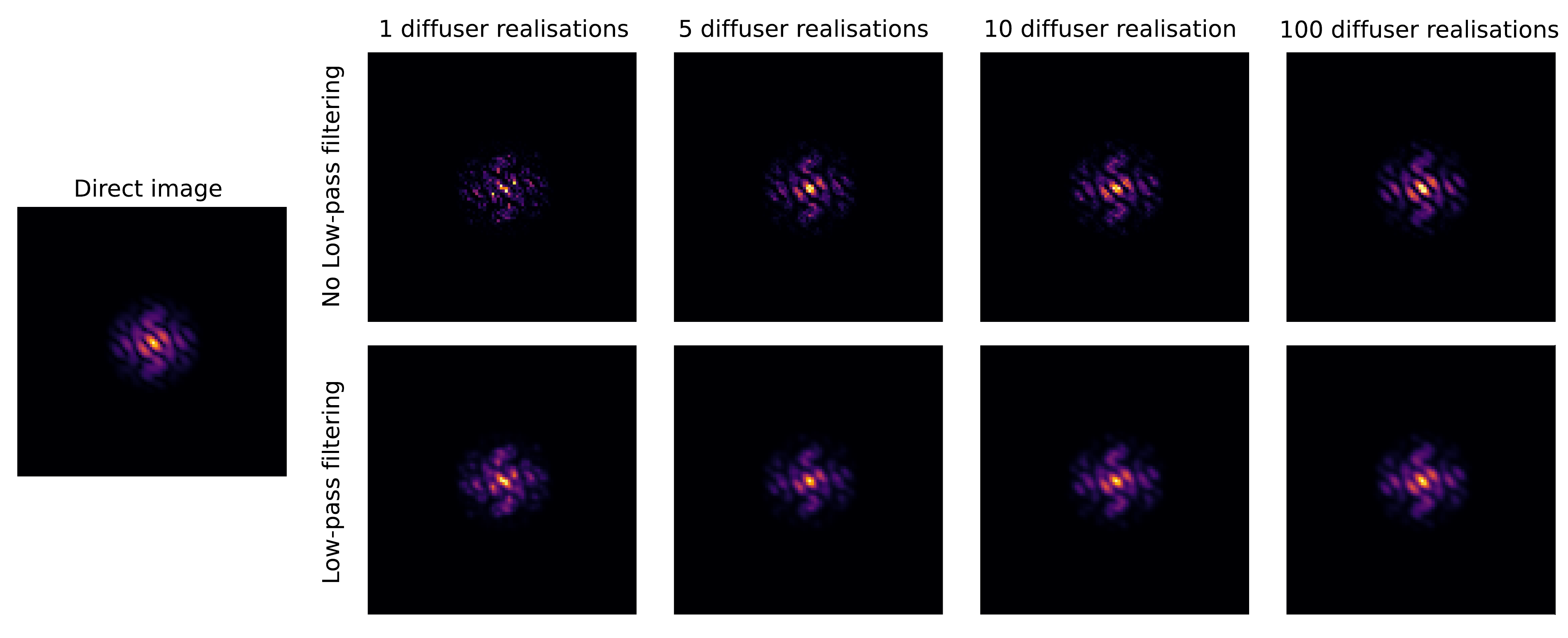} 
    \caption{Comparison of recovered Fourier moduli for a limited number of diffuser realizations -- simulation results. The Fourier modulus of the direct image is shown in the leftmost column. The remaining columns present Fourier moduli recovered with our method for 1, 5, 10, and 100 diffuser realizations using 10 million photons distributed across 10 thousand frames. The first row shows the recovered Fourier moduli without Gaussian smoothing. The second row shows the same Fourier moduli after convolution with a 2D Gaussian kernel (\(\sigma = 0.8\)).
}
    \label{sim2-b}
\end{figure}

In this simulation, we maintained a constant number of camera frames while limiting the number of possible diffuser realizations. Consequently, some diffuser realizations were used multiple times. Similar to the previous section, we calculated the modulus of the Fourier transform for each frame, averaged these moduli with RMS, and compared the result to the Fourier transform modulus of a direct image obtained without a diffuser, using a large number of photons. The comparison was again quantified using the error metric:

\[
\text{Error} = \left\|\frac{F_{\text{RMS}}}{\|F_{\text{RMS}}\|_F} - \frac{F_{\text{direct}}}{\|F_{\text{direct}}\|_F}\right\|_F
\]

As in the previous section, no Gaussian envelope removal or Hann windowing was applied. In this simulation, however, we additionally investigated the effect of smoothing the recovered Fourier modulus by convolution with a narrow Gaussian function. The reconstruction quality was evaluated both without (Fig. \ref{sim2}a) and with (Fig. \ref{sim2}b) Gaussian smoothing, implemented by convolving the recovered Fourier modulus with a 2D Gaussian kernel ($\sigma = 0.8$).

Figures \ref{sim2}a and \ref{sim2}b show the results of this simulation for 2, 4, 6, 8, and 10 million photons distributed across 10 thousand camera pixels. The X-axis represents the number of diffuser realizations, while the Y-axis shows the corresponding error values.

The results show that smoothing the recovered Fourier modulus significantly improves reconstruction quality (Fig. \ref{sim2}b). Increasing the total number of detected photons leads to lower error values and improved reconstruction quality. The results also demonstrate that a larger number of diffuser realizations further improves the accuracy of the estimated Fourier modulus.

Figure \ref{sim2-b} compares the recovered Fourier moduli obtained with our method for 1, 5, 10, and 100 diffuser realizations using 10 million photons distributed across 10 thousand frames. The results are shown together with the Fourier modulus of the corresponding direct image. The first row shows the recovered Fourier moduli without Gaussian smoothing, while the second row shows the same results after convolution with a Gaussian kernel $\sigma=0.8$.

It is worth noting that even when the diffuser remains unchanged and only a single diffuser realization is present, the recovered Fourier modulus after Gaussian smoothing remains similar to the Fourier modulus of the direct image. In this simulation, 10 million photons were distributed across 10 thousand frames, all recorded through the same diffuser realization. Although such signal splitting is inefficient and a single long-exposure frame would be preferable for a static diffuser, in practice the diffuser dynamics may be unknown. One may therefore choose to acquire short-exposure frames assuming that the diffuser is changing, while in reality it remains stationary. This example shows that the method still provides a meaningful estimate of the Fourier modulus even in such a non-optimal scenario.

\section{Imaging with photon-tagging cameras}\label{piccolo}

\begin{figure}[h!]
    \centering
    \includegraphics[width=\textwidth]{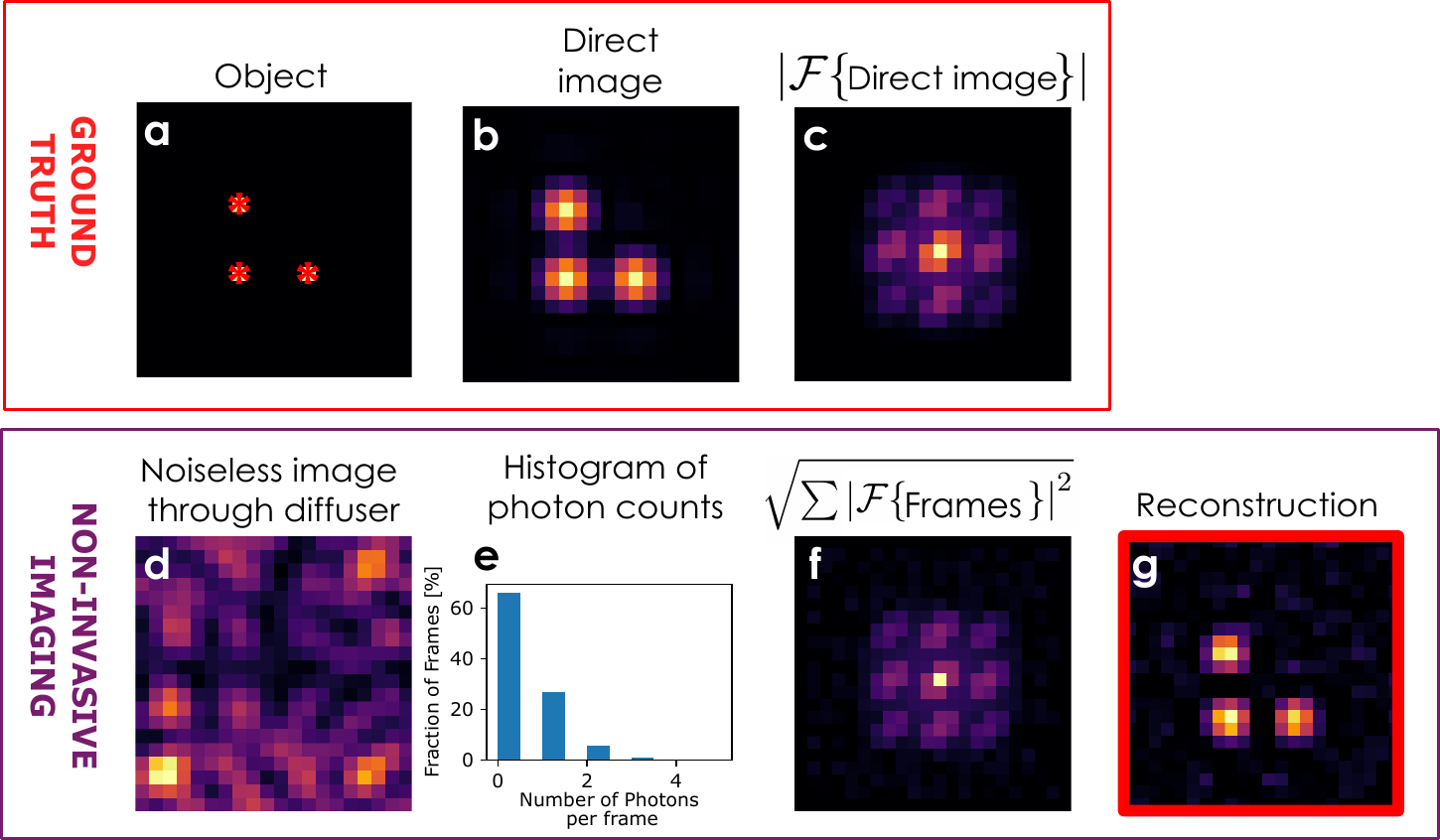} 
    \caption{Simulation of imaging with a photon-tagging camera under extremely photon-starved conditions. (a) Simulated object consisting of three emitters. (b) Image of the object recorded by a finite-resolution imaging system without a diffuser and recorded on a ($20 \times 20$) pixel detector. (c) Fourier modulus of the image shown in (b). (d) Example speckle pattern generated by imaging the object through a static diffuser. (e) Histogram of detected photon counts per frame for a simulation consisting of 600,830,000 frames and a total of 251,418,252 detected photons, corresponding to an average of 0.418 photons per frame. (f) Fourier modulus recovered by RMS averaging of the Fourier transform moduli calculated on each simulated frame. (g) Object image reconstructed from the recovered Fourier modulus using the phase retrieval algorithm.}
    \label{sim_piccolo}
\end{figure}

In this section, we discuss the potential use of fast photon-tagging cameras in our experiment. To push the temporal performance of the method towards its fundamental limits, one could combine a picosecond pulsed laser with single-photon detector arrays such as an intensified Tpx3Cam camera \cite{supp:Nomerotski2023, supp:Ianzano2020} or SPAD arrays \cite{supp:Bruschini2019}.

Such cameras are capable of recording the arrival times of individual photons with temporal resolutions down to hundreds of picoseconds \cite{supp:Bruschini2019}. Combined with a pulsed laser source, this approach could extend our method towards the fundamental limit imposed by the dynamics of the scattering medium. In such a configuration, photons would be collected within a time window comparable to the fluorescence lifetime of the dye after each laser pulse, and the only assumption would be that the scattering medium remains unchanged during the time that the molecules remain in the excited state, which is typically on the order of a few nanoseconds.

A single laser pulse can contain a large amount of energy and excite many fluorophores simultaneously, while the time between pulses allows the fluorophores to recover before the next excitation event. As a result, a large number of emitters could contribute fluorescence photons within each acquisition window while reducing the effects of photobleaching.

Figure \ref{sim_piccolo} presents a simulation of such an enhanced implementation of our method under extremely photon-starved conditions using a very small detector consisting of only ($20 \times 20$) pixels, simulating a photon-tagging array such as a SPAD array \cite{supp:Bruschini2019}. Due to the limited number of pixels, we considered a simple object composed of only three emitters. Figure \ref{sim_piccolo}a shows the object to be imaged. Figure \ref{sim_piccolo}b shows the image of this object recorded by a finite-resolution imaging system without a diffuser. This panel therefore represents the image that would be obtained in a conventional wide-field microscope. Figure \ref{sim_piccolo}c shows the Fourier modulus of the image from Fig. \ref{sim_piccolo}b.

If the object from Fig. \ref{sim_piccolo}a were imaged through a static diffuser, the resulting speckle pattern recorded by the camera would look like the example shown in Fig. \ref{sim_piccolo}d. For ultrafast acquisition windows with durations of only a few nanoseconds, the detector may collect very few photons per frame. To investigate this regime, we simulated an extreme case consisting of 600,830,000 frames containing a total of 251,418,252 detected photons. This corresponds to an average of only 0.418 photons per frame. Accumulating more than 600 million frames may seem impractical at first glance. However, if one frame is acquired following each laser pulse, a 20 MHz pulsed laser would produce 20 million frames per second. Under these conditions, 600 million frames could be collected in only about 30 seconds.

Figure \ref{sim_piccolo}e shows the histogram of photon counts per frame. Most simulated frames (66.0\%) contain no photons at all. Approximately 27.0\% of frames contain a single photon. However, a single photon does not carry information about the sample structure in our method, since the relevant information is encoded in photon-pair correlations. The modulus of the Fourier transform of a single-photon event, represented by a Dirac delta function, is constant. Only 5.7\% of frames contain two photons and therefore carry structural information about the object, while only 1.3\% of frames contain more than two photons.

Even in this extreme scenario, averaging the Fourier transform moduli of all 600,830,000 frames using RMS produced a meaningful estimate of the Fourier modulus of the direct image, which closely resembles Fig. \ref{sim_piccolo}c. The recovered Fourier modulus shown in Fig. \ref{sim_piccolo}f was used as the input for the phase retrieval algorithm, yielding the reconstructed image shown in Fig. \ref{sim_piccolo}g. The reconstructed image closely resembles the direct wide-field image shown in Fig. \ref{sim_piccolo}b.

SPAD detector arrays implemented in CMOS technology are rapidly advancing. Currently, SPAD arrays capable of recording photon arrival times are available with sizes on the order of ($10^3$) pixels, typically consisting of a few tens of pixels in each dimension \cite{supp:Bruschini2019}. At the same time, megapixel-scale SPAD arrays are also available, although these devices typically operate in a gated mode rather than recording the arrival time of every detected photon \cite{supp:Bruschini2019}.

\endgroup

\end{document}